**Predicted decline in common bird and butterfly species despite conservation policy scenarios in Europe**

Stanislas Rigal[1,2], Maxime Lenormand[1], Léa Tardieu[1,3], Hany Alonso[4], Ainārs Auniņš[5,6], Tristan R. M. Bakx[7], Michiel P. Boom[8], Mattia Brambilla[9], Lluis Brotons[7,10,11], Tomasz Chodkiewicz[12,13], Benoît Fontaine[14], Zdenek Faltynek Fric[15], Anna Gamero[16], Sergi Herrando[7,10,17], John Atle Kålås[18], Johannes Kamp[19,20], Petras Kurlavičius[21,22], Mikko Kuussaari[23], Aleksi Lehikoinen[24], Dirk Maes[25,26], Xavier Mestdagh[27], Martin Musche[28], Ingar Jostein Øien[29], Lars B. Pettersson[30], Jiří Reif[31,32], Johannes Rüdisser[33], Martina Šašić[34], Reto Schmucki[35], Josef Settele[28], Constanti Stefanescu[36], Bård Gunnar Stokke[18], Nicolas Strebel[37], Nicolas Titeux[27], Sven Trautmann[20], Chris Van Swaay[38], Sandra Luque[1]

[1] INRAE, National Research Institute on Agriculture, Food & the Environment, TETIS Unit, Maison de la télédétection, 34090 Montpellier, France
[2] CESAB, Foundation for Research on Biodiversity, Montpellier, France
[3] CIRED, Ecole des Ponts ParisTech, AgroParisTech, université Paris-Saclay, Cirad, CNRS, EHESS, Nogent-sur-Marne, France
[4] SPEA - Portuguese Society for the Study of Birds
[5] Department of Ecology, Faculty of Medicine and Life Sciences, University of Latvia, Latvia
[6] Latvian Ornithological Society, Latvia
[7] CREAF, 08193, Cerdanyola del Vallès, Spain
[8] Sovon, Dutch Centre for Field Ornithology, Nijmegen, The Netherlands
[9] Department of Environmental Science and Policy, Milan University, Italy
[10] Catalan Ornithological Institute, Natural Science Museum of Barcelona, Barcelona, Spain
[11] CSIC, 08193, Cerdanyola del Vallès, Spain
[12] Museum & Institute of Zoology, Polish Academy of Sciences, Warszawa, Poland
[13] Polish Society for the Protection of Birds (OTOP BirdLife Poland), Warszawa, Poland
[14] Patrinat & UMR7204 Centre d'Écologie et des Sciences de la Conservation (CESCO), MNHN: Muséum national d'Histoire naturelle (MNHN)-CNRS-SU, Paris 75005, France
[15] Biology Centre CAS, Branisovska 31, CZ-37005 Ceske Budejovice, Czechia,
[16] Czech Society for Ornithology/Pan-European Common Bird Monitoring Scheme, Prague, Czechia
[17] European Bird Census Council (EBCC), Prague, Czech Republic
[18] Norwegian Institute for Nature Research, 7485 Trondheim, Norway
[19] University of Göttingen, Department of Conservation Biology, Bürgerstr. 50, 37073 Göttingen, Germany
[20] Dachverband Deutscher Avifaunisten (DDA), An den Speichern 2, 48157 Münster, Germany
[21] Vytautas Magnus University, Kaunas, Lithuania
[22] Lithuanian Ornithological Society (LOD), Vilnius, Lithuania
[23] Finnish Environment Institute, Helsinki, Finland
[24] Finnish Museum of Natural History, University of Helsinki, Finland
[25] Research Institute for Nature and Forest (INBO), Herman Teirlinckgebouw, Havenlaan 88 PO Box 73, B-1000 Brussels, Belgium
[26] Radboud Institute for Biological and Environmental Sciences (RIBES), Radboud University, PO Box 9010, NL-6500 GL Nijmegen, the Netherlands

[27] Luxembourg Institute of Science and Technology, Environmental Sensing and Modelling unit, Biodiversity Monitoring and Assessment group, 41 rue du Brill, L-4422 Belvaux, Luxembourg
[28] Helmholtz Centre for Environmental Research - UFZ, Department of Conservation Biology & Social-Ecological Systems, Theodor-Lieser-Str. 4, 06120 Halle, Germany
[29] BirdLife Norway Sandgata 30B NO-7012 Trondheim, Norway
[30] Biodiversity and Evolution, Department of Biology, Lund University, Sweden
[31] Institute for Environmental Studies, Faculty of Science, Charles University, Prague, Czechia
[32] Department of Zoology, Faculty of Science, Palacký University, Olomouc, Czechia
[33] Department of Ecology, Universität Innsbruck, Innsbruck, Austria
[34] Zoological department, Croatian Natural History Museum Demetrova 1 HR-10000 Zagreb
[35] UK Centre for Ecology & Hydrology, Maclean Building Benson lane, Wallingford, OX10 8BB, UK
[36] Natural Sciences Museum of Granollers, BiBio Research Group, Granollers, Spain
[37] Swiss Ornithological Institute, Seerose 1, 6204 Sempach, Switzerland
[38] Vlinderstichting, P.O. Box 506, NL-6700 AM Wageningen, Netherlands

## Abstract

In response to increasing threats to biodiversity, conservation objectives have been set to halt biodiversity decline by reducing direct anthropogenic drivers. However, the potential effects of these objectives on common species remain rarely studied. We analyse the effect of a range of drivers related to climate, land use and land use intensity, on 265 common bird and 144 common butterfly species from more than 20,000 sites between 2000 and 2021 across 26 European countries. We use land-use and land-use intensity scenarios produced previously using the IPBES Nature Futures Framework, and climate change scenarios in order to project biodiversity drivers in Europe up to 2050. We translate these driver changes into abundance variations for common bird and butterfly species, and for multi-species indicators used to monitor common biodiversity status in Europe. The projected trends relatively improve, while still declining for birds, notably farmland species, under the scenarios meeting conservation objectives, with few effects on butterflies. No scenario shows a stop or a reversal in the average decline in abundance of bird and butterfly species. Our results therefore question the common biodiversity future under current conservation policies and highlight the need for other anticipatory frameworks, not implicitly based on a growing need for natural resources.

## 1. Introduction

To halt species abundance declines driven by the consequences of human activities[1–3], international conservation efforts have proposed tools, such as the establishment of protected areas and more appropriate management of ecosystems[4]. These tools have been translated by international conservation bodies into protection and restoration measures and goals, such as the European Union (EU) Birds (2009/147/EC) and Habitat (92/43/EEC) directives or 30% target for area-based nature conservation in the Kunming-Montreal Global Biodiversity Framework[5,6].

While conservation measures such as protected areas are an essential conservation tool in reducing these direct drivers on biodiversity[7], their effect alone cannot reverse the decline in common biodiversity[8–10]. Common species, abundant through habitat generalism or a wide distribution of their preferred habitat, are not particularly targeted by those conservation actions[11,12]. Yet, they are often directly impacted by human activity for instance by agriculture, forestry or urbanisation[13] and the depletion of populations observed in several taxa is a first step towards the status of threatened species[14,15]. This massive decline in abundance has repercussions for the functioning of ecosystems[16]. Conservation objectives must, however, comply with society's needs that influence direct drivers on biodiversity through material and land-use requirements[17,18]. These requirements are modulated by socio-political choices which include conservation policies (e.g. the establishment of new protected areas) and production methods (e.g. sustainable agriculture and forestry)[19]. Increased societal needs, expected in a context of economic and demographic growth[20], have so far been matched by intensifying land use (e.g. industrialisation of agriculture and forestry, densification) or by land consumption (e.g. urban sprawl, expansion of agricultural land), affecting common biodiversity[21,22].

Forward-looking approaches have emerged to better anticipate changes in societal needs (e.g. energy, material, food and land) and their potential effects on the achievement of conservation targets. The shared socio-economic pathways (SSPs) are a common scenario framework to depict changes impulsed by society under different assumptions on socio-political choices[23]. These different assumptions may be nuanced by making societal preferences and values more explicit. The IPBES Nature Future Framework (NFF) provides an operational tool allowing biodiversity positive futures under different value settings to be identified[24]. The wide range in expected land-use and land-use intensity changes across scenarios[19,25] is therefore likely to influence the future dynamics of common biodiversity. In particular, the trade-off between an increase in natural lands combined with higher intensive practices in productive areas versus landscapes where human activities and non-human dynamics coexist is expected to result in contrasted dynamics for species[26]. The majority of common birds, for instance, depend on open land vs. forests[27] or are sensitive to agricultural practices[22]. However, biodiversity in forward-looking approaches remains estimated via aggregated biodiversity indices (e.g. Living Planet Index, Biodiversity Intactness Index)[28–30] which are challenging to interpret[31,32]. Additionally, the consequences for biodiversity of diverse scenarios might also be only estimated from the achievement of conservation targets[33], without explicitly accounting for common biodiversity.

Anticipating the consequences of socio-economic trajectories on common biodiversity requires the ability to translate changes in human drivers and pressures and environmental constraints driven by human activities into their effects on common species. Until recently, multi-taxa approaches of common biodiversity response to global change have mostly been restricted to meta-analyses[34], but monitoring data enabling direct multi-taxa analyses are increasingly becoming available[35–37]. Multi-taxa approaches are essential for a deeper understanding of how the composition of biotic communities and diverse taxa can be affected by a given set of anthropogenic pressures[3,38]. Large-scale monitoring of common species in Europe particularly focuses on birds and butterflies[39–41] with long term national programs contributing to the PanEuropean Common Bird Monitoring Scheme (PECBMS) and the European Butterfly Monitoring Scheme (eBMS). These data have been used to identify the turnover in common biodiversity and quantify the decline in most common species at regional scales[1,42,43] and the effect of direct drivers[22,44] and conservation measures[45] on common species. In Europe, each region faces specific levels of land-use change[46], land-use intensity such as agricultural intensification[47] or land abandonment[48], and climate change[49], which requires a spatially explicit approach to assess the effect of drivers on species[50,51]. In addition to specific responses, the use of multi-species indicators helps to reflect differences in species sensitivity, such as farmland and forest indices for birds[52] and grassland and woodland indices for butterflies[40]. Increases of these indices are also biodiversity targets in the EU Nature Restoration Regulation (NRR)[53].

To analyse whether socio-economic scenarios designed with specific conservation objectives and sustainability policies may benefit common biodiversity, we assess in this study the predicted effects on common bird and butterfly species of Europe-wide sustainability scenarios. We use data from the monitoring of common breeding birds and butterflies in Europe between 2000 and 2021 at over 20,000 sites in 26 European countries to estimate the correlative observed effect of climate change, land use change and land use intensity change on 265 species of common birds and 144 species of common butterflies. We use spatial scenarios of changes in drivers under business-as-usual and scenarios reaching several international and EU conservation objectives, in order to anticipate the effect of these changes on common biodiversity and more specifically on indicator species for forest and open habitats.

## 2. Results

### 2.1 Overall approach

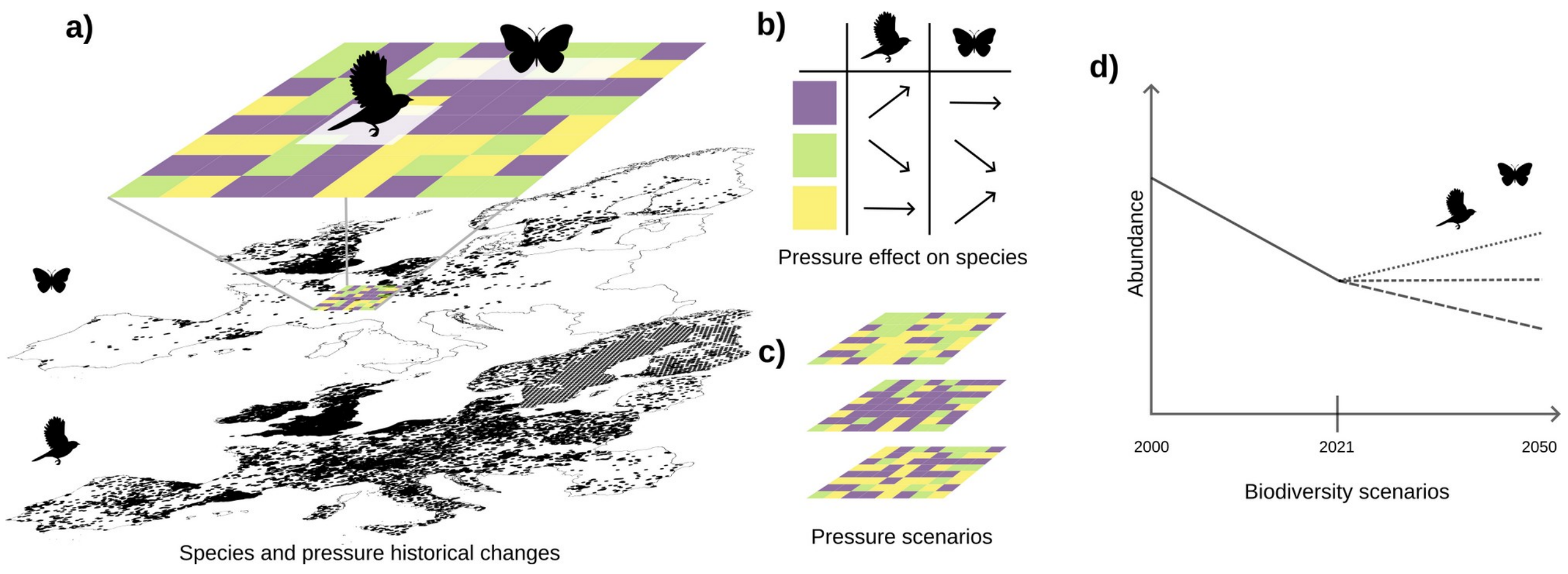

Figure 1: Workflow of the analyses. a) We combined European-scale monitoring data for common birds and butterflies with 1 × 1 km data on land use, land-use intensity and climate change. b) We estimated effects of land-use, land-use intensity and climate change on individual species trends. c) We combined land-use, land-use intensity and climate scenarios with estimated effects of direct drivers on each species to d) produce trends in multi-species indices based on birds and butterflies in each scenario.

We used monitoring data of common bird and butterfly species from 2000 to 2021 in Europe together with observed changes in climate, land use and land-use intensity over the same period to quantify the effect of these direct drivers on species trends (see Methods, Fig. 1a, b and Fig. 2). From the estimated effects of climate, land-use and land-use intensity changes on species trends and the predicted changes in these drivers up to 2050 (Fig. 1c and Fig. 3), we assessed changes in common biodiversity in terms of bird and butterfly relative abundance at the European level (Fig. 1d and Fig. 4) and by biophysical regions (Fig. 5 and 6). The scenarios of driver changes are based on the similar climatic intermediate scenario RCP 4.5 (leading to a 2°C increase in global temperature by mid-century[54]) and Land-use scenarios for 2050[55] produced in line with the IPBES promoted Nature Future Framework (NFF)[19]: one Shared Socio-economic Pathway "Taking the Green Road" (SSP1), and three NFF scenarios corresponding to major pathways along which biodiversity positive values can be implemented (Nature for Nature, Nature as Culture, Nature for Society). The NFF[56] derived scenarios extend SSP1 and provide a flexible tool designed to support the development of scenarios and models of desirable futures for people and nature. In our approach this allowed incorporating conservation and sustainability targets in line with European Union Green Deal targets and post-2020 Global Biodiversity Framework targets: reduced use of nitrogen fertiliser, increased tree cover, no-net-loss of natural areas and expanded protected areas[19].

## 2.2 Climate, land-use and land-use intensity effects on bird and butterfly trends

Concerning the effects of climate change at the European scale, we found a negative effect of spring temperature change on a majority of forest bird species (16 negative effects [-] and 13 positive effects [+], Fig. 2c and Fig. S1a, S1b) and butterfly species (83 -, 52 +, Fig. 2d). Increase in spring precipitation has a negative effect on a majority of bird species (119 -, 104 +, Fig. 2a), in particular farmland birds (25 -, 9 +, Fig. 2b) and butterfly species (81 -, 54 +) both in grassland (10 -, 2 +, Fig.

2e) and woodland species (17 -, 11 +, Fig. 2f). Increase in spring temperature variation has a positive effect on a majority of bird species (109 -, 114 +) and a contrasted effect across butterfly species (67 -, 68 +). Overall, the share of bird species negatively affected by climate variables are higher in the southern and eastern parts of Europe (Fig. S1c).

Land-use effects of urbanisation are negative on a majority of bird species (126 -, 97 +), visible both in farmland (21 -, 13 +) and forest species (22 -, 7 +) but not on butterfly species (57-, 78 +). Increases in tree cover density have a negative effect on a majority of birds (130 -, 93 +), in particular farmland species (23 -, 11 +), but a positive effect on a majority of butterfly species (61 -, 74 +). Increase in agricultural area has a negative effect on butterfly species (79 -, 56 +), in particular woodland butterflies (17 -, 11 +), whereas the effect is positive on a majority of grassland butterflies (4 -, 8 +) and farmland birds (14 -, 20 +). An increase in landscape diversity has a positive effect on a majority of butterfly species (40 -, 95 +), in particular woodland species (8 -, 20 +), but the effect is more balanced on bird species (110 -, 113 +). The negative effects of land use change are evenly spread across Europe for birds and even more visible in Eastern Europe for butterflies (Fig. S1d).

The effects of land-use intensity are particularly visible in agricultural areas, with a negative effect of high intensive farmlands across taxa relative to low intensive farmland (for birds: 123 -, 99 + in high intensive, 91 -, 132 + in low intensive; for butterflies 77 -, 58 + in high intensive, 71 -, 64 + in low intensive). This effect is not only visible in open land species (for birds: 17 -, 17 + in high intensive, 15 -, 19 + in low intensive; for butterflies 11 -, 1 + in high intensive, 7 -, 5 + in low intensive), but also in forest species (for birds: 26 -, 3 + in high intensive, 9 -, 20 + in low intensive; for butterflies 18 -, 10 + in high intensive, 16 -, 12 + in low intensive). For forest management, the difference is mostly visible for butterfly species (75 -, 59 + in highly managed forest vs. 69 -, 66 + in less intensively managed forest) and forest birds (21 -, 8 + in highly managed forest vs. 13 -, 16 + in less intensively managed forest). An increasing percentage of protected area has a positive effect on the trend of a majority of farmland bird species (13 -, 21 +) but a negative effect on a majority of forest bird species (17 -, 12 +) and butterfly species (79 -, 56 +), in particular woodland butterflies (18 -, 10 +). The negative effects of land-use intensity are evenly spread across Europe for birds and more visible in Eastern Europe for butterflies.

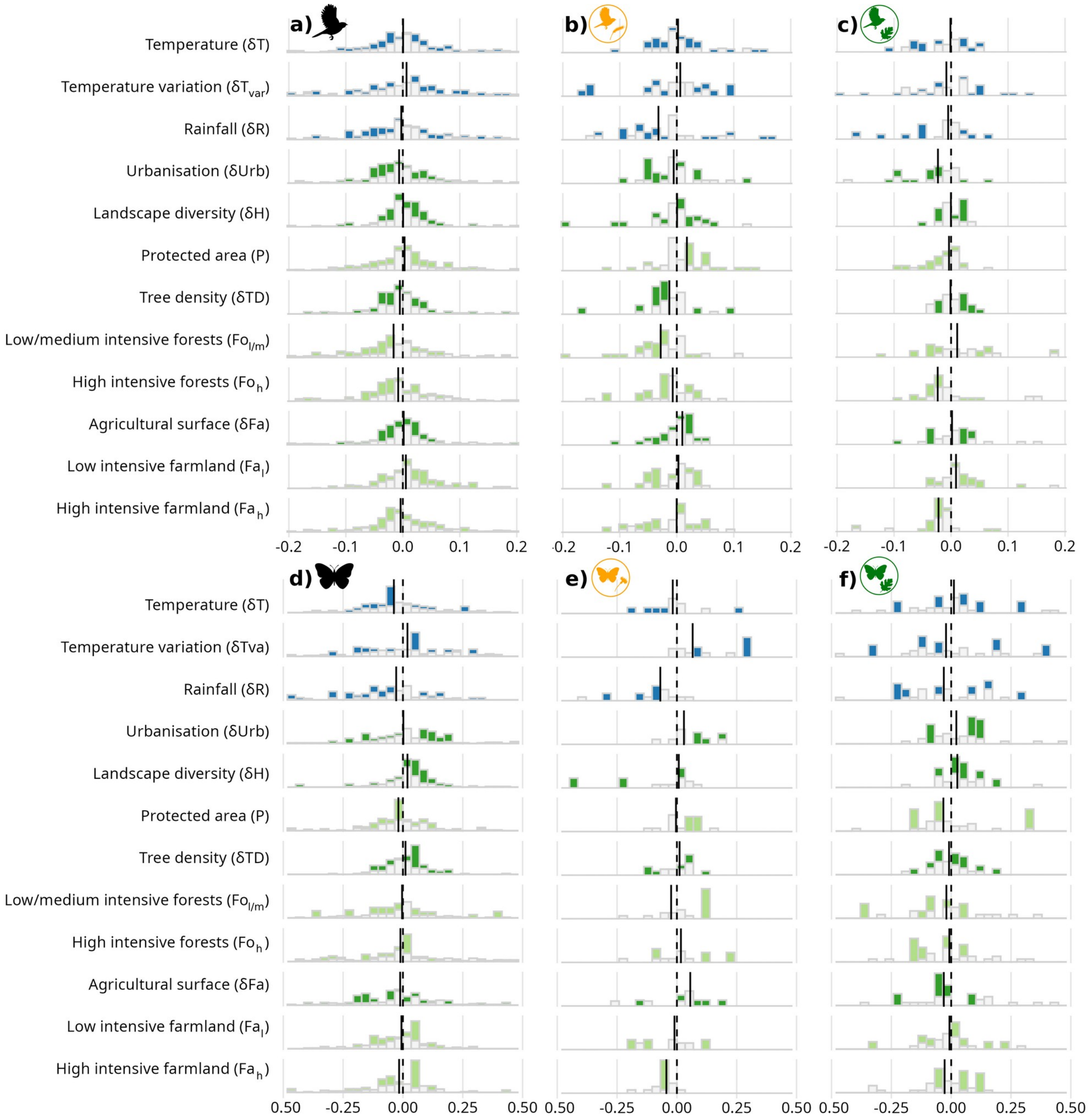

Figure 2: Effects of change in climate variables, change in land-use variables and of land-use intensity variables on temporal trends at the European level for a) Common bird species (n = 223), b) farmland bird species (n = 34), c) forest bird species (n = 19), d) Common butterfly species (n = 135), e) grassland butterfly species (n = 12) and f) woodland butterfly species (n = 28). Non-significant estimates are in grey, and significant estimates in blue for climate, in dark green for land use and in light green for land-use intensity. Values on the x-axis correspond to the values of α coefficients in Eq. 5. Bar height corresponds to the number of species, y-axes are standardised for each group of species.

## 2.3 Effects of conservation-targeted scenarios on common biodiversity

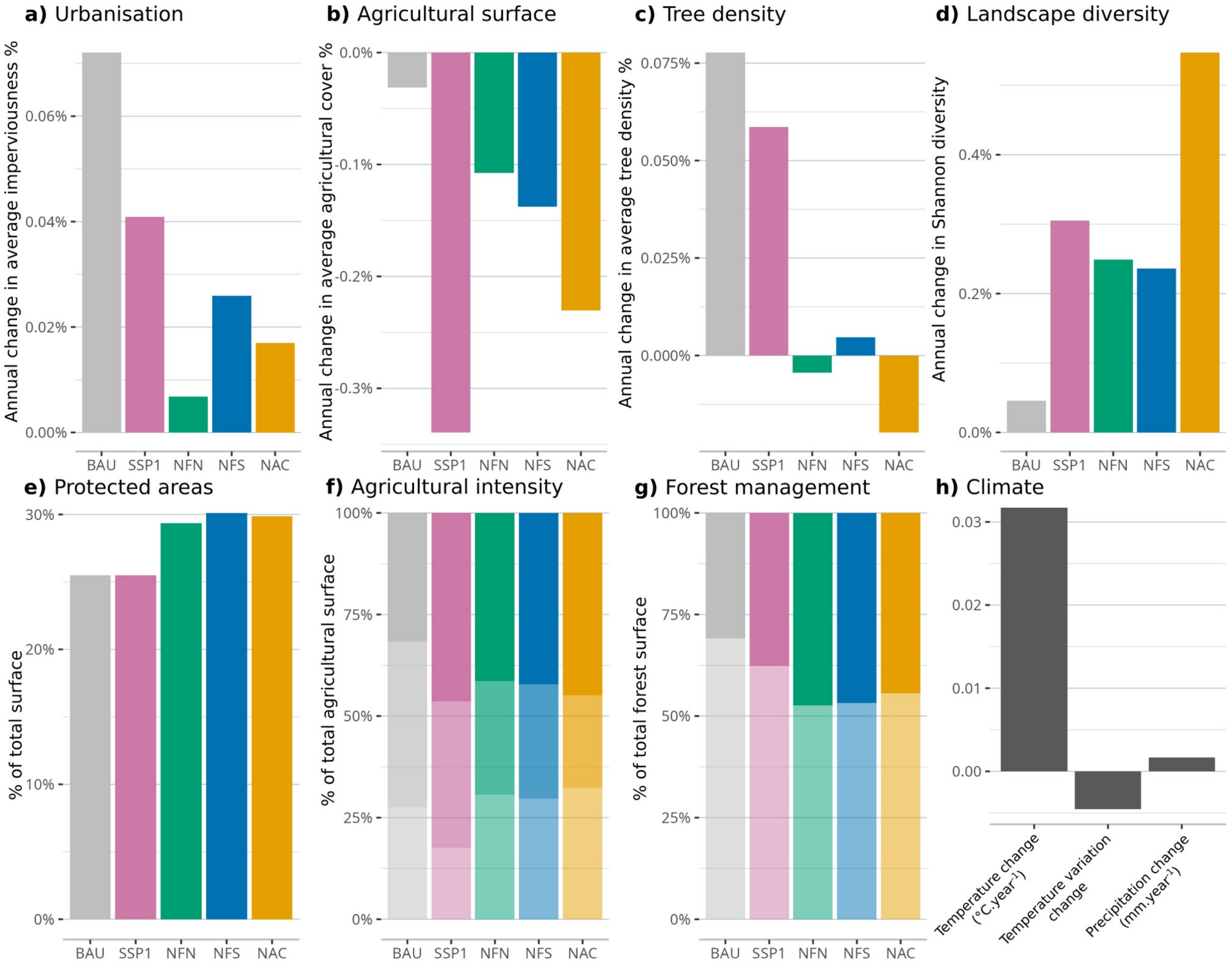

Figure 3: Predicted changes in scenarios in 2050 at the European scale in land use: a) urbanisation (annual rate of change in the average percentage of imperviousness), b) agricultural surface (annual rate of change in the the average percentage of agricultural cover), c) tree density (annual rate of change in the average percentage of tree density), d) landscape diversity (annual rate of change in the average landscape diversity); in land-use intensity: e) protected areas (average percentage of protected areas), f) production intensity in agricultural areas (low intensive in shaded colour, medium intensive in intermediate colour and high intensive in full colour), g) production intensity in forest areas (low/medium intensive in shaded colour, high intensive in full colour), and h) in climate (temperature change in °C.year$^{-1}$, and precipitation change in mm.year$^{-1}$). BAU: Business-As-Usual, SSP1: Shared Socioeconomic Pathway 1, NFN: Nature For Nature, NFS: Nature For Society, NAC: Nature As Culture. Note that in addition to the quantitative differences between scenarios, the location of the changes varies (Extended Fig. 3-7).

The observed declines in Common Bird Index, Farmland Bird Index, Common Butterfly Index, Grassland Butterfly Index and Woodland Butterfly Index are expected to continue under the different scenarios (Fig. 4a, d, e, and Extended Fig. 1 a, d, e for pair-wise comparison of trends). The three scenarios with additional conservation targets (Nature For Nature (NFN), Nature For Society (NFS) and Nature As Culture (NAC)) result in a smaller decline in Common Bird Index and Farmland Bird Index compared to SSP1 (Fig. 4a, b, Extended Fig. 1a, b). NFN, NAC and, to a smaller extent, NFS also improve the trend projected under the Business As Usual (BAU, adding the expected impact of climate change to the past trend) for birds. In these three scenarios, the combined effect of higher share in low intensive farmlands, higher landscape diversity and increase in protected areas (Fig. 4b and f and Fig. S1e-g) is expected to compensate for the negative effect of temperature increase and forests with high intensive management more than in other scenarios.

For butterflies, NAC enables the lowest decline in Common Butterfly Index and Woodland Butterfly Index, while NFN and NFS results in trends similar to BAU. In NAC, the combined effect of a higher increase in landscape diversity, a higher decrease in agricultural cover and a lower share of forests with high intensive management is expected to compensate for the negative effect of temperature increase more than in other scenarios. SSP1 is also expected to lower the decline compared to NFN and NFS in Common Butterfly Index, in particular due to the decrease in agricultural cover, which is the highest across the studied scenarios (Fig. S1h-j).

Two of the scenarios with additional conservation targets (NFN and NFS) do not deviate from the expected trend in the Forest Bird Index (Fig. 4c, Extended Fig. 1c), in Common Butterfly Index and Woodland Butterfly Index (Fig. 4d, f, Extended Fig. 1d, f). None of the three scenarios with additional conservation targets deviate from the expected trend in Grassland Butterfly Index (Fig. 4e, Extended Fig. 1e). For forest birds, NAC and SSP1 are expected to result in a lower increase compared to BAU. The increase in forests with high intensive management while changes in tree density remain limited, in NFN, NFS and NAC prevents the expected trend to be more positive than in BAU. The reduced share of low intensive farmland in SSP1 is expected to result in an even lower trend compared to BAU (Fig. S1g).

The expected effect of climate change, independently of changes in land use and land-use intensity, is reflected in the difference between the observed past trend and the BAU, namely a positive effect on farmland birds (Fig. 4b) and a negative effect on Common Bird Index and woodland butterflies (Fig. 4a, f).

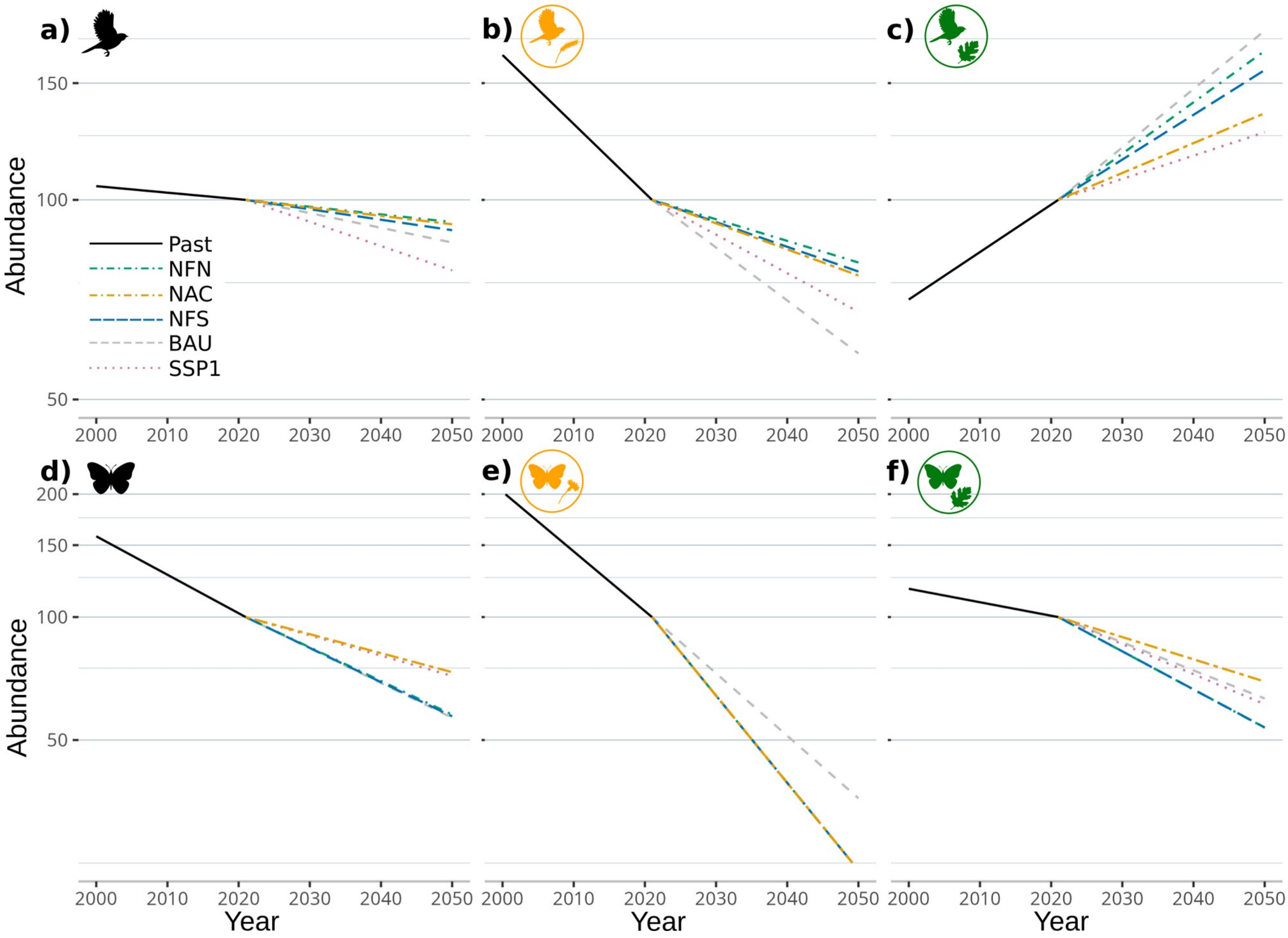

Figure 4: Predicted abundance trends of a) Common Bird Index, b) Farmland Bird Index, c) Forest Bird Index, d) Common Butterfly Index, e) Grassland Butterfly Index and f) Woodland Butterfly Index. Observed trend (2000-2021) is displayed in black, Business As Usual (BAU) in dotted grey, the "Taking the Green Road" Shared Socioeconomic Pathway (SSP1) in purple, Nature For Nature (NFN) in green, Nature For Society (NFS) in blue and Nature As Culture (NAC) in yellow (see geometric standard errors in Extended Fig. 1). We use 2021 as a pivotal year, i.e. a reference to express the relative population abundance with a value of 100 in 2021.

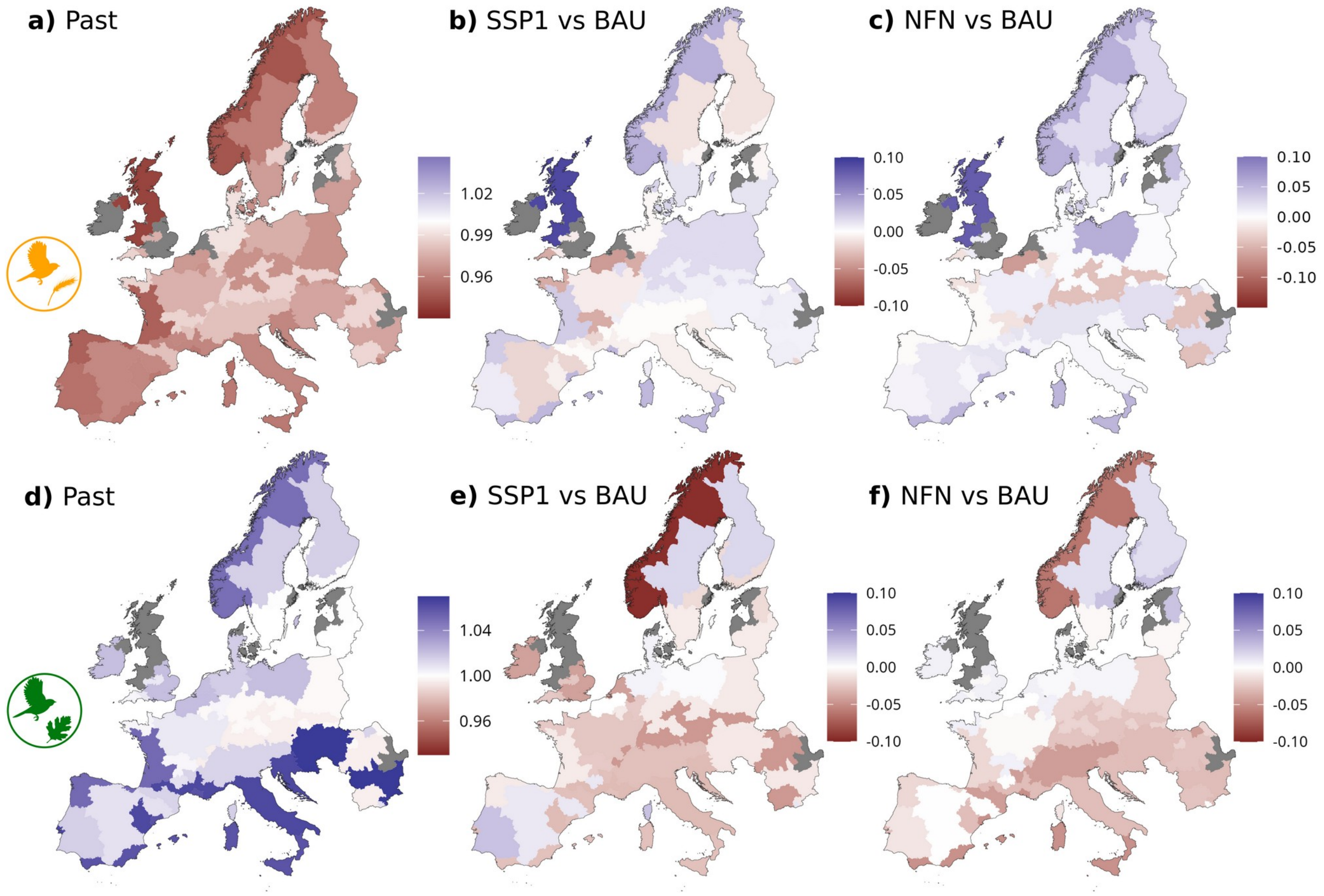

Figure 5: Mean annual trend in abundance by biophysical region for a) farmland bird species between 2000 and 2021, b) farmland bird species between mean trends in SSP1 and BAU, c) farmland bird species difference between mean trends in NFN and BAU, and d) for forest bird species between 2000 and 2021, e) difference between mean trends in SSP1 and BAU and f) difference between mean trends in NFN and BAU. Regions in grey have either not enough data for trend estimate or trend models with a low fit.

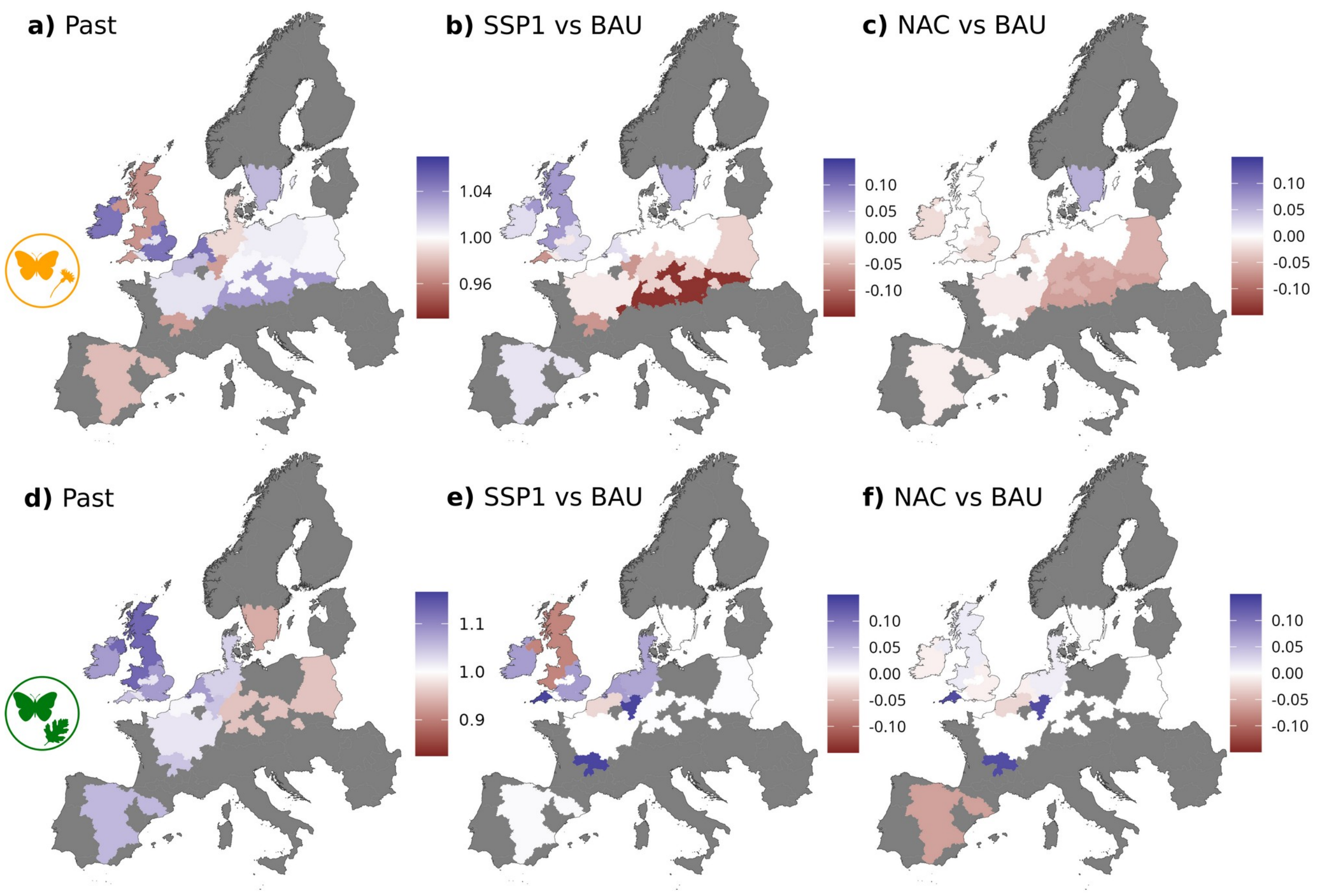

Figure 6: Mean annual trend in abundance by biophysical region for a) grassland butterfly species between 2000 and 2021, b) difference between mean trends in SSP1 and BAU, c) difference between mean trends in NAC and BAU, and d) for woodland butterfly species between 2000 and 2021, e) difference between mean trends in SSP1 and BAU and f) difference between mean trends in NAC and BAU. Regions in grey have either not enough data for trend estimate or trend models with a low fit.

The pattern of observed and expected changes in bird and butterfly indices is spatially heterogeneous (Fig. 5 and 6 and Fig. S1n-s), following the spatial distribution of direct drivers (Extended Fig. 2-7 and Fig. S1k-m). In the scenario with the most positive trend for Farmland birds at the European level (NFN), Farmland Bird Index is expected to show more negative trends in biophysical regions with an increase in forest cover (e.g. Western part of France, North-western Spain), while more positive trends are expected in biophysical regions with an increase in low intensity farmland, protected areas and temperature (e.g. Northern UK, Scandinavia) (Fig. 5c and Fig. S1t).

In the scenario with the most positive trend for FoBI at the European level (NFN), forest birds are expected to show more positive trends in biophysical regions with positive forest cover change, in particular with low to medium management intensity (e.g. North-western Europe). In contrast, forest birds are expected to exhibit lower trends under NFN in biophysical regions with an increase in high intensive farmland, precipitation increase and rising urbanisation (e.g. Eastern Europe and Southern Europe) (Fig. 5f and Fig S1t).

For butterflies, the scenario with the most positive trend at the European level is NAC. Grassland butterfly species are expected to show more negative trends in this scenario in biophysical regions with increasing high intensive farmland and higher share of forests under low to medium intensity management (e.g. Central Europe). In contrast, grassland butterflies are expected to show more positive trends in NAC in biophysical regions with increasing temperature and higher share of intensive forests (e.g. Southern Sweden and Finland) (Fig. 6c and Fig. S1t).

Woodland butterfly species are expected to show more negative trends in NAC in regions with increasing precipitation (e.g. Spain) and higher levels of protected areas. In contrast, woodland butterflies are expected to show more positive trends under NAC in regions with increasing low intensive farmland and landscape diversity (e.g. Northern UK) (Fig. 6f and Fig. S1t).

## 3. Discussion

In this study, we explore the effects of Europe-wide sustainability scenarios based on conservation targets, on common biodiversity using abundant and widespread species from two different groups, birds and butterflies. We show that scenarios considered to be ambitious in terms of conservation, involving the implementation of strategies related to biodiversity conservation and restoration in Europe, are not expected to be sufficient to halt population declines in common species.

The observed decline in common birds, particularly those associated with agricultural environments, is consistent with findings that have accumulated over several decades in Europe[22,52,57]. A comparable decline is visible in butterflies at the European level[40,58]. This decline is especially pronounced in species linked to grassland habitats, echoing previous results from the European Grassland Butterfly Indicator[59,60]. Our results show that the intensification of agricultural production and vegetation encroachment related to land abandonment are linked to these declines, in line with direct drivers that have already been identified on these groups[44,61–66].

Our results also indicate that climate change, particularly the increase in average spring temperatures, has so far had a balanced effect on bird species. In contrast, the effect on butterfly species appears predominantly negative, adding to the range of direct drivers already affecting them. The consequences of climate change are yet complex, including changes in precipitation and extreme event regimes, and species responses are diverse and may even be opposite between cold and hot dwellers[67–70]. For species depending on forest environments, however, our analysis confirms that the trends for birds was positive over the past two decades, and the decline in butterflies is less pronounced than in grassland species, in line with previous results[40,71]. Recent increase in forest cover might explain the increase observed in forest birds in Europe[22], still heterogeneously distributed across Europe[52], as well as, for woodland butterflies, the positive influence of urban environment[40] and forest disturbance[40,72] providing new habitats.

In addition, our results show that birds and butterflies respond differently to forest management. Whereas for forest birds lower management has a more positive effect, the results are more contrasted for woodland butterflies. Afforestation and management practices affect both woodland parcel size[73], light availability and plant community composition, which is critical for many specialised butterfly species[74], although such practices may temporarily provide habitat for edge and open-land species[75]. Protected areas were found to have a generally beneficial effect on a majority of bird species by mitigating land-use intensification and change, a pattern previously documented[9,76]. A significant share of species yet experience a negative effect of protected areas, which has been related to insufficient conservation effort and an inadequate management for birds and butterflies[10,77]. This is particularly the case for woodland butterflies that show overall more negative trends in protected areas, in line with previous results on Natura 2000 areas[45], as most butterfly species depend on open space in forests[59], whereas protected areas can experience a canopy closure in a context of decreased forestry. This relative negative effect of protected areas on population dynamics is also visible for a majority of forest birds and must be interpreted in the light of a more marked increase in tree density outside protected areas than within them, whilst the abundance of these species remains higher within protected areas.

Although these results are correlational, many of them have therefore based on a plausible causal hypothesis, given their close alignment with mechanisms identified in previous studies. These results thus provide useful insights into the future of common biodiversity in socio-economics scenarios. The scenarios analysed in this study are based on the shared socio-economic pathway (SSP1) that is designed to take the most account of environmental issues, compared to other SSPs[78]. The implementation of SSP1 analysed in this study relies however on specific choices, in particular adopting a more land-sparing approach, illustrated by a strong decline in agricultural areas and a higher share of high intensive farmland compared to the current situation. In contrast, an alternative SSP3 which is expected to lead to highest removal of natural land at the global level[79], would mostly reduce landscape diversity in Europe, with less consequences on the agricultural surface. Compare to SSP3, the implementation of SSP1 held more positive trends in birds and butterflies (except forest birds and grassland butterflies, see comparison between SSP1 and SSP3 in Fig. S1u, v) in line with the a priori expectation. However, the SSP1 scenario does not appear to show a clear trend in terms of overall biodiversity, as it predicts an improvement for butterflies through the increase in landscape diversity, but the less favourable outcome for birds through reduced agricultural area and highest share of high intensive farmlands.

The three other scenarios (NFN, NFS and NAC) include additional conservation targets that align with some of the objectives of the EU Biodiversity Strategy for 2030, the Forest Strategy, and the Farm to Fork Strategy[19]. They envisage expanding protected terrestrial areas to 30%, increasing natural area of 1% per year between 2030 and 2050, reducing the use of synthetic nitrogen fertilisers in agriculture by 30%, and planting 3 billion trees across urban environments and less intensively managed farmland. Taken together, these measures constitute a transformative strategy for reversing biodiversity loss, strengthening ecosystem resilience, and supporting the EU’s wider climate and sustainability goals, while aligning with Shared Socioeconomic Pathway 1 (SSP1). Our results show that the different variants of three scenarios with additional conservation targets (that vary in terms of protected areas localisation, urbanisation types and land-use intensity) benefit common birds overall, compared with the BAU scenario and the SSP1 scenario. This is the case for NFN in particular, which combines an increase in low intensive farmland also visible in NFS and NAC with the lowest reduction in agricultural area and the lowest rate of urbanisation. For common butterflies, of the three scenarios with additional conservation-targets, only NAC improves the BAU trend, through the strongest increase in landscape diversity.

Yet, in none of the three scenarios with additional conservation targets the decline in common birds and butterflies seems to be halted. The most significant positive effect in these scenarios is expected for bird species depending on open environments, for which the decline is slowed. A central feature of these three scenarios is the limitation of open-habitat loss, preventing the projected reduction of about one-third of European grasslands by 2050 under SSP1[19], which is a key for the conservation of those species[80]. But the type of agriculture, apart from the reduction in nitrogen inputs, does not undergo any profound change in these scenarios. Aside from the reduction in nitrogen inputs, the agricultural model remains largely unchanged and the share of high intensive farmlands increased in all scenarios. Intensive practices, characterised by pesticide use and large-scale monocultures, are among the main drivers of biodiversity loss[81] and continue to exert strong pressures on bird and butterfly populations, as confirmed in our results on the past two decades. Similarly, European forests face a dual constraint[82]: they must provide habitat while also serving as a production resource. Even in the three scenarios with additional conservation targets, the proportion of intensively managed forests increases compared to BAU, highlighting a key policy trade-off between biodiversity conservation and forestry production. Generalist forest species may benefit from forest expansion regardless of management type, whereas specialist species continue to decline or have relative lower trends (Fig. S1w, x), as observed in the past decades[83–85], underscoring the need for targeted management strategies that support habitat specialists. For butterflies, climate effects—particularly changes in precipitation and temperature—are crucial, leading to projected continuous declines across all scenarios, despite a recent regain after years of decline for common and woodland species[40]. This suggests that effective conservation policies must integrate both adaptive forest management and climate mitigation measures to safeguard sensitive species.

Furthermore, the beneficial effect for common species, observed on a continental scale, of scenarios with additional conservation targets compared to SSP1 and BAU (but also SSP3, Fig. S1u, v), is not evenly distributed. The improvement in trends for common birds is concentrated in western and northern Europe, where the baseline is already very low after decades of decline[22,52]. Conversely, in Eastern Europe, where the decline has been accentuated more recently by the agricultural changes brought by the CAP[86–88], the improvement in projected trends is less visible despite conservation policies. Similarly, in southern Europe, declines or lower increase in farmland and forest birds are expected as low-intensity agricultural and forestry systems are increasingly replaced by more intensive, high-productivity methods. For butterflies, projected population increases are heterogeneously distributed, largely driven by rising temperatures and changes in precipitation patterns[70], in line with the variable sensitivity to weather conditions of the different populations across Europe[89].

The work of forecasting remains limited notably by uncertainty related to the choice of conservation strategies, their translation into spatial modifications of direct drivers, the limited set of drivers available, the link to responses of common biodiversity and the absence of stochastic processes[90]. Using observed effects of climate change to anticipate the effects of future climate change, in particular, has to be interpreted with caution, as this driver is likely to reach levels of change that go beyond the ones observed in the past decades (i.e. the period for which biodiversity data are available)[91]. This may also be the case for other drivers, such as land-use intensity. Moreover, the focus on indicators based on a fixed set of species may not represent the changes expected in currently rare species whose future distribution could extend in Europe[92]. Consequently, anticipating the future of common species populations provides indications of species trends, allowing pairwise comparisons between scenarios, rather than a detailed quantification. Nevertheless, using common species helps to reduce prediction uncertainty and robust multi-species indices derived from these species facilitate the interpretation of expected change per habitat, highlighting differences between scenarios. Notably, the continued decline of common biodiversity—even under scenarios incorporating a range of conservation measures or production changes—raises important questions: although the range of conservation targets implemented in the scenarios used in this study is limited, current measures are already being scrutinised and challenged[93], prompting reflection on the potential impact of more transformative conservation actions. This issue underscores both the societal acceptability of the drastic changes needed to halt biodiversity loss in Western societies[2,94], and the limitations inherent in the SSP framework. If even SSP1, designed to focus on environmentally-friendly priorities[78,79], enhanced by additional conservation measures in the three variant scenarios explored here, is insufficient to stop declines in common species, more ambitious regulations and alternative anticipatory frameworks may be required—ones that reconsider assumptions about the ongoing demand for natural resources to support food, energy, and economic growth[20,95]. These findings highlight the need for bold, integrated policy strategies that balance resource use with biodiversity conservation to achieve meaningful outcomes for common biodiversity, in particular by enhancing the landscape diversity and the transition of the agricultural model. Global scenarios can provide an insight into this trajectory, characterised by low demand for energy and raw materials, coupled with a food transition and a reduction in pollution from the agricultural system[96]; however, spatially detailed scenarios aimed at reversing the decline of common species and identifying compatible socio-economic pathways remain an area for further research.

## 4. Online methods

### 4.1 Biodiversity data

#### 4.1.1 Bird data

We use annual count data of bird species at the site level from 27 European countries collated by the Pan-European Common Bird Monitoring Scheme (PECBMS). We focused on data between 2000 and 2021 to ensure a spatial coverage large enough at the start of the time period, as some national surveys were started in the last two decades. At each site, skilled volunteers record the number of individuals for each bird species during the breeding season, following three types of standardised schemes: point counts, line transects or territory mapping. In order to harmonise abundance data among schemes, we followed PECBMS guidelines, multiplying by two when pairs or territories are reported by a scheme. To extract common bird species, we selected the 265 most frequent species (see species lists in supplementary material 2) present in at least 1% of the visited sites, out of the 450 species recorded. Counts from these schemes may not be reliable for large groups of individuals of colonial species. We therefore capped outlier records of large groups at the upper 1% counts (i.e. 55 individuals). Not all sites are visited every year, thus, to be able to conduct the temporal analysis, we focused on sites visited at least five years between 2000 and 2021, with the most recent monitoring in the second part of the time period, i.e. in 2011 or after. At the end of the filter process, we obtained 6,364,720 annual measures of species abundance across 15,731 sites in 27 European countries (extended Fig. 8a and Fig. S3a-c).

#### 4.1.2 Butterfly data

We used annual count data of butterfly species at the transect level from 15 European countries collated through the European Butterfly Monitoring Scheme (eBMS). We choose the same time period as for bird data (2000-2021). Each country conducts monitoring under a similar protocol based on the Pollard transect[97]. Transects correspond to permanent monitoring routes ranging from 200 metres to 3 kilometres long, divided into sections of different habitat cover, where the abundance of each butterfly species is regularly recorded several times a year (weekly, fortnightly or monthly) by skilled volunteers. Butterflies present in a cube of 5-meters ahead, 5-meter above and 2.5-meters on each side of the observer are counted[97]. We used the annual aggregated abundance (sum of count records) provided by eBMS and we divided it by the number of visits to control for sampling effort. Similarly to the bird data, we focused on the 144 most frequent species (see species lists in supplementary material 2) monitored in at least 1% of the transects among the 318 monitored species. We capped outlier records of large groups at the upper 1% counts (i.e. 31 individuals). Since not all transects are visited every year, we applied the same filter as for bird data, retaining only transects visited at least during 5 years between 2000 and 2021 and for the last time in 2011 or after. This resulted in a dataset with 901,949 annual measures of species abundance across 4,959 transects in 15 European countries (extended Fig. 8b and Fig. S3d, e).

### 4.2 Environmental and pressure data

#### 4.2.1 Biophysical regions

The magnitude and direction of response to anthropogenic drivers may vary among different types of ecosystems. It is important to account for this variation in continental-wide analyses in which biophysical variables have a wide range. Different biophysical classifications have been proposed for Europe but most of them include land-cover or habitat in the stratification of the biophysical regions, potentially confounding the analysis. We therefore use 24 biophysical regions based on the primary landscape structure (climate, topography and soil) from the Spatial Regional Reference Framework, available at NUTS3 scale[98] (Fig. S3f).

4.2.2 Climate

Past. We used the E-OBS dataset[99] providing daily historical climate data for Europe at a 0.1° resolution (all drivers and environmental data used are listed in Table S4). We computed the average and variance of the mean temperature and the sum of precipitation during spring (1st March to 30th June, i.e., the main breeding period for birds in most areas) for each year between 2000 and 2005 and between 2016 and 2021. Increases in mean temperature, changes in precipitation, and in temperature variability have indeed been found to influence population dynamics of bird and butterfly species[100–102]. To project all the environmental and driver data at the same scale, we used the 1 × 1 km grid of Europe[103]. We employed the *exactextractr* R package[104] to extract the temperature average and variance and the sum of precipitation of each grid cell in the periods 2000-2005 and 2016-2021.

We used the average of 2000 to 2005 values as the baseline climate values (Extended Tab. 1). To assess the changes in the average temperature and the annual variation of temperature, as well as the sum of precipitation at each monitoring site, we first averaged the values of the historical period (2000 to 2005) and the recent period (2016 to 2021) in a 2.5km-radius circular buffer around sites (designed to both reflect local conditions and encompass the monitored areas of most schemes, and see Fig. S4a for the effect of buffer size). We then calculated the rate of change between the recent period and the historical period as follows (Eq. 1):

$$\delta X_{obs} = \frac{\overline{X_{obs\,recent}} - \overline{X_{obs\,historical}}}{2018 - 2002} \quad \text{(Eq. 1)}$$

with $X_{obs}$ the observed variable of interest.

Future. We obtained estimates of the future climate up to 2050 from projections of climate indicators for Europe (1 × 1 km grid) from the Climate Data Store[105] available from five regional projection models (CCLM4-8-17, HIRHAM5, RACMO22E, RCA4, WRF381P) under the intermediate climatic scenario RCP 4.5 (leading to a 2°C increase in global temperature by mid-century[54]). We focused at two spatial scales, the biophysical region and the European scale (comprising all 27 countries considered). For each biophysical region and for Europe, we estimated the 2050 mean temperature as the monthly mean temperature predicted for March to June 2050 across the five models. We also recorded the monthly mean temperature from the models for the recent period (2016 to 2021). To assess temperature variability, we calculated the difference between the averaged maximum and minimum spring temperature (March to June) for the five models and then averaged these values across the five models. We follow the same approach to retrieve the model estimates of temperature variability during the recent period (2016-2021). For precipitation, we summed the projected precipitations from March to June 2050 and averaged the results across the five models. We also retrieved the model estimates of spring precipitation for the recent period by averaging values from 2016 to 2021 across the five models.

The model estimates for the recent period allowed to adjust the 2050 projections by accounting for discrepancies between modelled and observed data (Eq. 2):

$$X_{\exp 2050} = X_{mod\,2050} \cdot \frac{\overline{X_{obs\,recent}}}{\overline{X_{mod\,recent}}} \qquad \text{(Eq. 2)}$$

with $X_{exp}$ the corrected predicted climatic variable of interest, $X_{obs\ recent}$ the observed climatic variable of interest (average between 2016 and 2021) and $X_{mod}$ the model estimate of the climatic variable of interest.

We finally computed the expected rate of change between 2050 and the recent period (Eq. 3)

$$\delta X_{\exp} = \frac{X_{\exp 2050} - \overline{X_{obs\,recent}}}{2050 - 2018} \qquad \text{(Eq. 3)}$$

4.2.3 Land use and land cover

Past. We assessed urbanisation using the imperviousness data provided at a 100 × 100 m resolution for Europe in 2006 and 2018[106]. The change in tree cover density was obtained using tree cover density data provided at a 100 × 100 m resolution in 2012 and 2018[107] (see relationships between High Resolution Layer and CORINE classification in Fig. S4b-e). For each available year, we averaged the imperviousness and tree cover density values within each cell of the 1 × 1 km European grid. The change in agricultural cover was retrieved using the 100 × 100 m CORINE Land Cover accounting layers (designed for land cover change accounting) for 2000 and 2018[108]. We aggregated, for each 1 × 1 km European grid cell, the proportion of land classified as arable land, permanent crops, pastures and heterogeneous agricultural areas. These values were then averaged within the 2.5km buffer around each monitoring site/transect. The rates of change for urbanisation, forest cover density and agricultural areas were calculated using Eq. 1, with 2006, 2012 and 2000 used as the historical period respectively (Extended Tab. 1).

We used the 2018 100 × 100 m CORINE Land Cover accounting layer to assess the main land-use type around each monitoring site/transect. We attributed to each 1 × 1 km European grid cell the main represented land-cover among the 44 CORINE land-cover types. We then used the upper aggregation level of CORINE with five land-cover types (urbanised areas, agricultural areas, forest and semi-natural areas, wetland, water areas) to characterise the land use of each 1 × 1 km European grid cell. We then calculated the proportion of the five land-cover types within a 2.5 km buffer around each site/transect. Due to the low number of site/transect surrounded by wetland, water bodies, we finally grouped these two categories, ending up with four categories: urban, forest and semi-natural areas, agricultural areas and wetland/water areas.

We used the European Land System map[109] to assess the intensity of land use[46] at a 1 × 1 km resolution. This map classifies land into seven land systems—human settlements, forest, cropland, permanent cropland, grassland and agricultural mosaic—categorised by land-use intensity level (low, high and medium). It is representative on average for the year 2015, even though the year of the various datasets may vary around this date. For forest intensity, we merged low to medium management intensity as low intensity would correspond to primary forests that are not well represented in Europe. We calculated within a 2.5 km buffer around each monitoring site/transect, the percentage of forests under low to medium management intensity and under high management intensity. For farmland intensity, we calculated within a 2.5 km buffer around each monitoring site/transect, the share of permanent crops, grassland, cropland and agricultural mosaic under low and high intensive production systems.

Future. To assess the changes in land use and land-use intensity by 2050 at both the biophysical region and European scales, we used the European Land-use scenarios for 2050[55] produced in line with the IPBES promoted Nature Future Framework (NFF)[19]. These scenarios apply the same land system and intensity classification as in the European Land System map[46], and are available for four scenarios: one Shared Socio-economic Pathway "Taking the Green Road" (SSP1), and three NFF scenarios corresponding to major pathways along which biodiversity positive values can be implemented (Nature for Nature, Nature as Culture, Nature for Society)[24]. The NFF scenarios in this implementation extend SSP1 by incorporating explicit biodiversity positive values in the narratives. It also allows the inclusion of additional sustainability targets in line with European Union Green Deal targets and post-2020 Global Biodiversity Framework targets, such as reduced use of nitrogen fertiliser, increased tree cover and expanded protected areas[19]. We retrieved model estimates of land use and land-use intensity for each region and each scenario. The projected urbanisation cover for 2050 was obtained by aggregating the share of medium-urban (suburbs) and high-urban (cities) land systems (low urban correspond to villages). The projected open land cover combines areas of permanent crops, grassland, cropland and agricultural mosaic land systems. Projected forest cover was calculated by aggregating the areas classified as forests under low, medium and high management intensity. We also extracted projections of forests with low to medium wood production, forests with high wood production, and farmland/open land under low, medium and high intensive production systems. These model outputs were corrected using the land system map at the start of simulation (2015) in Eq. 2. The expected changes in urbanisation, tree cover density and open land cover by 2050 were calculated using Eq. 3.

### 4.2.4 Landscape diversity

Past. We computed the Shannon diversity index (Eq. 4) to estimate landscape diversity[110] in each cell of the 1 × 1 km European grid. To compute the landscape diversity, we focused on the 28 land-use classes from the 100 × 100 m CORINE Land Cover accounting layers in 2000 and 2018 belonging to agricultural areas, forest and semi-natural areas and wetland. The value of the Shannon diversity index for each monitoring site/transect was derived by averaging the index values within a 2.5km buffer. This index ranges from 0, in case of total homogeneity in the cell, to ln(28) which represents the maximum diversity when all the land-use classes are equally present.

$$H_i = -\sum_{j=1}^{LU} p_{ij} \ln(p_{ij}) \qquad \text{(Eq. 4)}$$

with $H_i$ the Shannon diversity index of cell $i$, $p_{ij}$ the proportion of land-use class $j$ in cell $i$ and $LU$ the number of different land-use classes.

Future. To assess changes in landscape diversity at the biophysical region and European scales, we aggregate the cover of forest/shrub and cropland mosaic, forest/shrub and grassland mosaics and agricultural mosaics of the European Land-use scenarios for 2050[55]. These model estimates were corrected using the land system map at the start of simulation (2015) in Eq. 2 and the expected change in landscape diversity was calculated using Eq. 3.

#### 4.2.5 Protected areas

Past. We used the World Database on Protected Areas[111] to compute the percentage of protected area in each cell of the 1 × 1 km European grid. As information on the management and the type of protection is not available for many regions, we computed the percentage of all protected areas within a 2.5 km buffer zone.

Future. To estimate the projected cover of protected areas at the biophysical region and European scales, we used the spatial priorities maps[112], which classify grid cells according to protection priority: vertebrate conservation, cultural Nature Contribution to People (NCP) and regulating NCP. These differences in protection priorities align with the different NFF European Land-use scenarios[113], with new protected areas focused on vertebrate conservation in the Nature for Nature scenario, on culturally valued agricultural landscape (cultural NCP) in the Nature as Culture scenario, and on carbon accumulation (regulating NCP) in the Nature for Society scenario. We used the version of the priority map that complements existing Natura 2000 areas with an additional 5% of priority sites to calculate the projected percentage of protected areas in 2050 for each scenario and region. The priority map also provides the current extent of Natura 2000 coverage used to correct the projections in Eq. 2.

#### 4.2.6 Ecosystem productivity

The primary energy available for bird and butterfly communities across the European continent differs across climates and influences the carrying capacity and thus the abundance of species. We therefore assessed ecosystem primary productivity by using the dry matter productivity data[114] available at spatial resolution of 1 × 1 km and a temporal resolution of 10 days. We averaged the dry matter productivity ($D$) for the spring season (March to June) of each year from 2000 to 2019. For each monitoring site/transect, we calculated the mean dry matter productivity within a 2.5 km buffer.

### 4.3 Statistical analysis

#### 4.3.1 Modelling climate, land-use and land-use intensity effect on species trends

We assessed the effect of each driver and environmental variable on the temporal trend of each species at both the biophysical region and the European levels (Fig. S1f). To do so, we modelled the observed abundance between 2000 and 2021 ($C$ counts) using environmental control variables ($LU$ main land-use type, $T_{init}$ mean spring temperature at the initial period, $R_{init}$ spring rainfall amount at the initial period, $H_{init}$ landscape diversity at the initial period, $D$ dry matter productivity) and monitoring control variables ($Effort_{area}$ sampling area surface or transect length and a random term of $Site_{id}$ site/transect identity nested in $Scheme_{id}$ scheme identity). We included the effect of time ($Y$ year) to estimate the temporal trend and interaction terms between year and driver variables to assess their influence on the temporal trend. These covariates included changes in mean spring temperature ($\delta T$), spring temperature variance ($\delta T_{var}$), spring rainfall amount ($\delta R$), urbanisation ($\delta Urb$), tree-cover density ($\delta TCD$), forest type (*Fo:* low/medium production or high production), farmland cover ($\delta Fa$), farmland intensity (*Fa:* low and high production system intensity), landscape diversity ($\delta H$) and percentage of protected area ($P$). All continuous variables were scaled and centred. Multicollinearity among variables was checked using the variance inflation factor (VIF) and was always below 2. We used a Generalised Additive Model (GAMM) (Eq. 5) to explicitly model the spatial auto-correlation between sites with a tensor product of a thin plate regression spline based on geographic coordinates of sites/transects[115]. A negative binomial regression with a log link was applied to account for over-dispersion in bird and butterfly count data.

The model was fitted for each of the 265 bird and 144 butterfly species at the European level (see a comparison with existing European temporal trends for some of those species in Fig. S4f, g). At the biophysical region level, a similar model was applied for species monitored at more than 60 sites/transects. We retained only combinations of species and biophysical regions for which the model explains more than 20% of the deviance (see details on model check in Fig. S4h, i, number of species by biophysical region in Fig. S4j-m and workflow across spatial levels in Fig. S3g), i.e. 223 bird species and 135 butterfly species.

$$\begin{aligned} C = {} & \beta_0 + \alpha_1 Y + \alpha_2 \delta Urb.Y + \alpha_3 \delta TCD.Y + \alpha_4 Fo_{low/medium}.Y + \alpha_5 Fo_{high}.Y \\ & + \alpha_6 \delta Fa.Y + \alpha_7 Fa_{low}.Y + \alpha_8 Fa_{high}.Y + \alpha_9 \delta T.Y + \alpha_{10} \delta T_{var}.Y + \alpha_{11} \delta R.Y \\ & + \alpha_{12} \delta H.Y + \alpha_{13} P.Y + \beta_1 LU + \beta_2 T_{init} + \beta_3 R_{init} + \beta_4 H_{init} + \beta_5 D + \beta_6 Effort_{area} \\ & + te(lon, lat, bs = \mathrm{tp}) + Site_{id} | S_{id} \end{aligned} \qquad \text{(Eq. 5)}$$

4.3.2 Future projection of changes in biodiversity indices

To estimate the projected temporal trend of the species between the recent period and 2050, we focused on the parameters influencing the temporal trend ($\alpha_2$ to $\alpha_{13}$ in Eq. 5, see Fig. S1e-j for the effect of each parameter on each trend and Fig. S1y for comparison of trends calculated only with significant parameters). We used the standard deviation of the observed variables (i.e. mean and σ of $X_{obs}$ and $\delta X_{obs}$) to scale and centre environmental and driver values ($X_{exp}$) and their changes ($\delta X_{exp}$) obtained from scenarios. For each species and by biophysical region, we calculated:
1) the temporal trend between 2000 and 2021 using the observed values of climate, land-use and land-use intensity changes;
2) the expected temporal trend between the recent period and 2050 under business as usual scenario (BAU) by assuming that the observed values of changes in land-use and land-use intensity changes remain the same and using the expected changes in climate variables (i.e. BAU corresponds to a climate reference scenario);
3) the expected temporal trends between the recent period and 2050 in the four alternative scenarios (SSP1, NAC, NFN and NFS) using the expected climate, land-use and land-use intensity changes.
These projections were derived using Eq. 6.

$$\frac{\partial C}{\partial Y} = \alpha_1 + \alpha_2 \delta Urb + \alpha_3 \delta TD + \alpha_4 Fo_{low/medium} + \alpha_5 Fo_{high} + \alpha_6 \delta Fa + \alpha_7 Fa_{low} + \alpha_8 Fa_{high} + \alpha_9 \delta T + \alpha_{10} \delta T_{var} + \alpha_{11} \delta R + \alpha_{12} \delta H . + \alpha_{13} P \quad \text{(Eq. 6)}$$

We finally computed the predicted bird and butterfly indices using the geometric mean of species temporal trend ($\mu_g$, Eq. 7) by biophysical region and for all Europe. We computed the multi-species indicators based on species' habitat preference, using for the Farmland Bird Index 34 species among the 38 listed by PECBMS at the European level, for the Forest Bird Index 29 species among the 34 listed by PECBMS at the European level, for the Grassland Butterfly Indicator 12 among the 17 used to produce the EU indicator[116] and for the Woodland Butterfly Indicator (Woodland Butterfly Index) 28 among the 33 used to produce the EU indicator[40] (species list in supplementary material 2). In addition, we computed the Common Bird Index with the 145 bird species for which the GAM explains more than 20% of the deviance among the 168 species from the original index and a Common Butterfly Index using the 135 butterfly species for which the GAM explains more than 20% of the deviance. To avoid unrealistic exponential increases or decreases in population that could affect multi-species indices, we bounded the observed and projected population increases and decreases. To do so, we used the maximum observed yearly increases and decreases of bird European populations between 1980 and 2024 produced by PECBMS, i.e. a multiplicative trend between 0.93 and 1.07, corresponding to a yearly decrease or increase of 7% per year.

$$\mu_g(N_I, s) = e^{\frac{1}{N_I}\sum_{n=1}^{N_I} \frac{\partial C(n,s)}{\partial Y}} \cdot 100 \quad \text{(Eq. 7)}$$

with $\frac{\partial C}{\partial Y}$ initially expressed in log scale from the GAM, for index $I$ (Common Bird Index, Farmland Bird Index, Forest Bird Index, Common Butterfly Index, Grassland Butterfly Index, Woodland Butterfly Index) and scenario $s$ (BAU, SSP1, NFN, NFS, NAC), and the $N$ species. We then compared the temporal trends of scenarios pair-wise using a two-sample t-test on their $\mu_g$ and geometric standard error.

Extended data

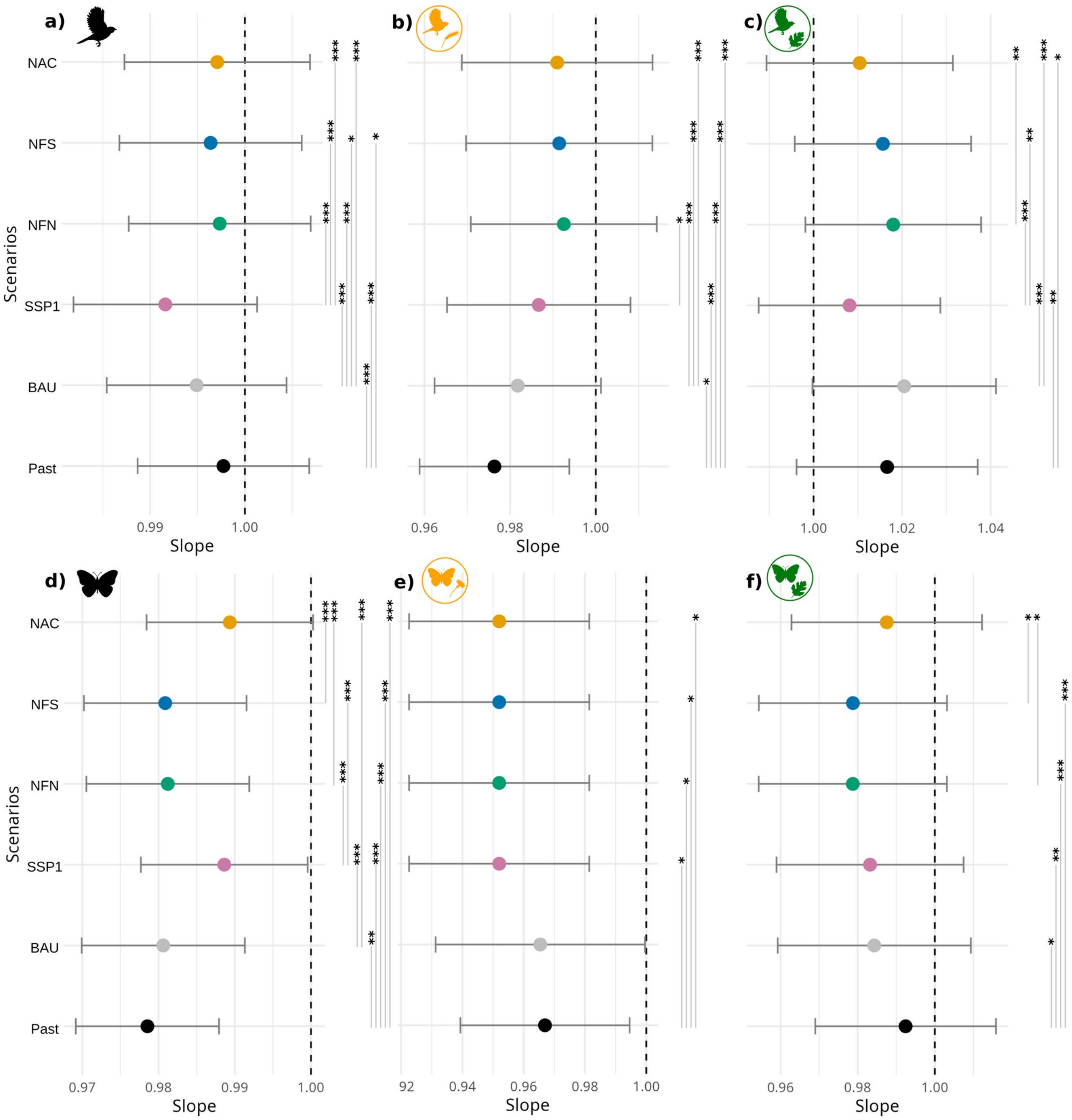

Extended Figure 1: Geometric mean of species temporal trend calculated with significant and non-significant parameters between 2000 and 2021 (Past) and in the different scenarios (BAU, SSP1, NFN, NFS, NAC), for a) Common Birds index, b) Farmland Bird index, c) Forest Bird index, d) Common Butterfly index, e) Grassland Butterfly index and d) Woodland Butterfly index. 95% confidence intervals from geometric standard error and the significant differences between scenarios are displayed (* p-value < 0.05, ** p-value < 0.01, *** p-value < 0.001).

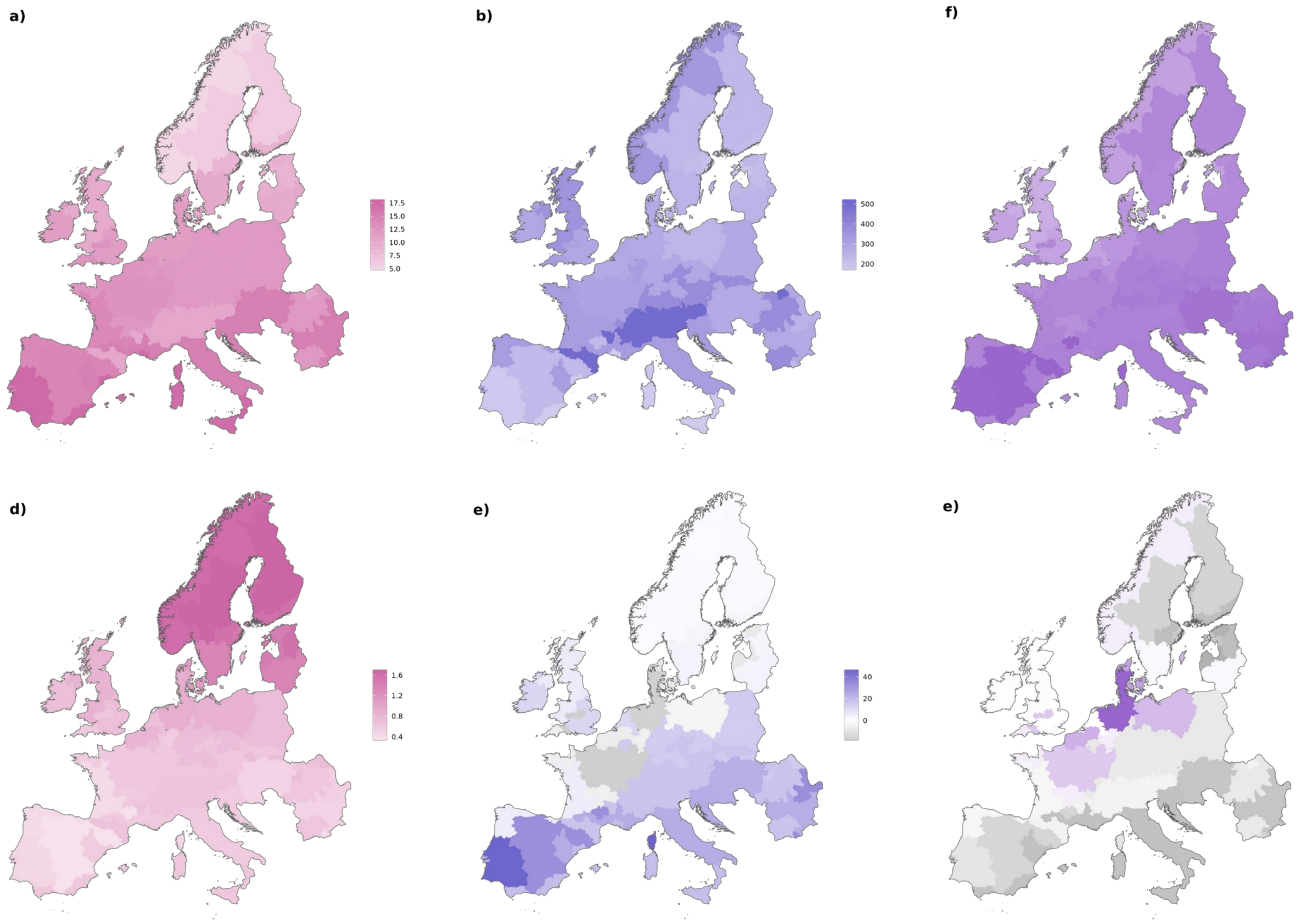

Extended Figure 2: Climate variable and changes by biophysical region: a) mean spring temperature (°C) in recent period (2016-2021), b) amount of spring precipitation (mm) in recent period, c) mean spring temperature variance in recent period, d) expected change in mean spring temperature (°C) between 2050 and recent period, e) expected change in amount of spring precipitation (mm) between 2050 and recent period, f) expected change in mean spring temperature variance between 2050 and recent period.

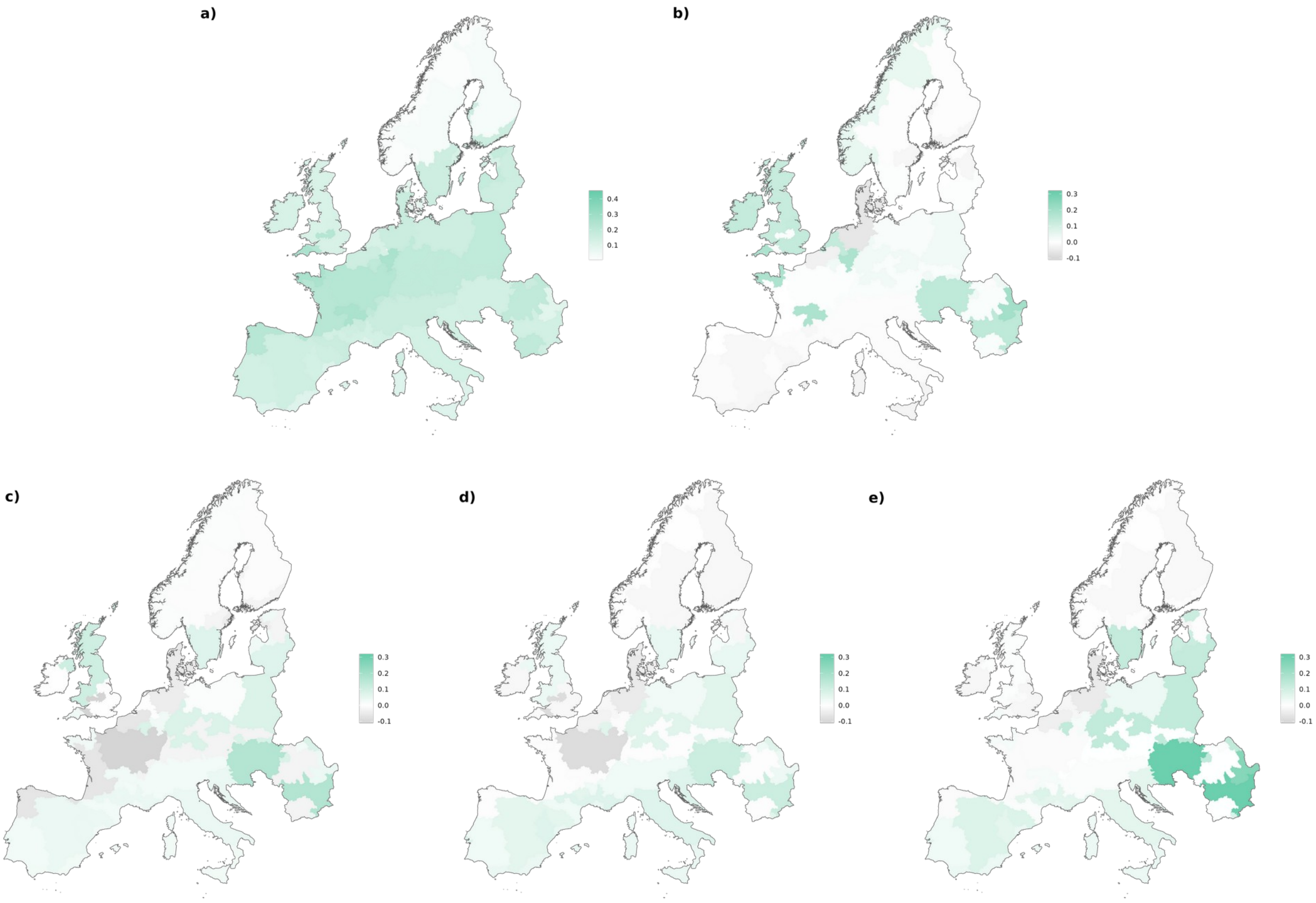

Extended Figure 3: Landscape diversity and changes by biophysical region: a) current landscape Shannon diversity, b) change in landscape diversity in SSP1, c) NFN, d) NFS, e) NAC.

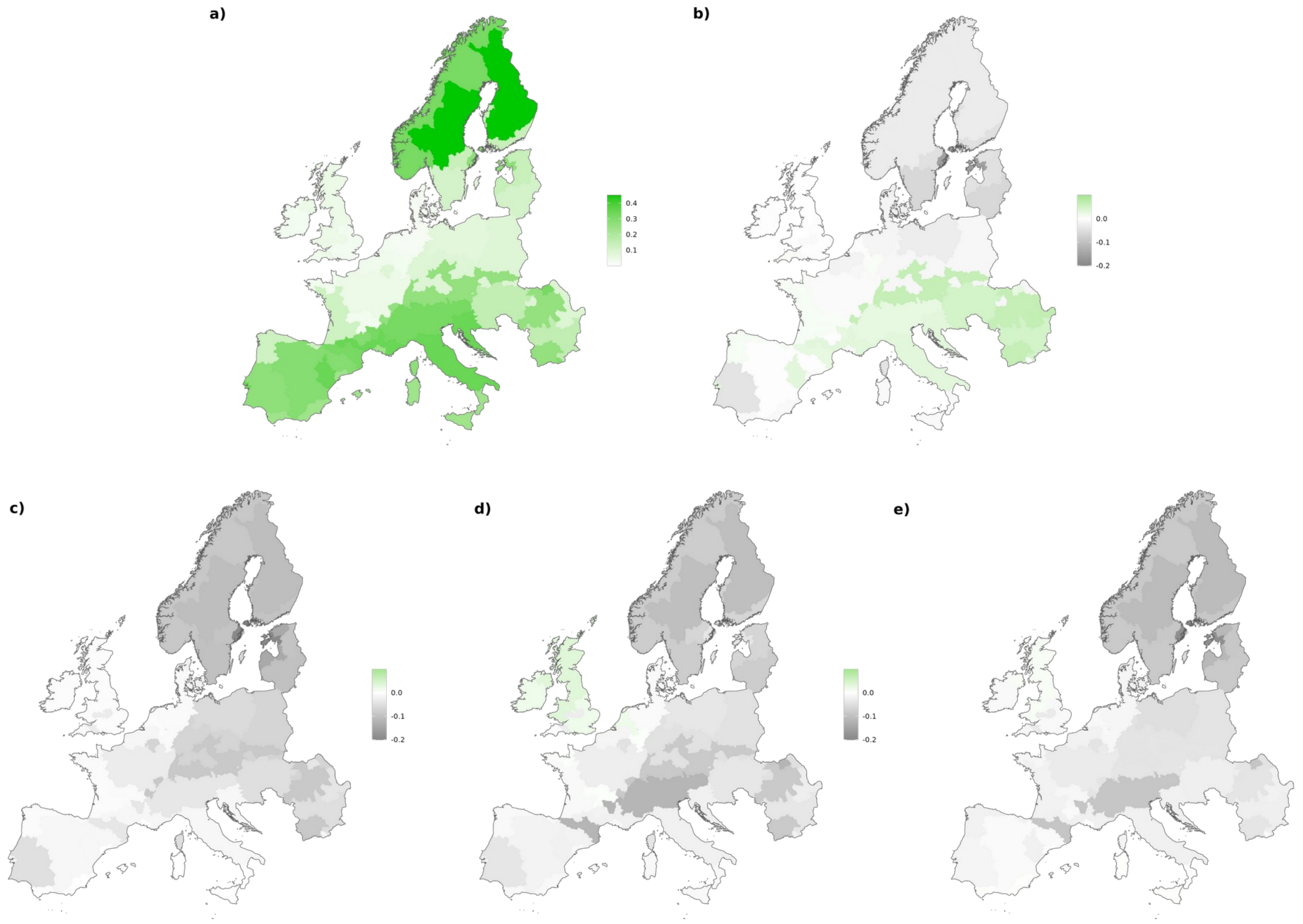

Extended Figure 4: Share of area cover by forests under low and medium intensity management and changes by biophysical region: a) current cover of low and medium intensity forest, b) change in cover of low and medium intensity forest in SSP1, c) NFN, d) NFS, e) NAC.

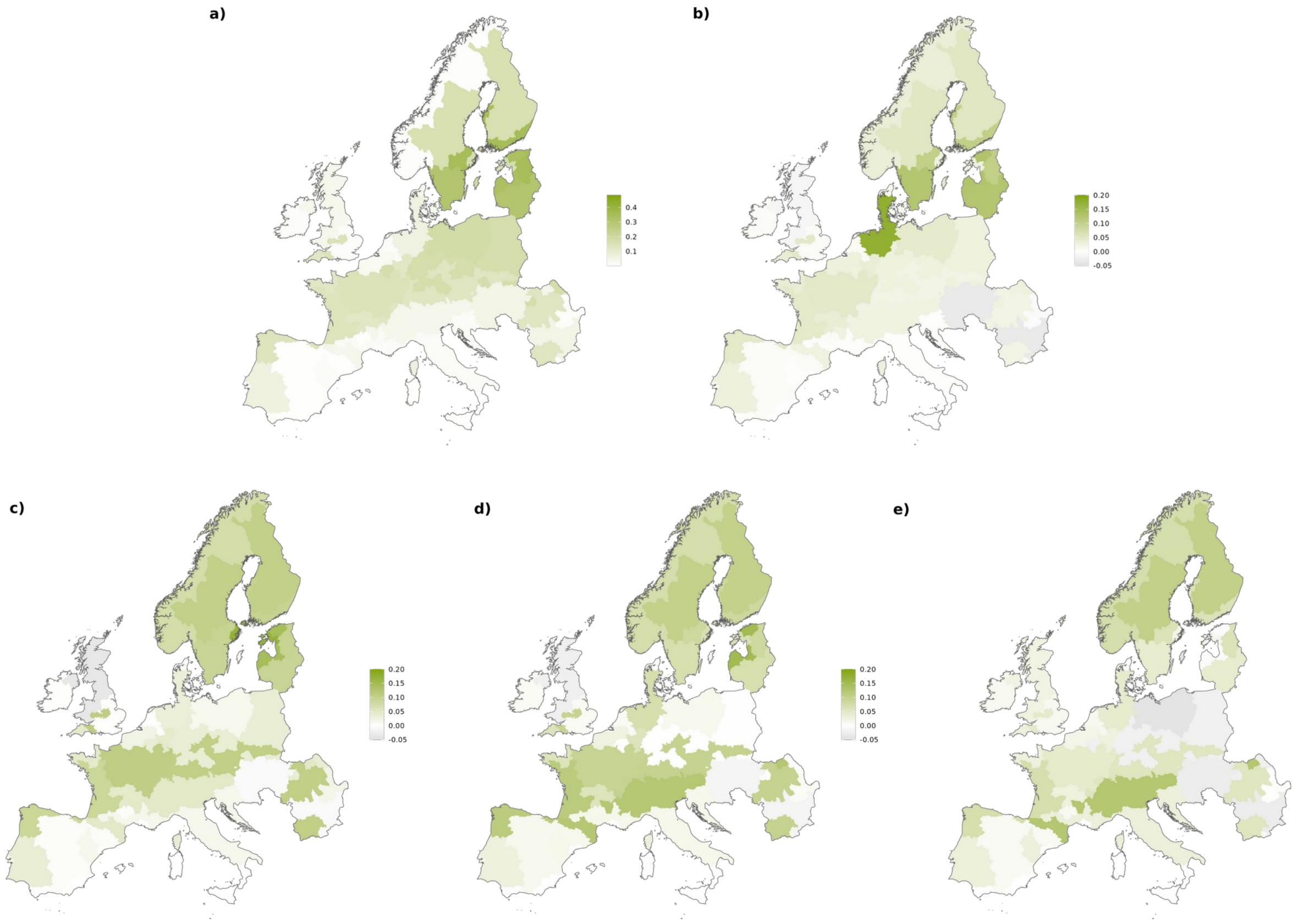

Extended Figure 5: Share of area cover by forests under high intensity management and changes by biophysical region: a) current cover of high intensity forest, b) change in cover of high intensity forest in SSP1, c) NFN, d) NFS, e) NAC.

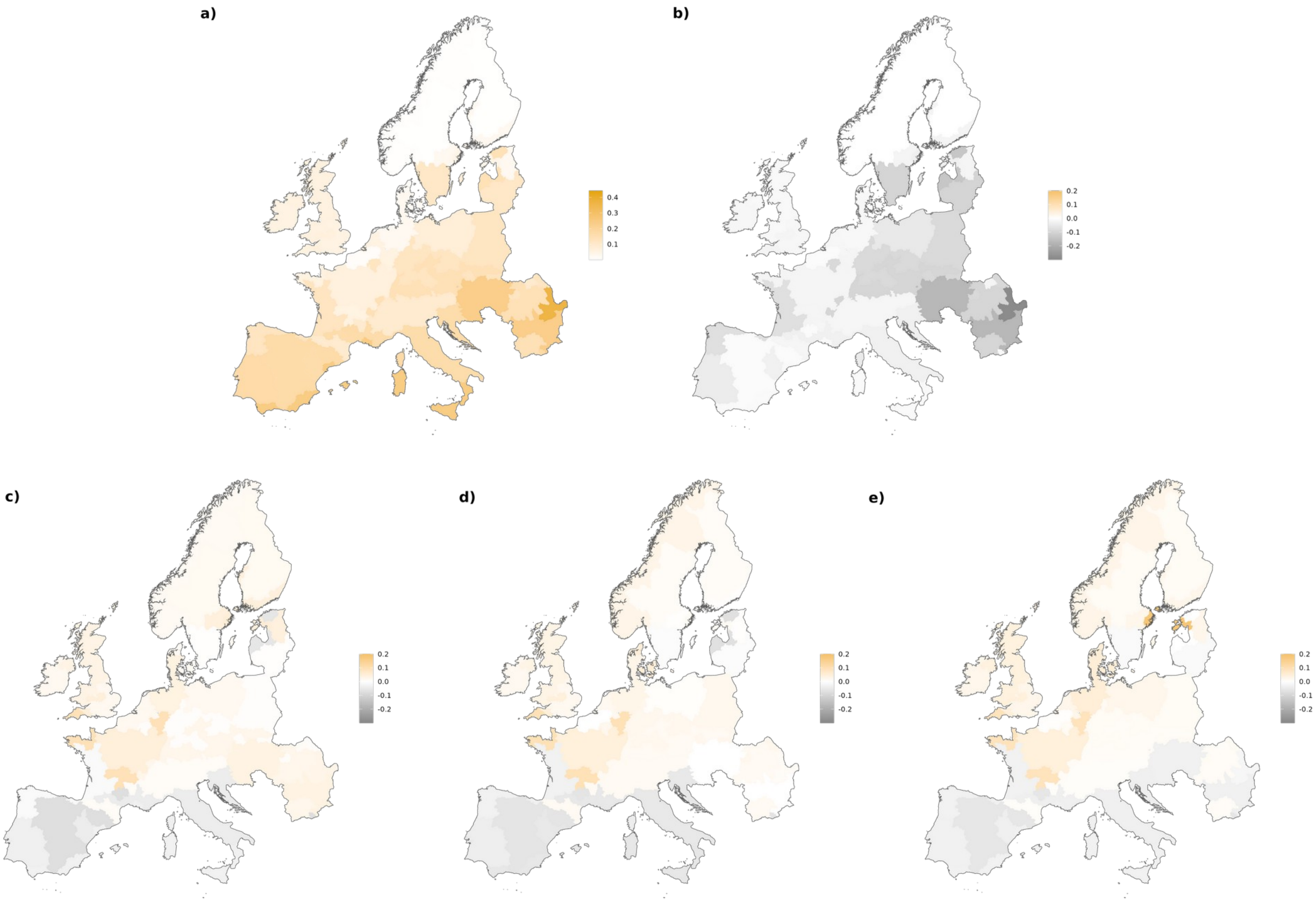

Extended Figure 6: Share of area cover by farmland under low intensity management and changes by biophysical region: a) current cover of low intensity farmland, b) change in cover of low intensity farmland in SSP1, c) NFN, d) NFS, e) NAC.

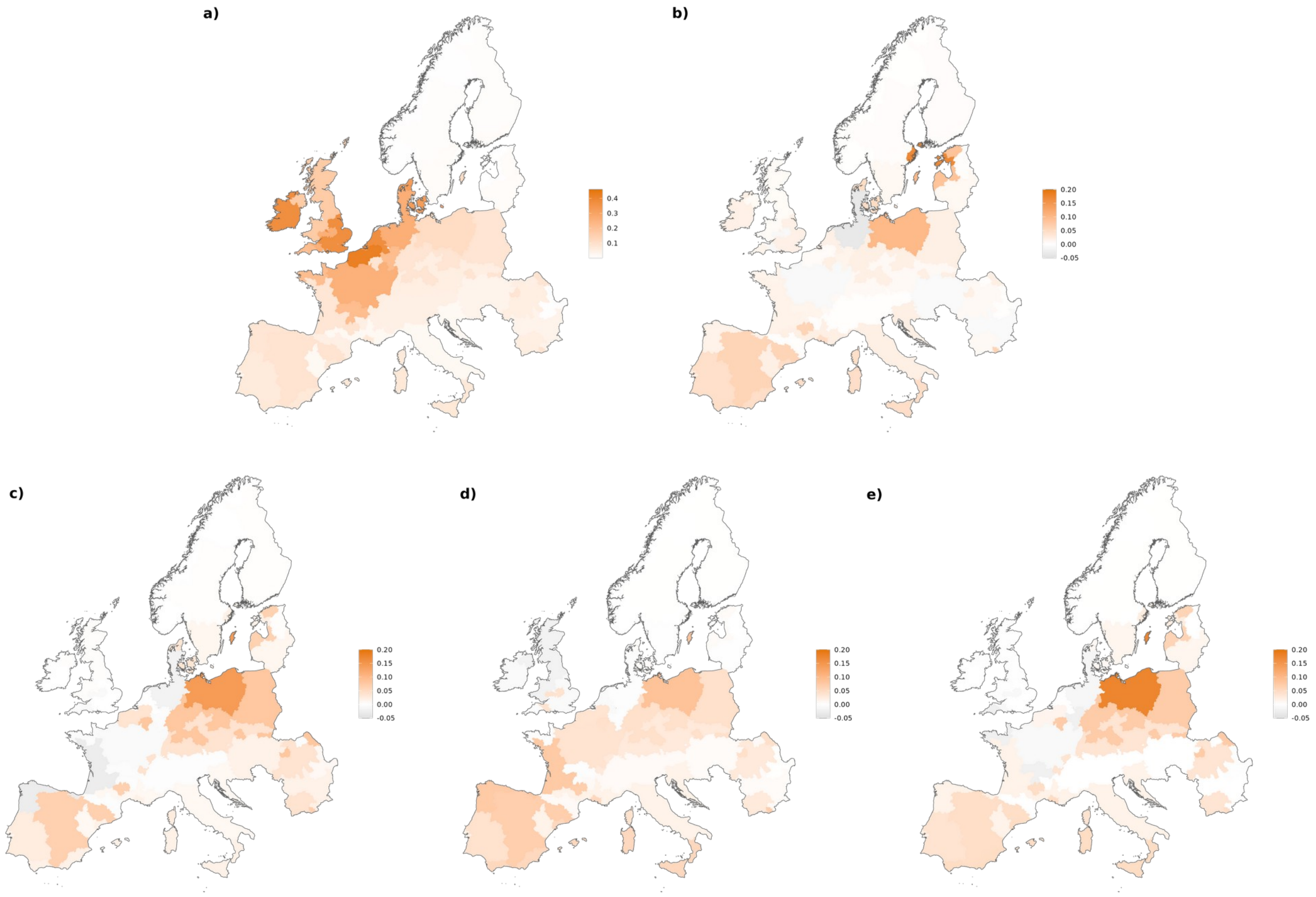

Extended Figure 7: Share of area cover by farmland under high intensity management and changes by biophysical region: a) current cover of high intensity farmland, b) change in cover of high intensity farmland in SSP1, c) NFN, d) NFS, e) NAC.

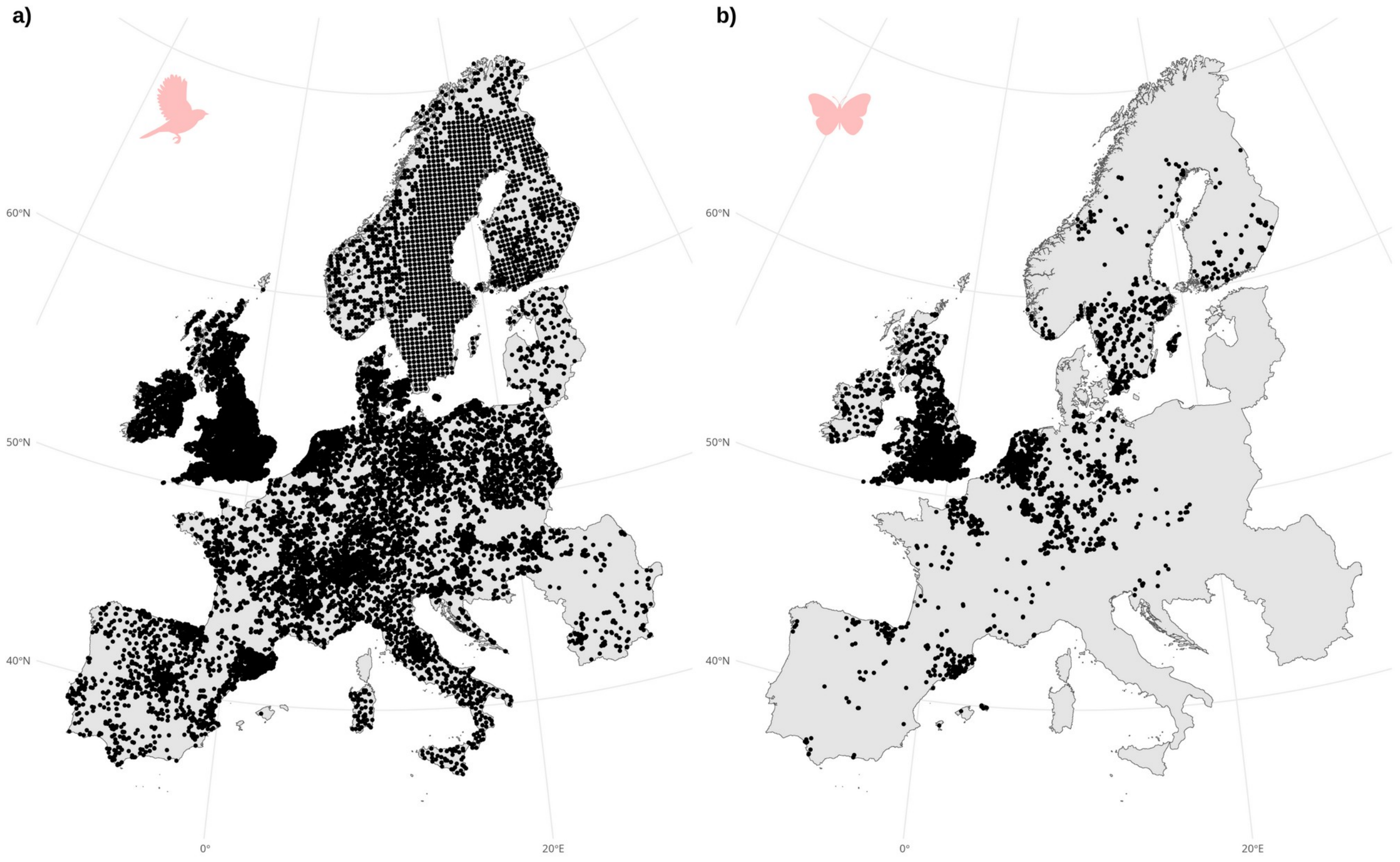

Extended Figure 8: Distribution of the sites included in this study of a) the Pan-European Common Bird Monitoring Scheme (15,731 sites in 27 countries) and b) the European Butterfly Monitoring Scheme (4,959 transects in 15 countries). This corresponds to sites and transects visited more than 5 years and for the last time in 2011 or later. The grey area represents the European countries included in this study.

| Data | Past | | Future |
| --- | --- | --- | --- |
| | Historical | Recent | |
| Birds | 2000 – 2021 (yearly) | | |
| Butterflies | 2000 – 2021 (yearly) | | |
| Climate | 2000 – 2005 (average) | 2016 – 2021 (average) | |
| Imperviousness | 2006 | 2018 | |
| Tree cover density | 2012 | 2018 | 2050 |
| Open land cover | 2000 | 2018 | |
| Land-use intensity | 2015 | | |
| Landscape diversity | 2000 | 2018 | |
| Protected areas | 2024 | | |
| Ecosystem productivity | 2000 – 2019 (average) | | |

Table 1: Details on the time period used for biodiversity, environmental and driver data

## CRediT authorship contribution statement

S.R.: Conceptualization, Methodology, Formal analysis, Writing - Original Draft, M.L.: Conceptualization, Methodology, Writing - Review & Editing, L.T.: Conceptualization, Writing - Review & Editing, S.L.: Conceptualization, Supervision, Writing - Review & Editing, Funding acquisition, T.B.: Methodology, Writing - Review & Editing, H.A., A.A., M.B., M.P.B., L.B., T.C., B.F., Z.F.F., A.G., S.H., J.A.K., J.K., P.K., M.K., A.L., D.M., X.M., M.M., I.J.Ø., L.P., J.R., J.M.R., M.Š., J.S., R.S., C.S., B.G.S., N.S., N.T., S.T. and C.v.S.: Provide national monitoring database, Writing - Review & Editing

## Declaration of competing interest

The authors declare that they have no known competing financial interests or personal relationships that could have appeared to influence the work reported in this paper.

## Acknowledgments

We would like to express our sincere gratitude to all the volunteers who provided data to the programme, without whom this research would not have been possible. We would also like to thank the European and national programme coordinators for organising the monitoring programmes in their countries and collecting the data that directly contributed to this research. We would finally like to thank M. Baragatti for helpful methodological discussions.

## Funding information

This study was partially funded by the BioAgora project funded by the European Union's Horizon Europe Research and Innovation Programme (grant agreement No. 101059438), the Horizon Europe project ECO2ADAPT (grant agreement No. 101059498) and wildE (GAP-101081251) and the Beyonds project funded by the French Research Foundation on Biodiversity.

## Data availability

Monitoring data used in this study are available upon request of the PanEuropean Common Bird Monitoring Scheme (https://pecbms.info/) and the European Butterfly Monitoring Scheme (https://butterfly-monitoring.net). The other data are used are publicly available and have been cited in the main text and references. The R script (R software version 4.4.3[117]) and data produced are available on Zenodo (https://zenodo.org/records/19218625) and the Shiny application (https://press-com-bio.sk8.inrae.fr/).

**Supplementary materials**

# Table of Contents

## Supplementary material 1 Additional results

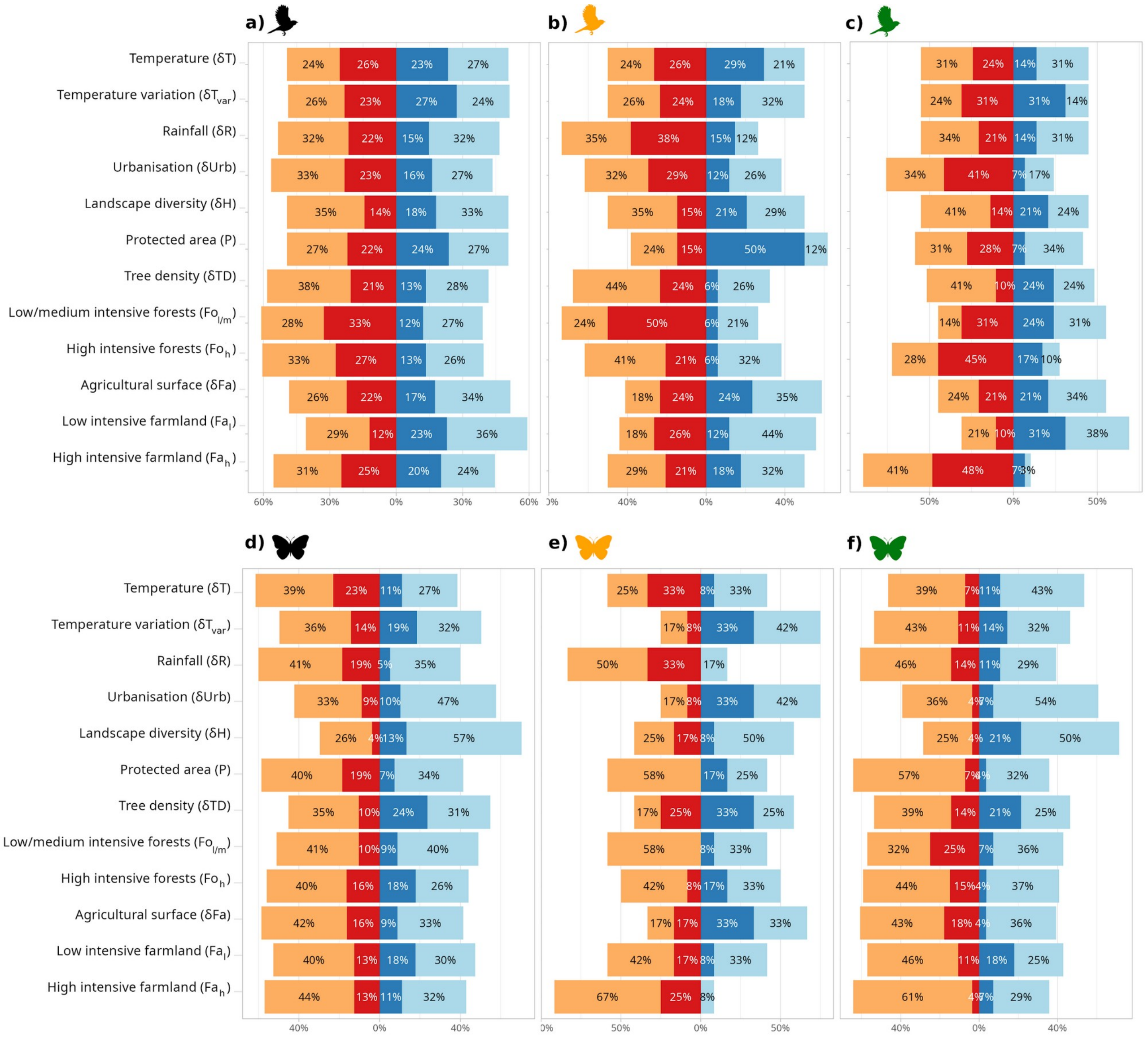

Figure S1a: Distribution of positive and negative effects of driver and environmental variables on temporal trends at the European level for all bird species (n = 223), b) farmland bird species (n = 34), c) forest bird species (n = 29), d) all butterfly species (n = 135), e) grassland butterfly species (n = 12) and f) woodland butterfly species (n = 28). Non-significant negative effects in orange, significant negative effects in dark red (p-value < 0.05), significant positive effects in dark blue (p-value < 0.05), non-significant positive effects in light blue.

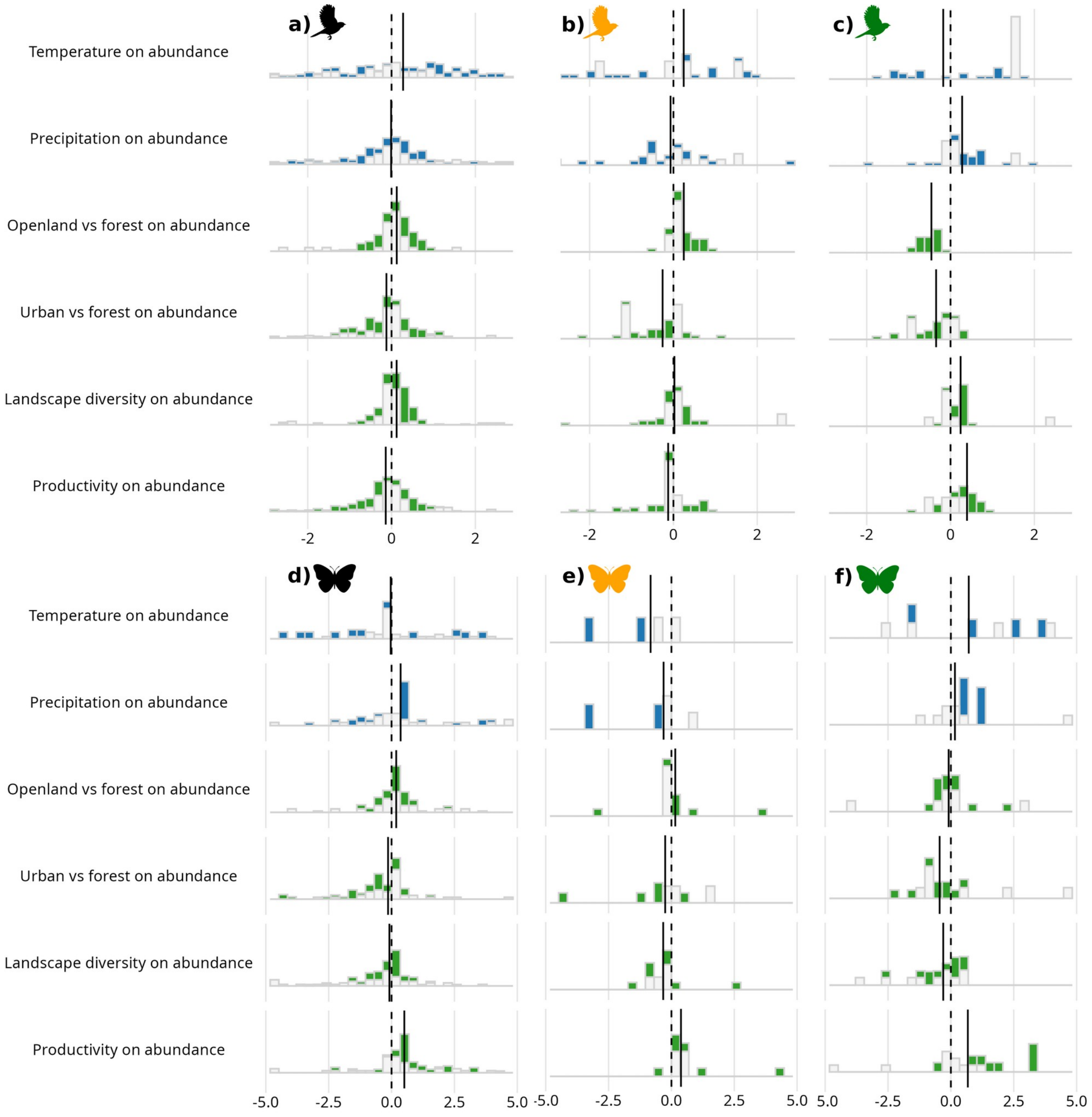

Figure S1b: Effects of climate variables (in blue), land-use variables (in dark green) and land-use intensity variables (in light green) on local abundance at the European level for all bird species (n = 223), b) farmland bird species (n = 34), c) forest bird species (n = 29), d) all butterfly species (n = 135), b) grassland butterfly species (n = 12) and c) woodland butterfly species (n = 28). Non-significant effects in grey.

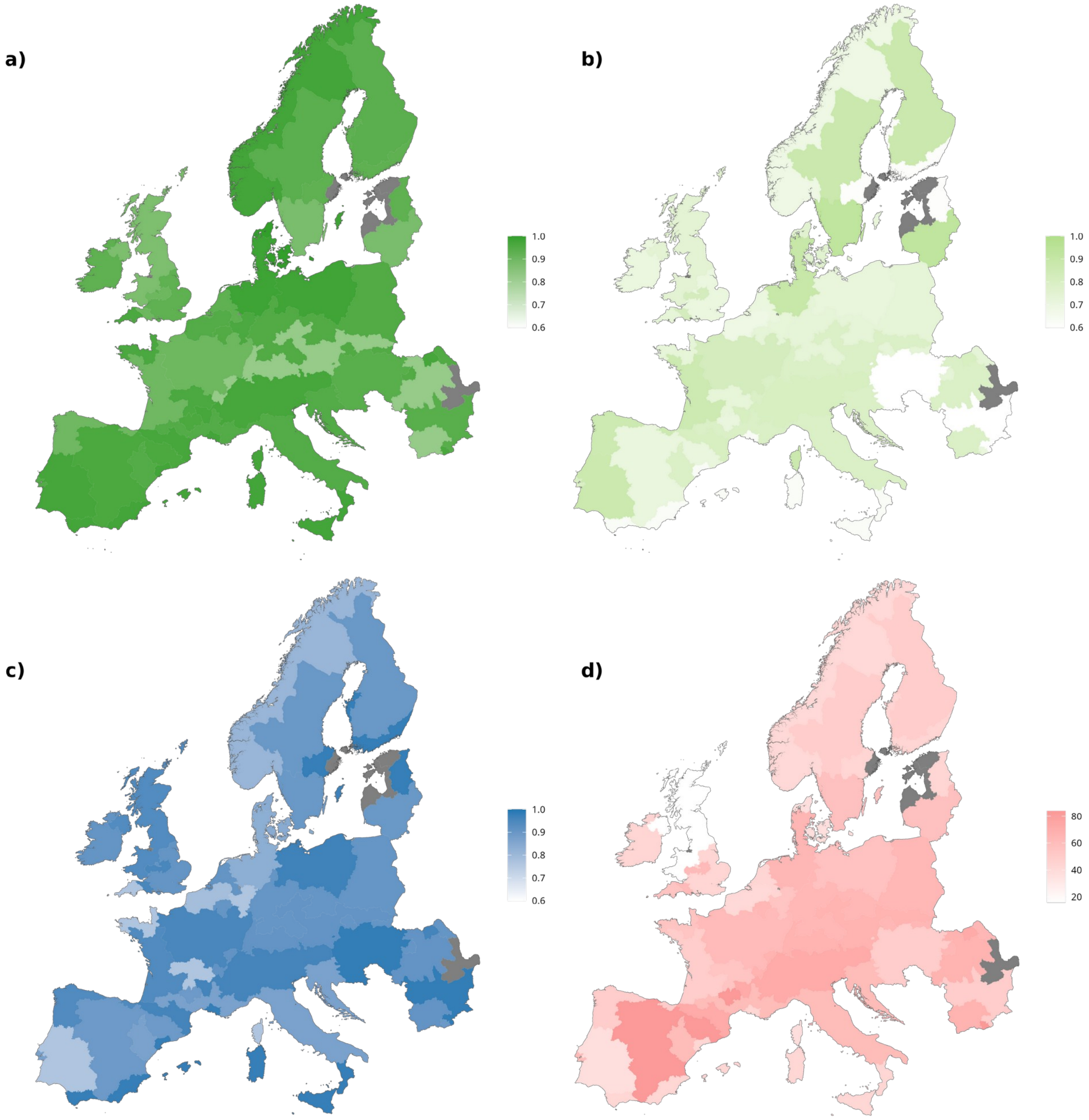

Figure S1c: Percentage of bird species trends negatively impacted by a) land-use change, b) high intensive forest or farmland, c) climate change in each biophysical regions, and d) total number of species analysed in each biophysical region. Regions with no data are in grey.

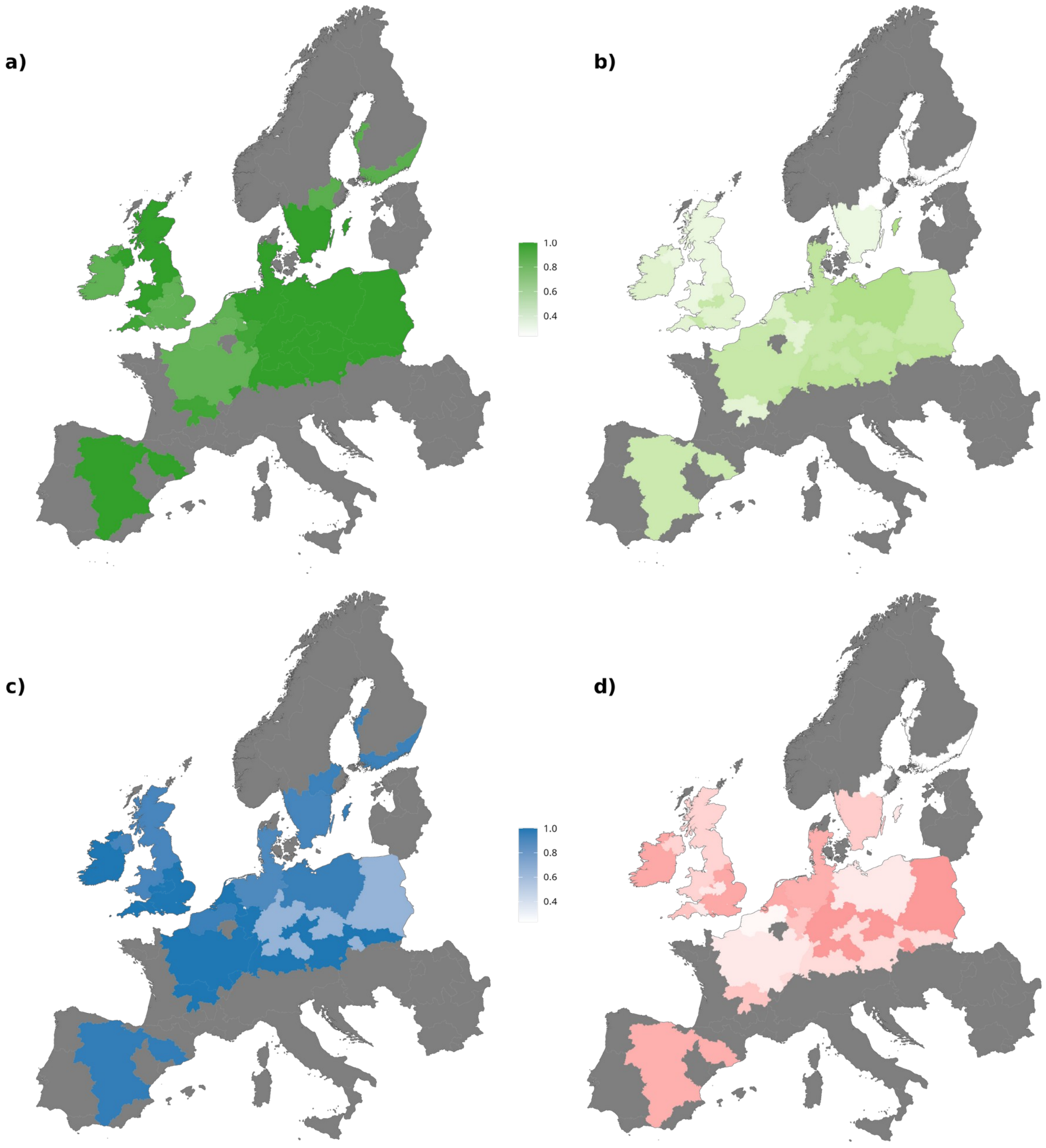

Figure S1d: Percentage of butterfly species trends negatively impacted by a) land-use change, b) high intensive forest or farmland, c) climate change in each biophysical regions, and d) total number of species analysed in each biophysical region. Regions with no data are in grey.

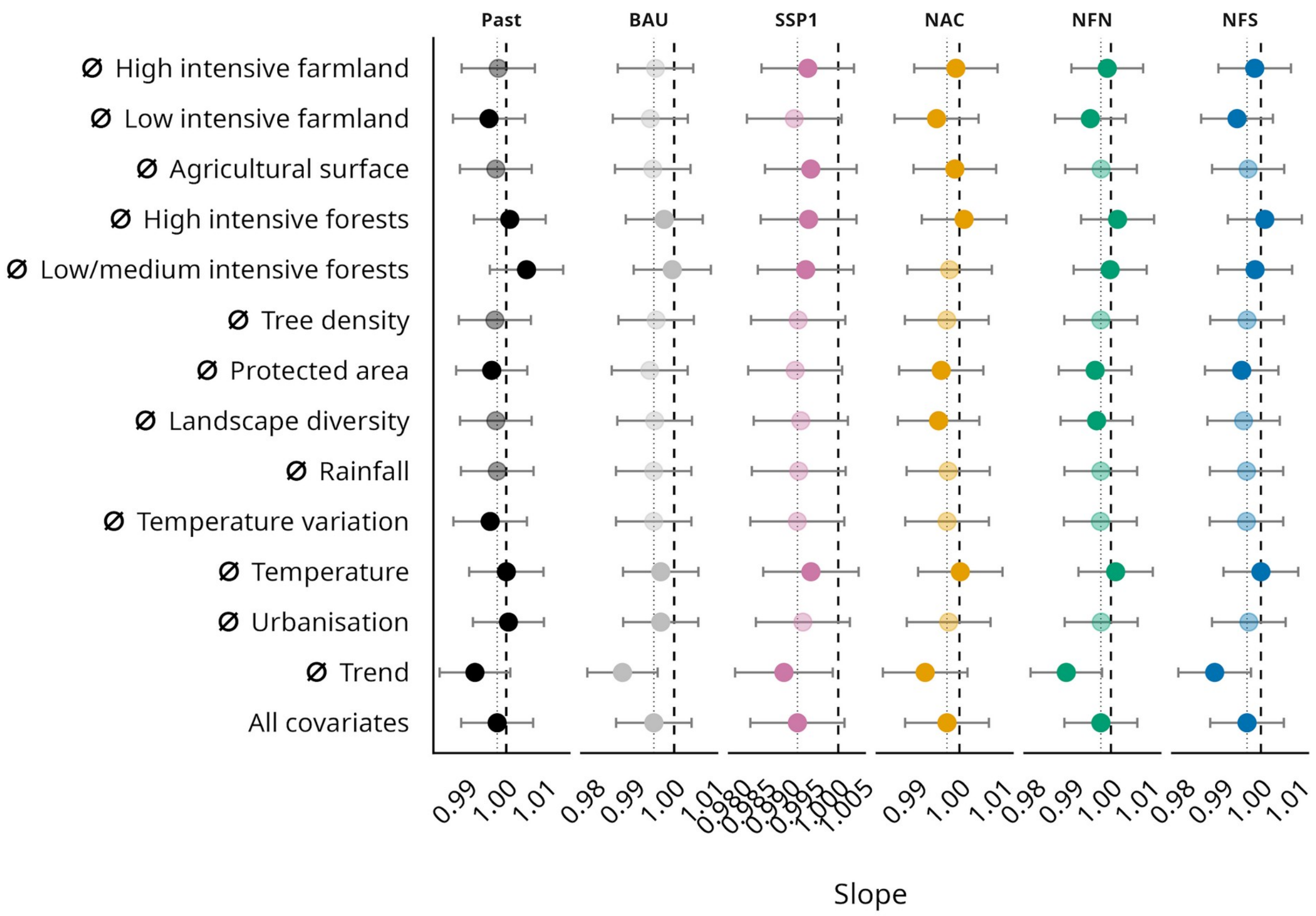

Figure S1e: Change in geometric mean of species temporal trend between 2000 and 2021 (past) and in the different scenarios (BAU, SSP1, NFN, NFS, NAC) for Common Bird index when removing each covariate. 95% confidence intervals from geometric standard error. Dots are shaded when not significantly different from the geometric mean calculated from all covariates. If the geometric mean is shifted to the right, the covariate has a negative effect, if it is shifted to the left, it has a positive effect.

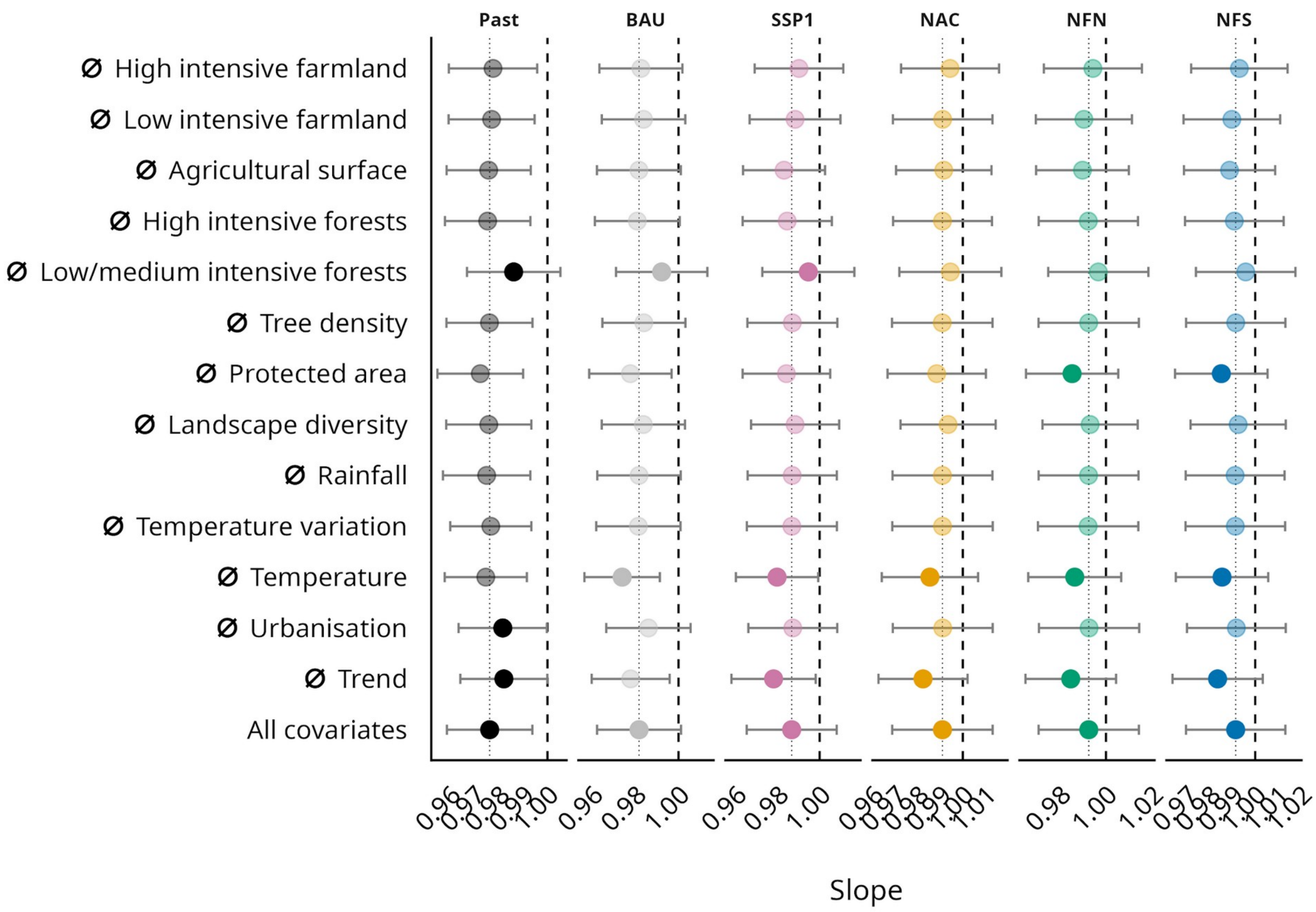

Figure S1f: Change in geometric mean of species temporal trend between 2000 and 2021 (past) and in the different scenarios (BAU, SSP1, NFN, NFS, NAC) for Farmland Bird index when removing each covariate. 95% confidence intervals from geometric standard error. Dots are shaded when not significantly different from the geometric mean calculated from all covariates. If the geometric mean is shifted to the right, the covariate has a negative effect, if it is shifted to the left, it has a positive effect.

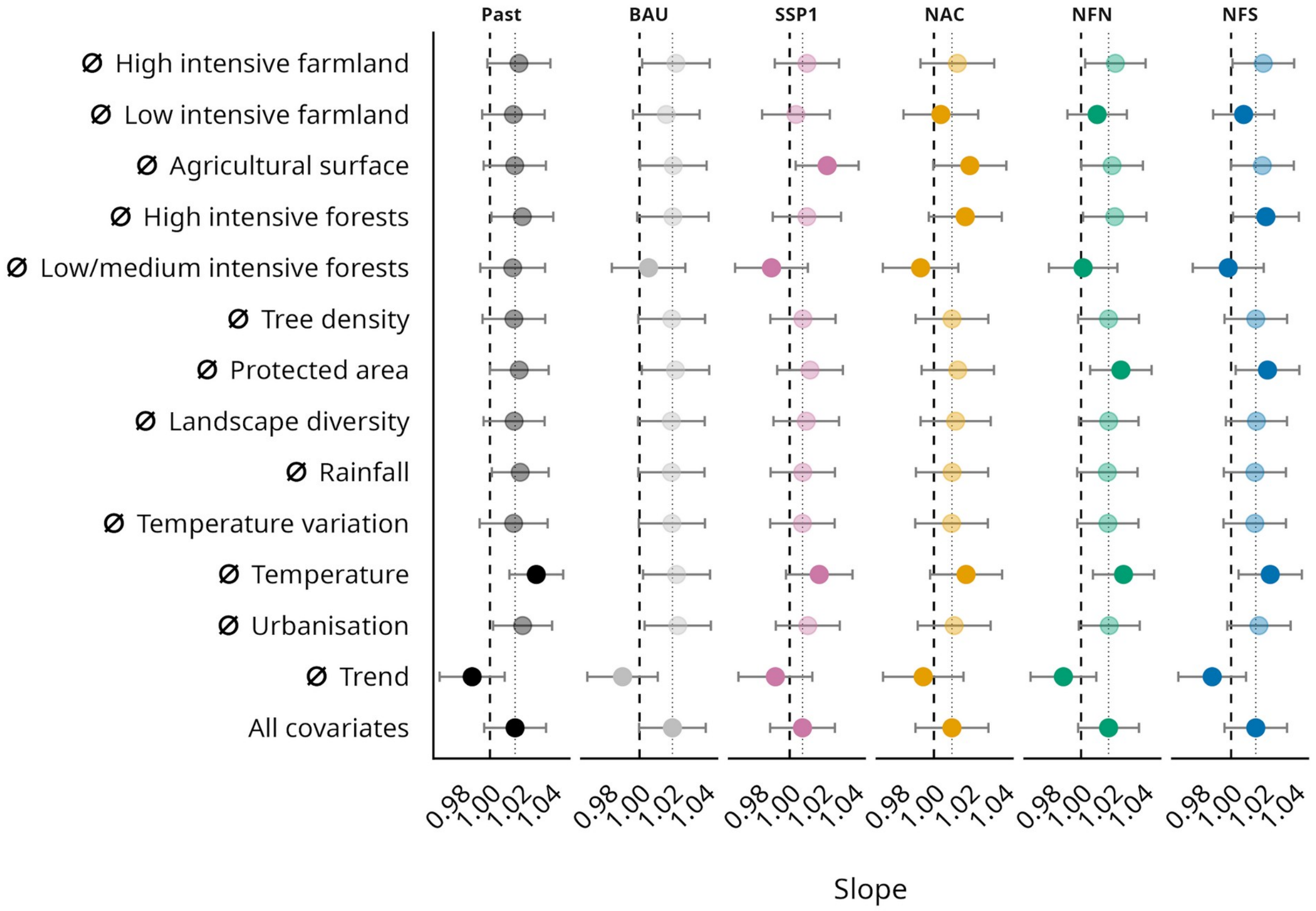

Figure S1g: Change in geometric mean of species temporal trend between 2000 and 2021 (past) and in the different scenarios (BAU, SSP1, NFN, NFS, NAC) for Forest Bird index when removing each covariate. 95% confidence intervals from geometric standard error. Dots are shaded when not significantly different from the geometric mean calculated from all covariates. If the geometric mean is shifted to the right, the covariate has a negative effect, if it is shifted to the left, it has a positive effect.

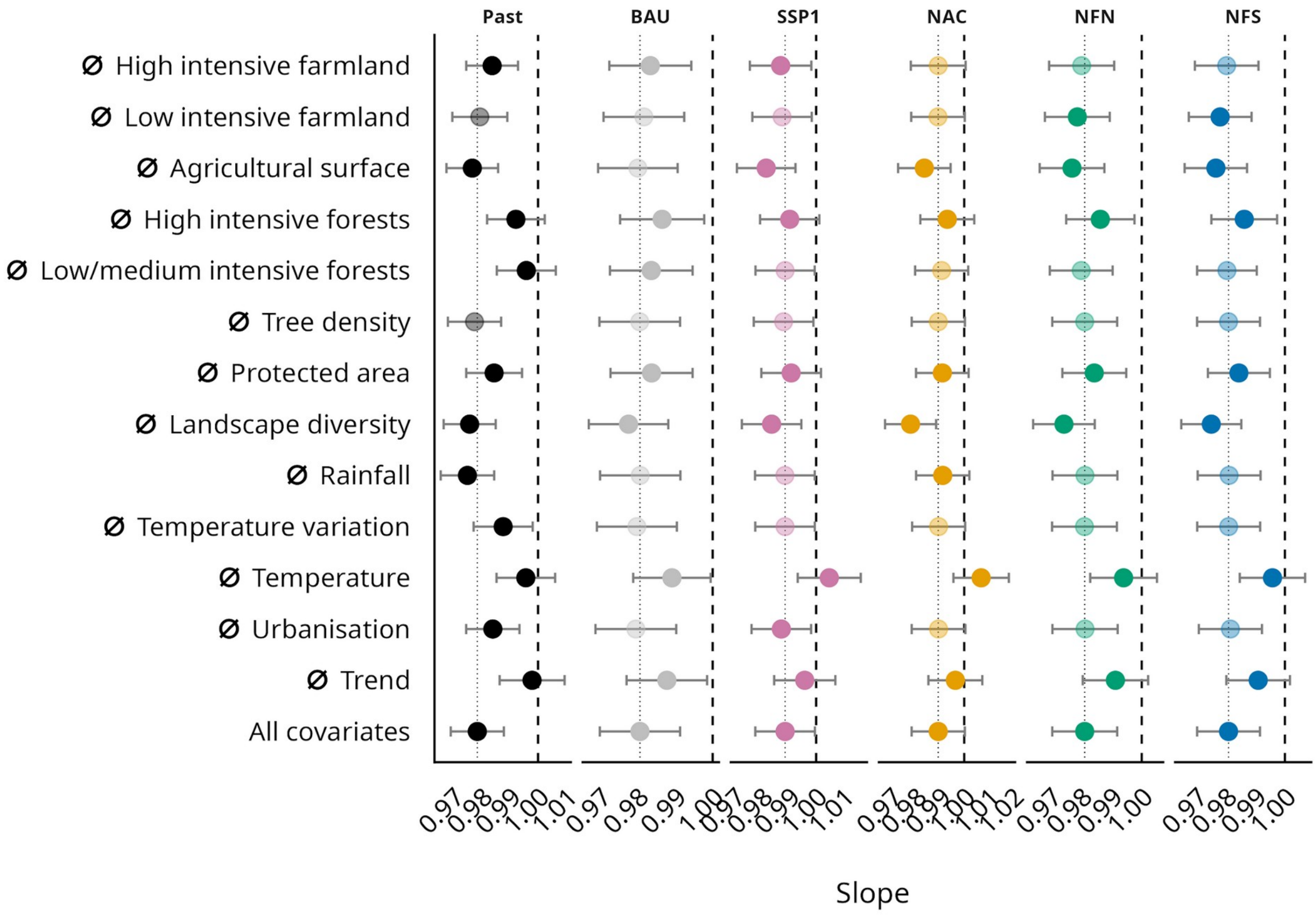

Figure S1h: Change in geometric mean of species temporal trend between 2000 and 2021 (past) and in the different scenarios (BAU, SSP1, NFN, NFS, NAC) for Common Butterfly index when removing each covariate. 95% confidence intervals from geometric standard error. Dots are shaded when not significantly different from the geometric mean calculated from all covariates. If the geometric mean is shifted to the right, the covariate has a negative effect, if it is shifted to the left, it has a positive effect.

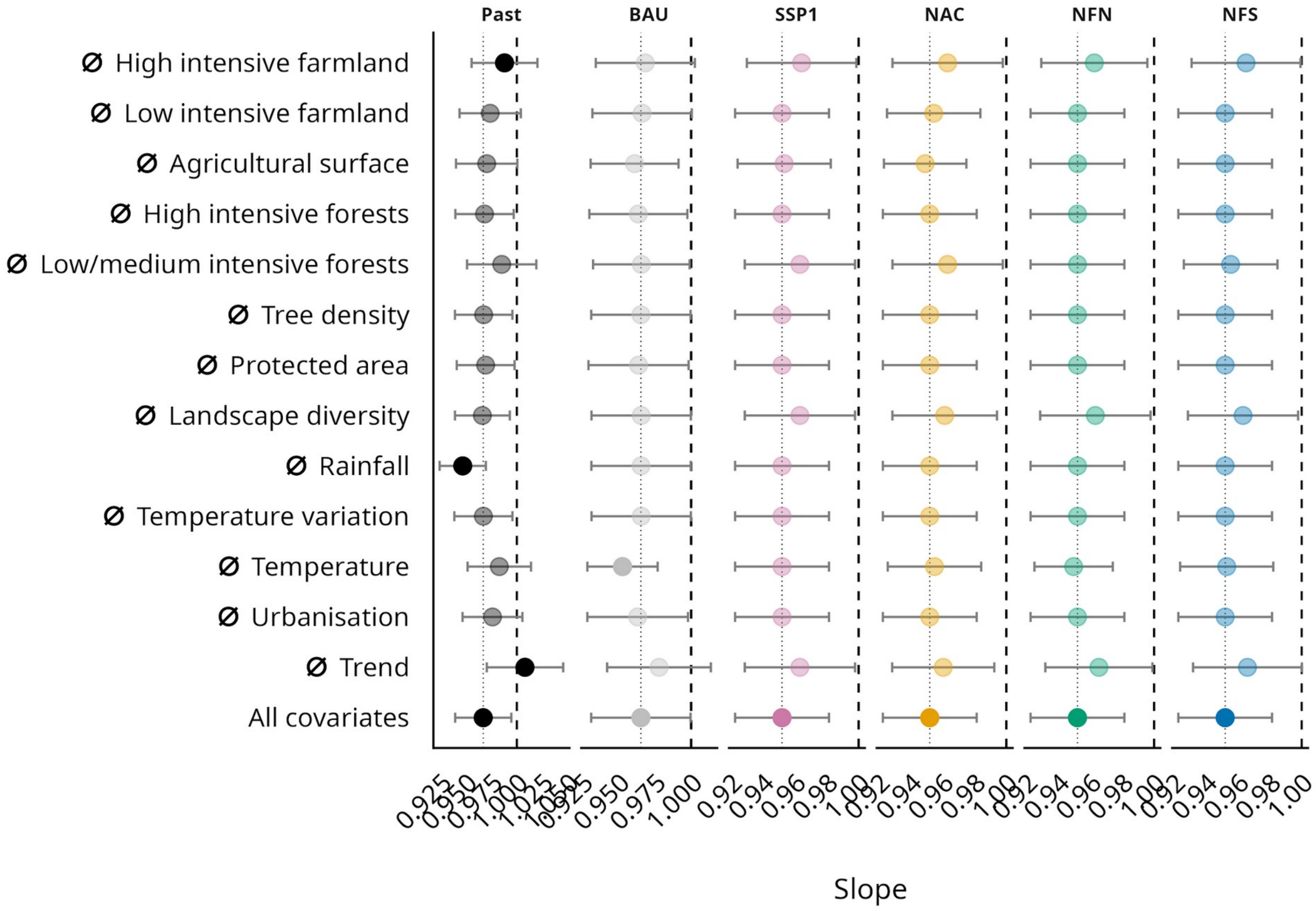

Figure S1i: Change in geometric mean of species temporal trend between 2000 and 2021 (past) and in the different scenarios (BAU, SSP1, NFN, NFS, NAC) for Grassland Butterfly index when removing each covariate. 95% confidence intervals from geometric standard error. Dots are shaded when not significantly different from the geometric mean calculated from all covariates. If the geometric mean is shifted to the right, the covariate has a negative effect, if it is shifted to the left, it has a positive effect.

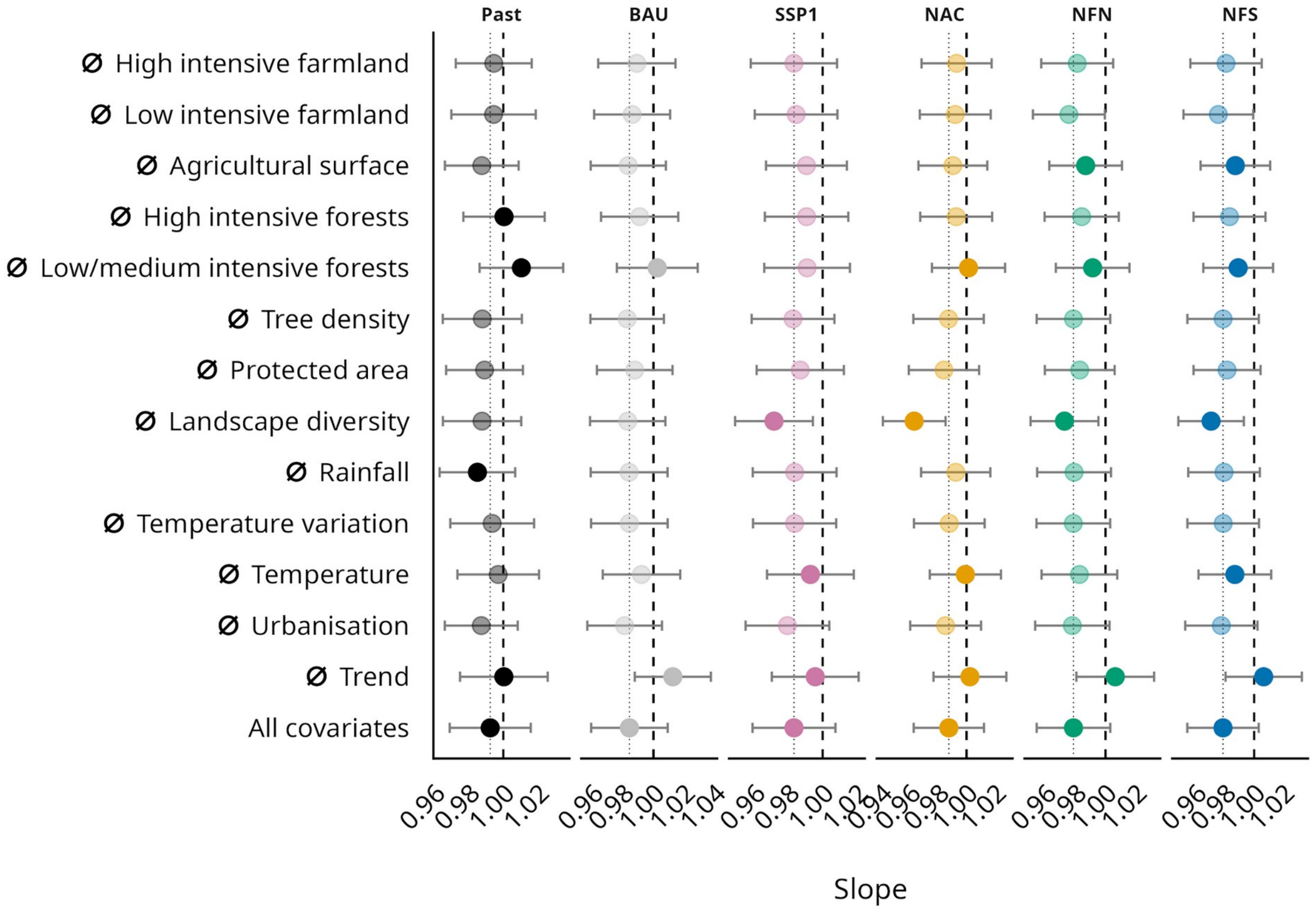

Figure S1j: Change in geometric mean of species temporal trend between 2000 and 2021 (past) and in the different scenarios (BAU, SSP1, NFN, NFS, NAC) for Woodland Butterfly index when removing each covariate. 95% confidence intervals from geometric standard error. Dots are shaded when not significantly different from the geometric mean calculated from all covariates. If the geometric mean is shifted to the right, the covariate has a negative effect, if it is shifted to the left, it has a positive effect.

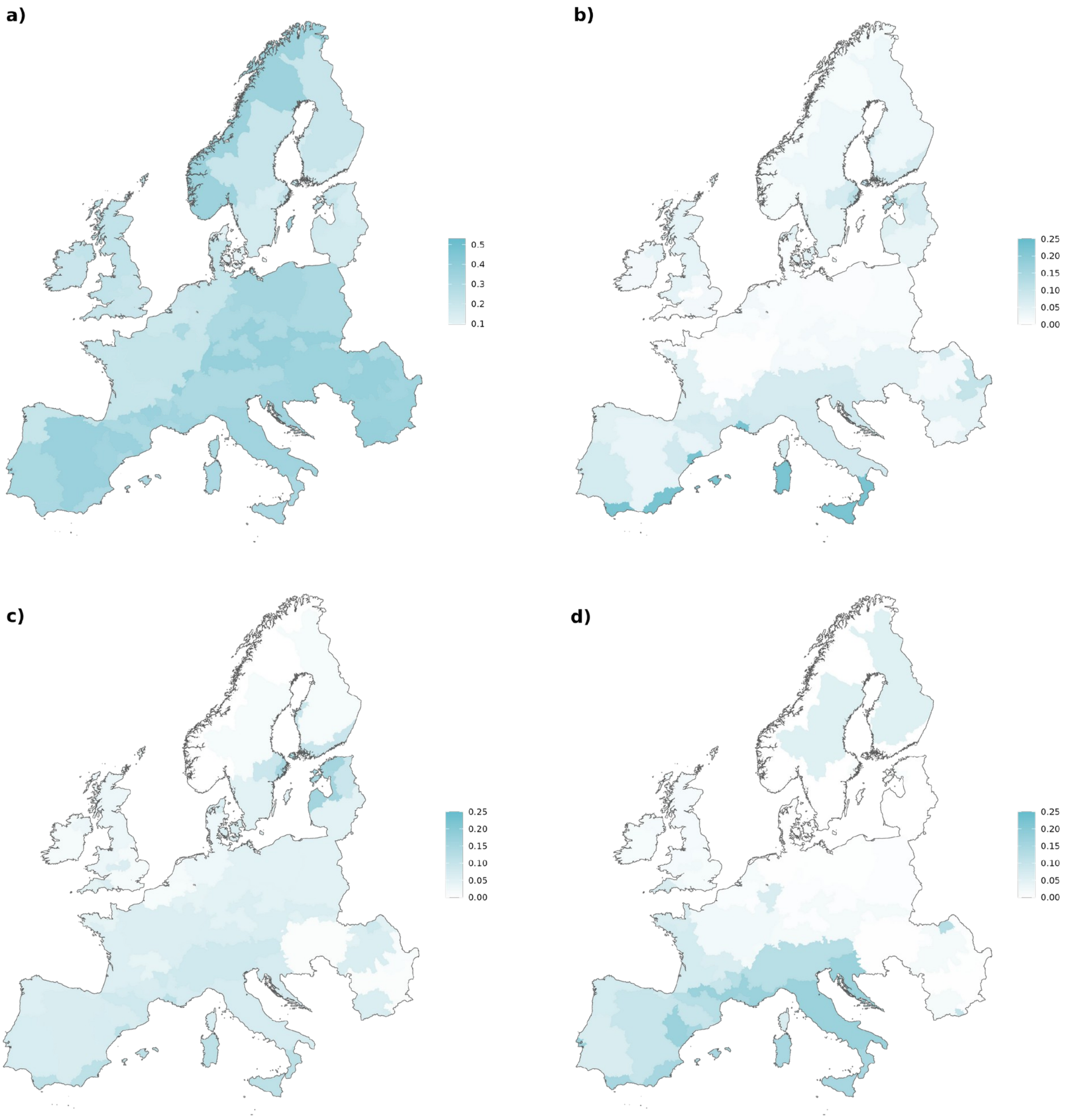

Figure S1k: Protected areas and changes by biophysical region: a) current proportion of Natura 2000 area, b) increase in proportion of Natura 2000 area in NFN, c) increase in proportion of Natura 2000 area in NFS, d) increase in proportion of Natura 2000 area in NAC.

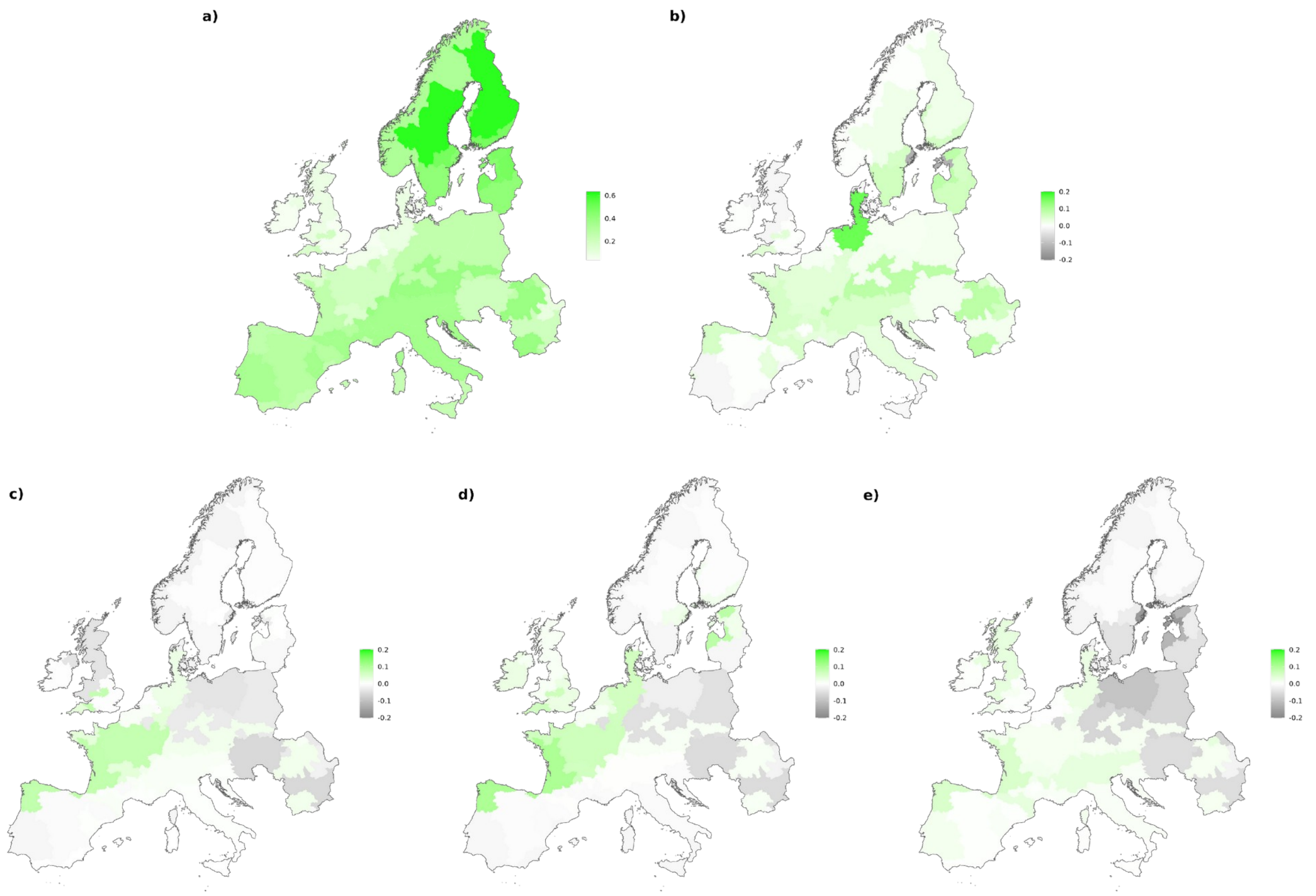

Figure S1l: Share of forest cover and changes by biophysical region: a) current forest cover, b) change in forest cover in SSP1, c) NFN, d) NFS, e) NAC.

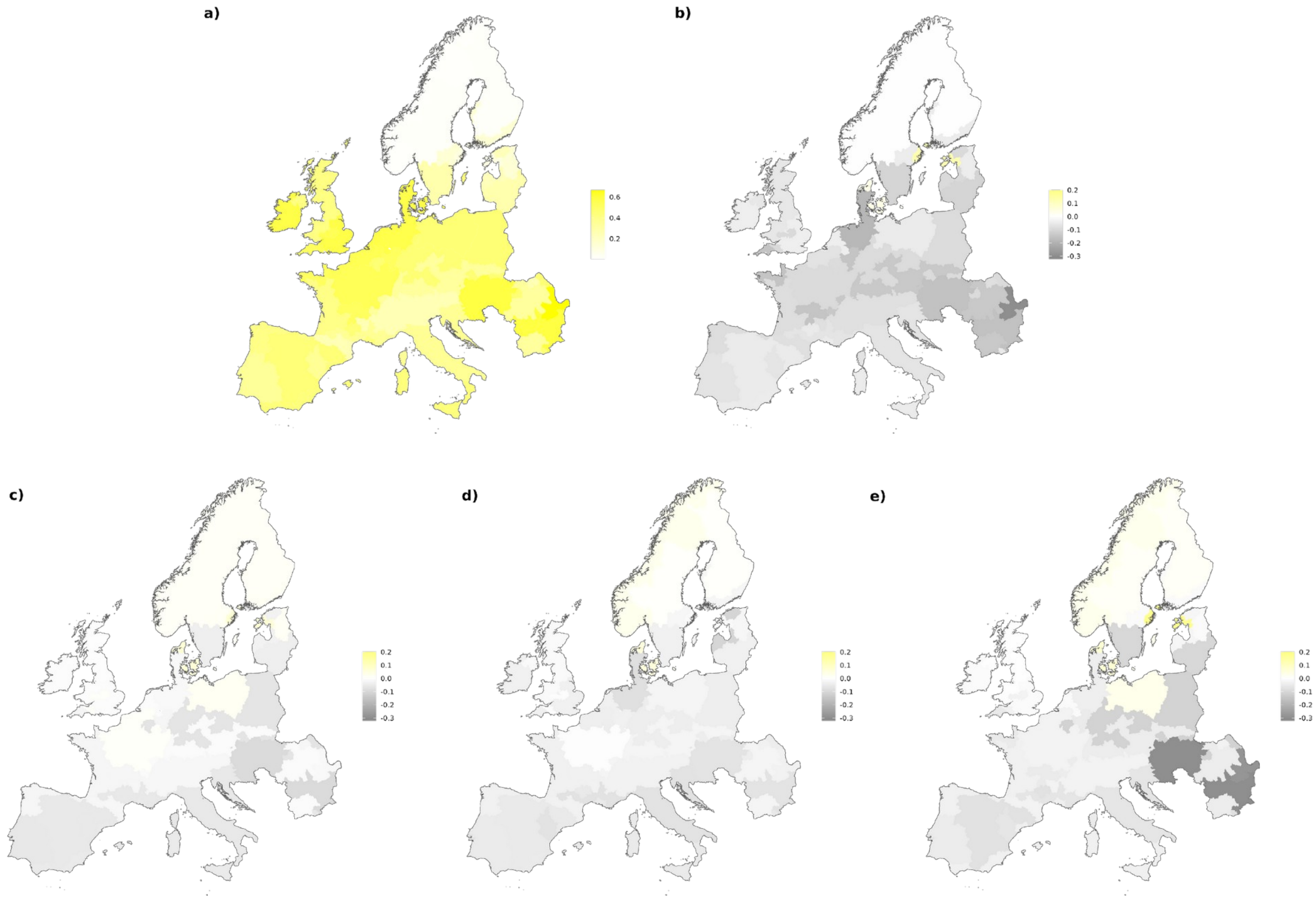

Figure S1m: Share of agricultural area and changes by biophysical region: a) current agricultural area, b) change in agricultural area in SSP1, c) NFN, d) NFS, e) NAC.

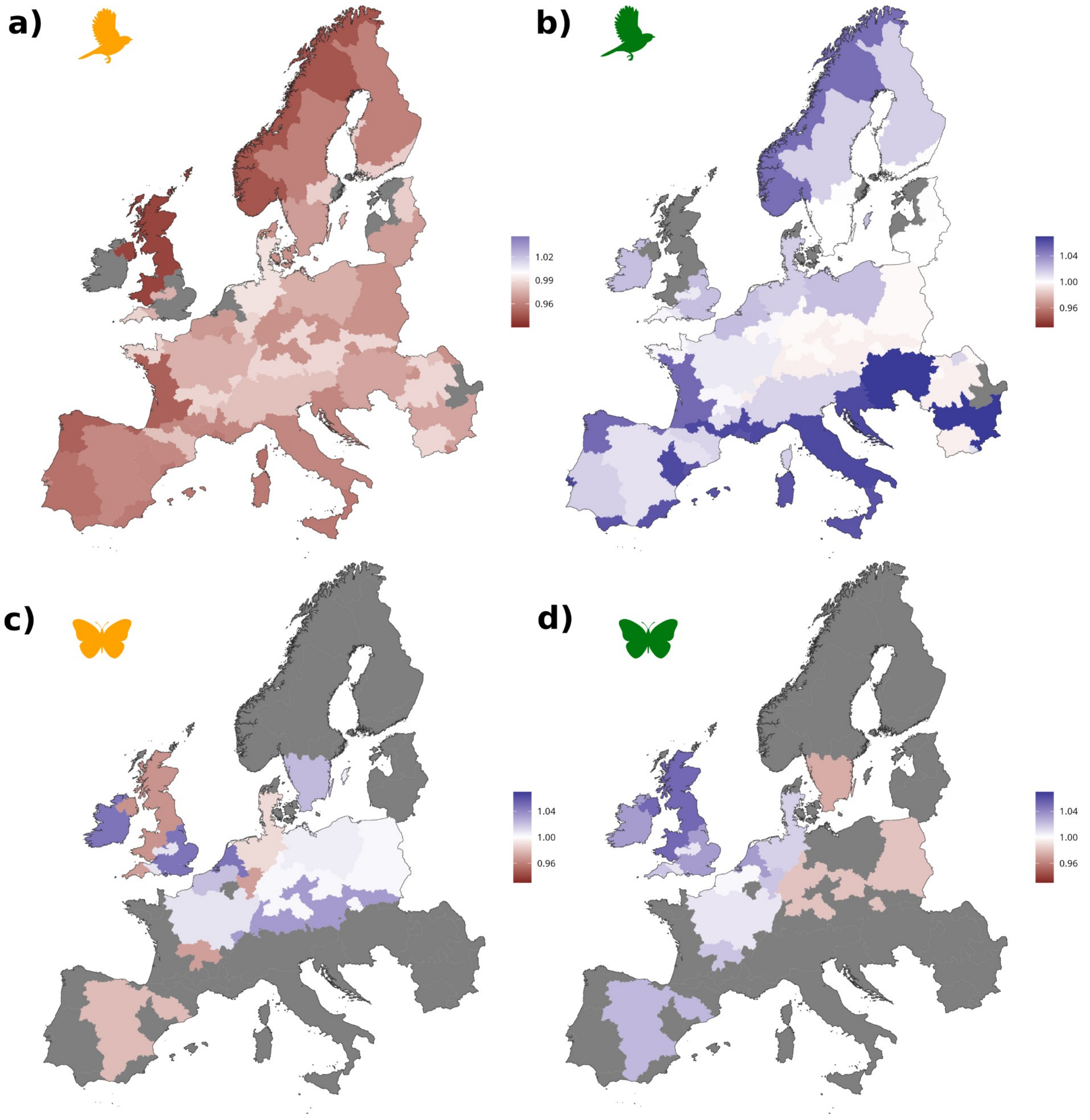

Figure S1n: Mean past annual trend (2000-2021) in abundance by biophysical region for a) farmland bird species, b) forest bird species, c) grassland butterfly species and d) woodland butterfly species. Regions in grey have either not enough data for trend estimate, or less than 3 species with trend models with a correct fit.

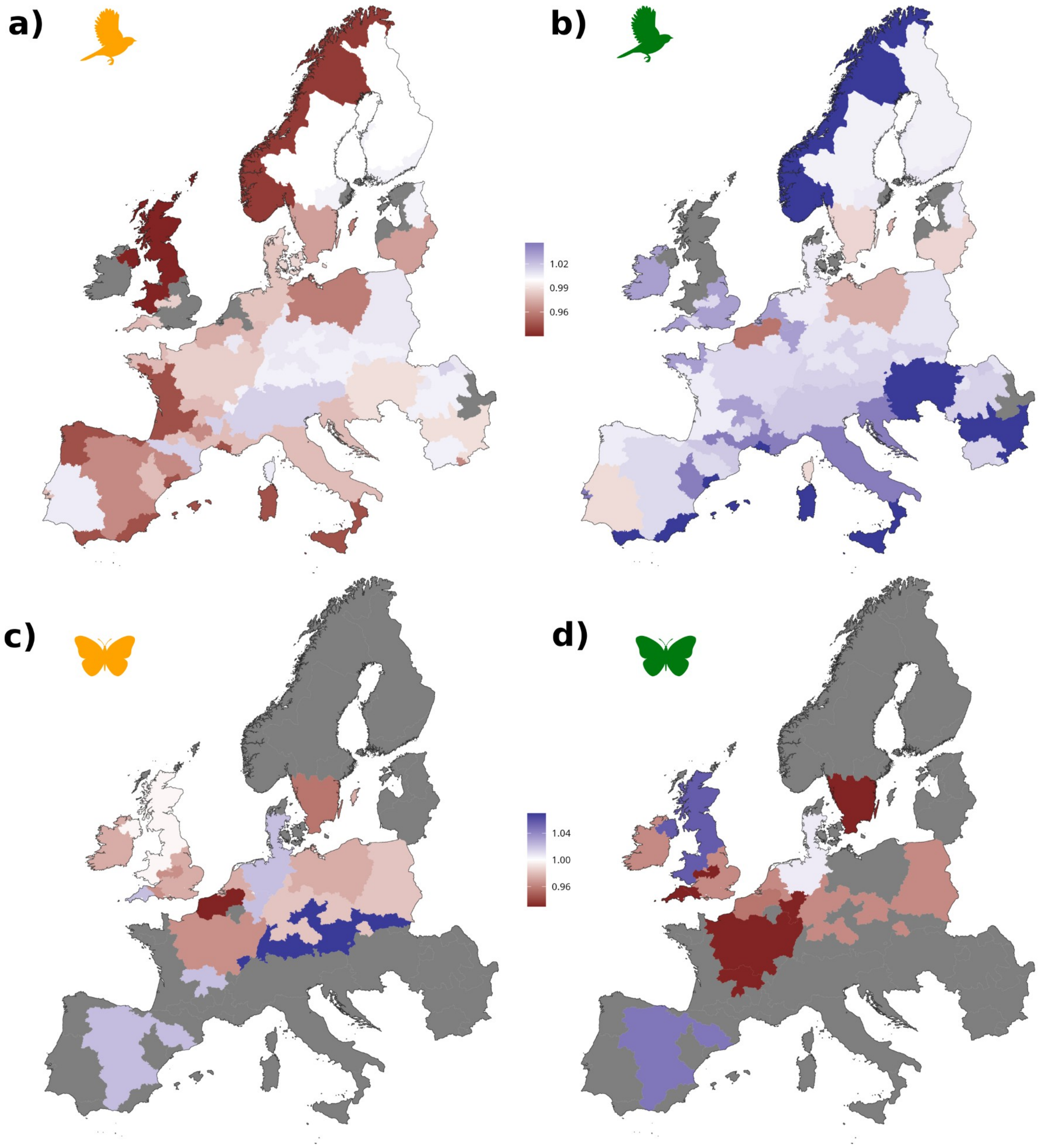

Figure S1o: Mean annual trend in abundance by biophysical region in the Business as Usual scenario for a) farmland bird species, b) forest bird species, c) grassland butterfly species and d) woodland butterfly species. Regions in grey have either not enough data for trend estimate, or less than 3 species with trend models with a correct fit.

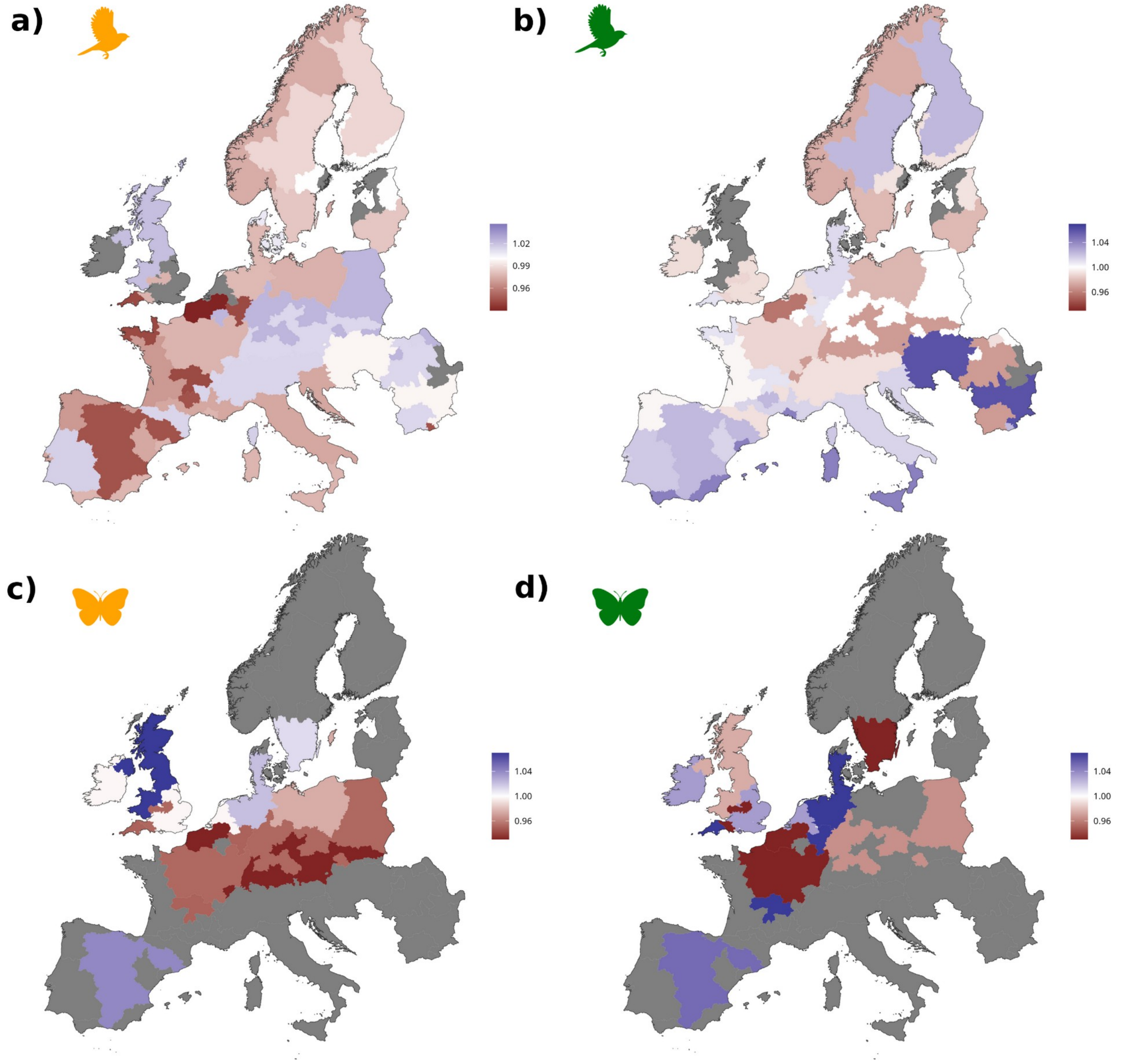

Figure S1p: Mean annual trend in abundance by biophysical region in the SSP1 scenario for a) farmland bird species, b) forest bird species, c) grassland butterfly species and d) woodland butterfly species. Regions in grey have either not enough data for trend estimate, or less than 3 species with trend models with a correct fit.

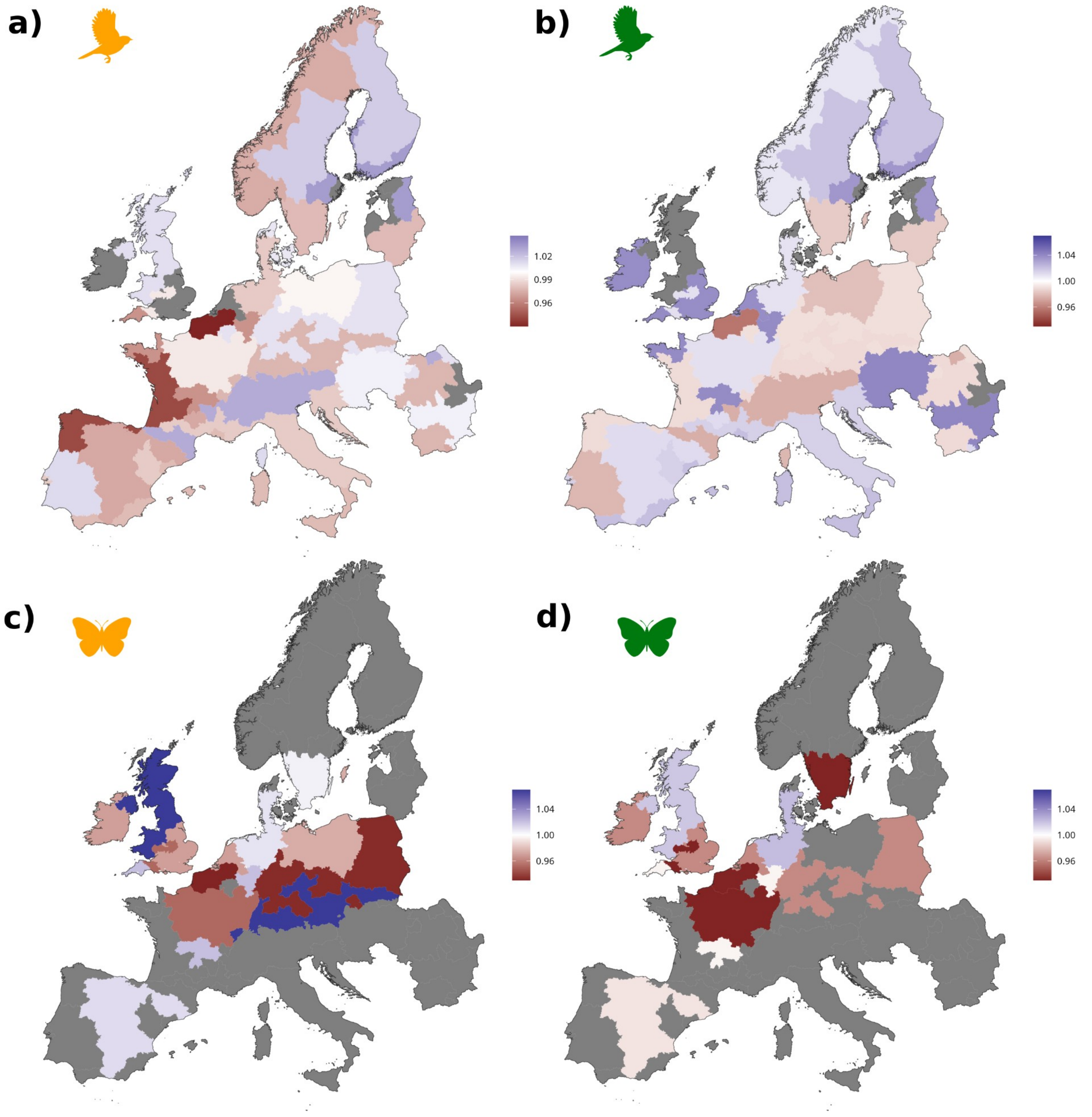

Figure S1q: Mean annual trend in abundance by biophysical region in the Nature For Nature scenario for a) farmland bird species, b) forest bird species, c) grassland butterfly species and d) woodland butterfly species. Regions in grey have either not enough data for trend estimate, or less than 3 species with trend models with a correct fit.

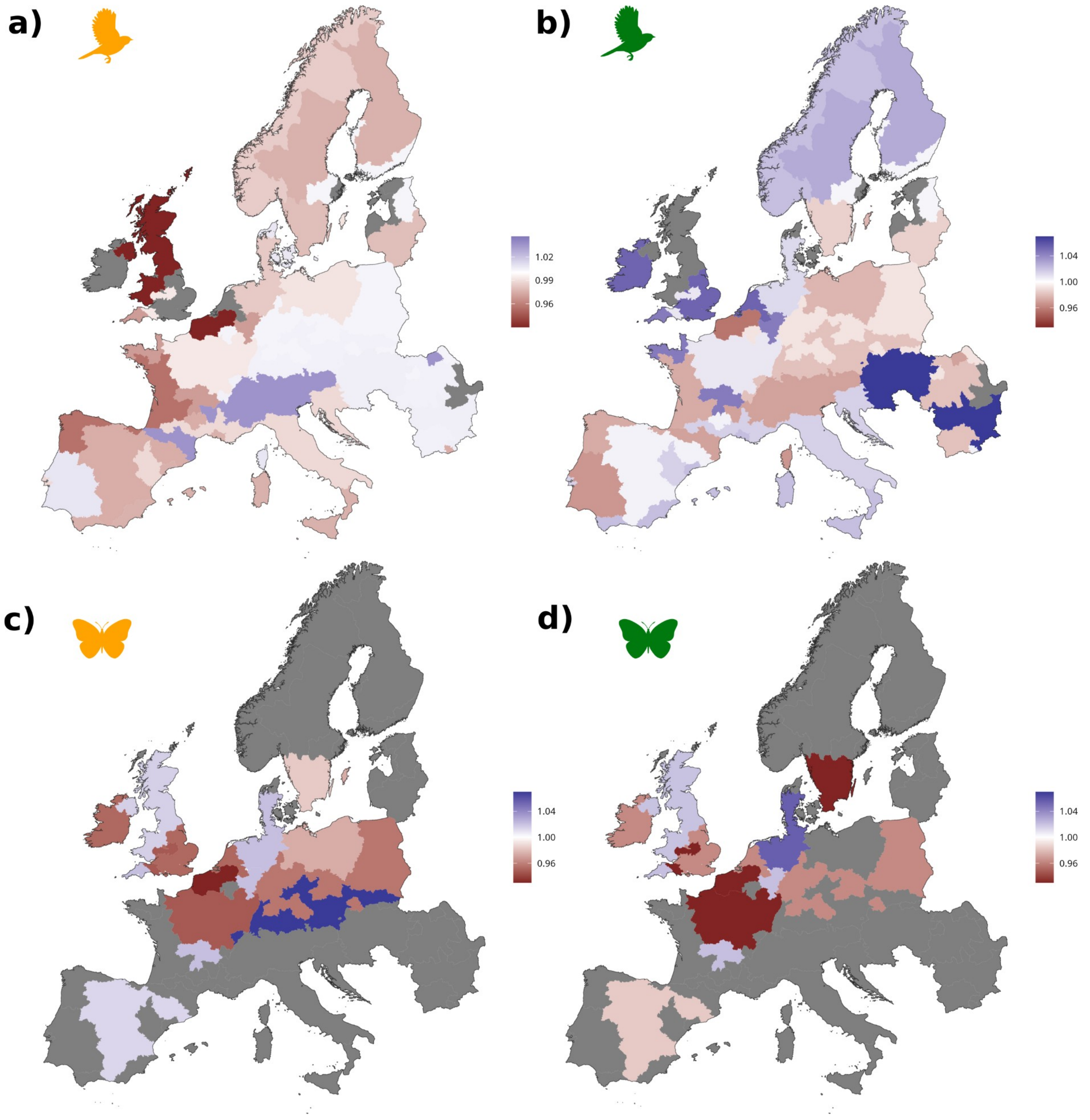

Figure S1r: Mean annual trend in abundance by biophysical region in the Nature For Society scenario for a) farmland bird species, b) forest bird species, c) grassland butterfly species and d) woodland butterfly species. Regions in grey have either not enough data for trend estimate, or less than 3 species with trend models with a correct fit.

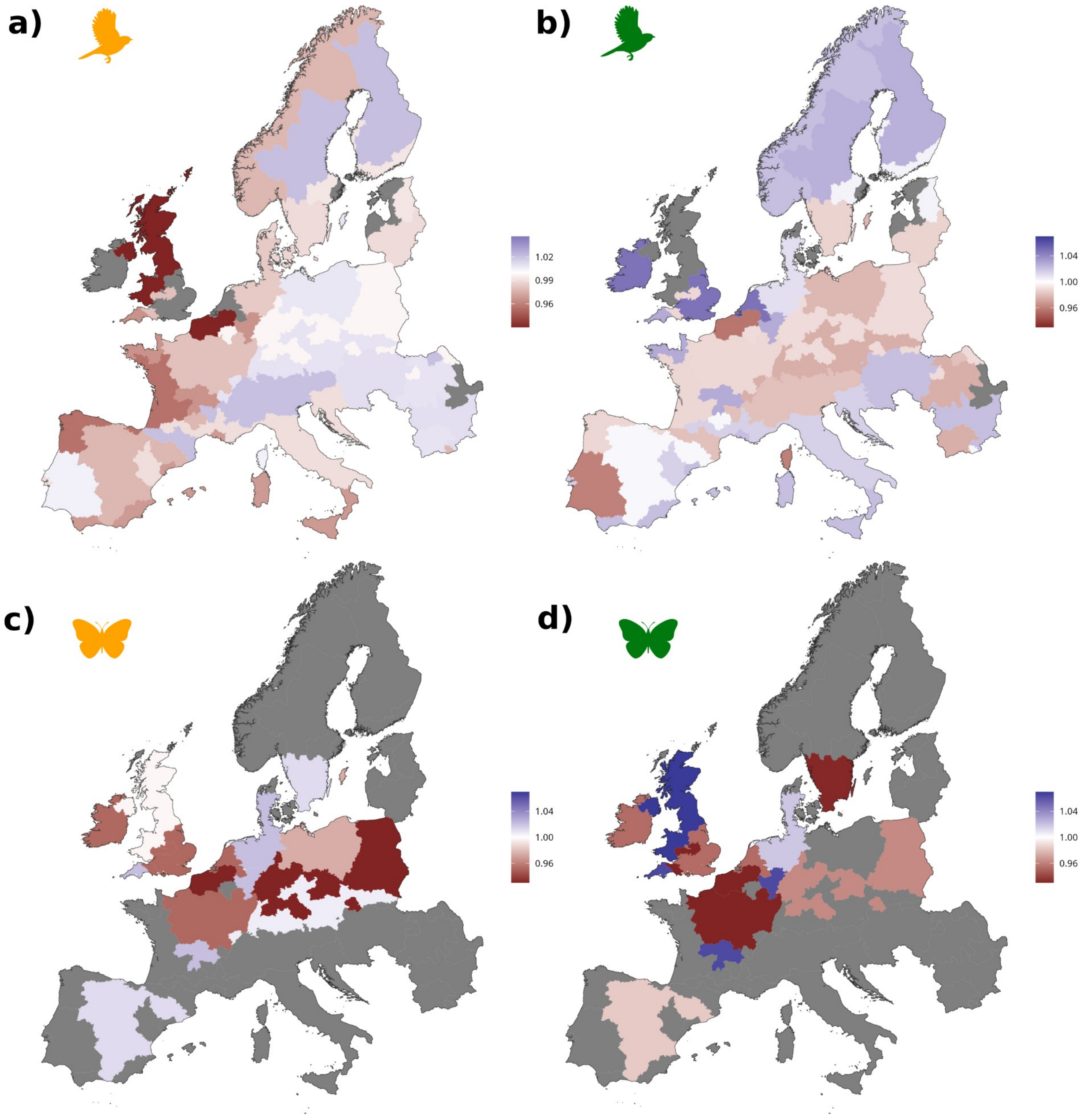

Figure S1s: Mean annual trend in abundance by biophysical region in the Nature As Culture scenario for a) farmland bird species, b) forest bird species, c) grassland butterfly species and d) woodland butterfly species. Regions in grey have either not enough data for trend estimate, or less than 3 species with trend models with a correct fit.

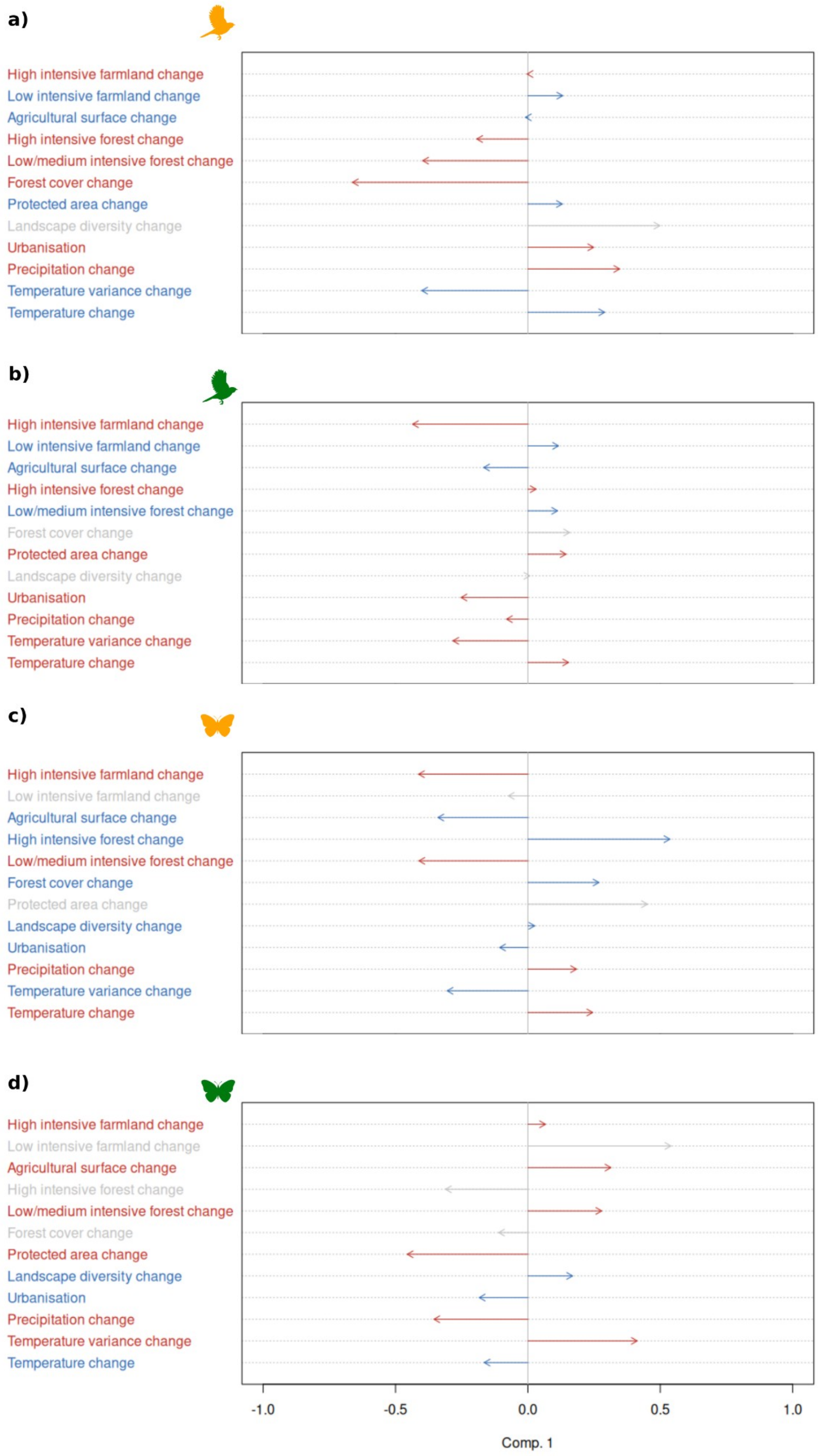

Figure S1t: Component of the Linear Discriminant Analysis (LDA) first axis separating biophysical regions by trend in the scenario with the most positive trend at the European level (NFN for birds, NAC for butterflies). Variables associated with positive trends on the right, and with negative trends on the left, for a) Farmland Birds, b) Forest Birds, c) Grassland Butterflies and d) Woodland Butterflies. Driver colour shows the sign of the effect on a majority of species at the European level (red for negative, blue for positive and grey for uncertain).

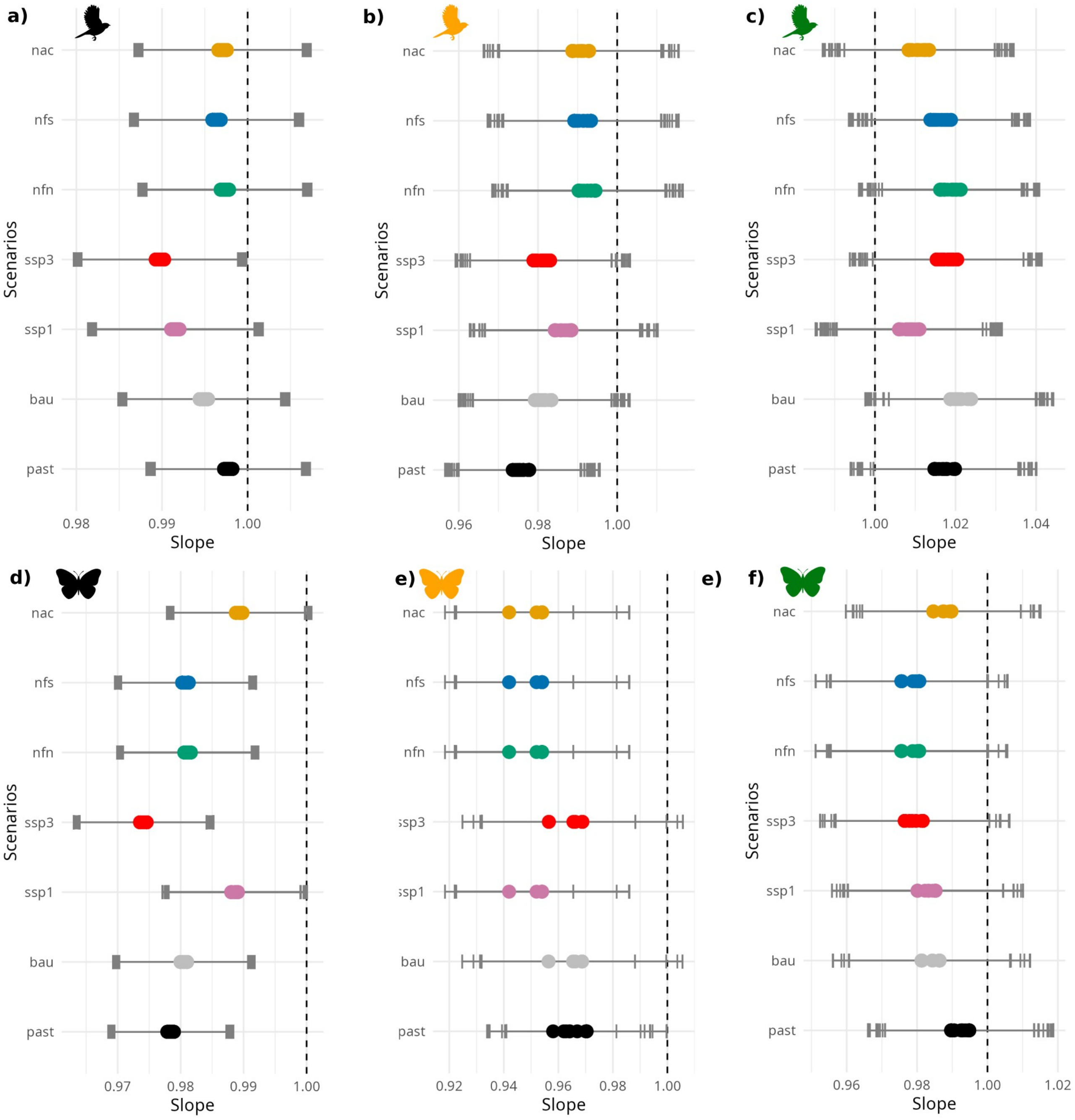

Figure S1u: Cross-validation of the geometric mean by leave-one-out species removal. Species temporal trend calculated with significant and non-significant parameters between 2000 and 2021 (past) and in the different scenarios including SSP3, for a) All Birds index, b) Farmland Bird index, c) Forest Bird index, d) All Butterfly index, e) Grassland Butterfly index and d) Woodland Butterfly index. 95% confidence intervals from geometric standard error.

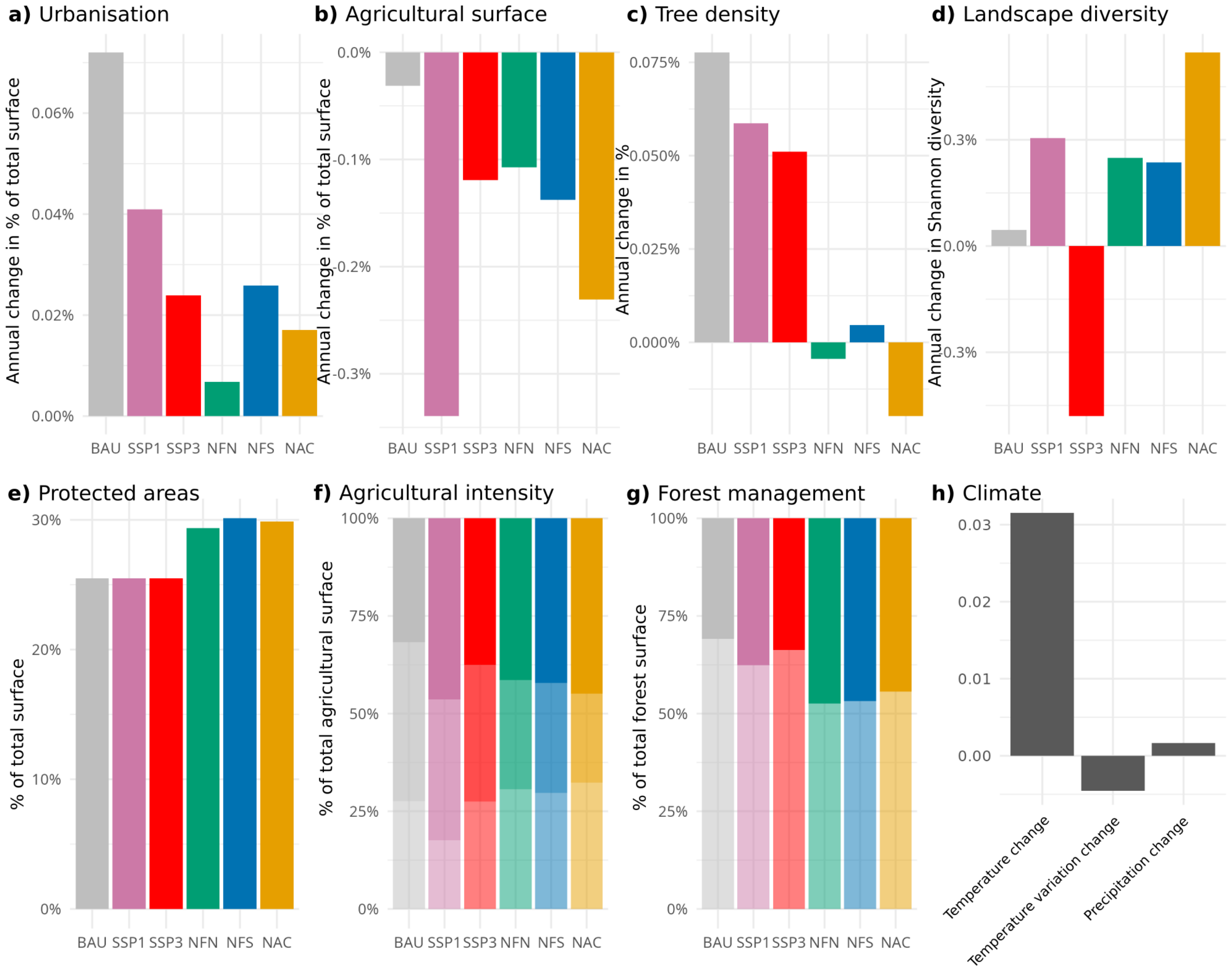

Figure S1v: Predicted changes in SSP3 compared to other scenarios at the European scale in land-use: a) urbanisation, b) agricultural surface, c) tree density, d) landscape diversity; in land-use intensity: e) protected areas, f) production intensity in agricultural areas (low intensive in full colour, medium intensive in intermediate colour and high intensive in shaded colour), g) production intensity in forest areas (low/medium intensive in full colour, high intensive in shaded colour), and h) in climate (temperature change in °C.year$^{-1}$, and precipitation change in mm.year$^{-1}$).

Assessing trend differences between habitat specialist and generalist species

We used the species generalism index (SGIc) provided by Godet et al.[3] for bird species to segregate between habitat specialist and generalist species while correcting for species range. We separate the most generalist species from the most specialist species using the median computed using the 195 available species (36.5). To compute the multi-species indices, species with a SGIc above the median are considered as generalists (Fig. S1w a-c) and species with a SGIc below the median are considered as specialists (Fig. S1x a-c). For butterfly species, we used the habitat specialisation index provided by Stefanescu et al (2011)[4]. This index segregates the species among four classes of habitat specialisation. We calculate the multi-species indices for common butterflies, grassland birds and woodland birds for the more generalist species belonging to the two last classes (Fig. S1w) and the more specialist species belonging to the two first classes of specialisation (Fig. S1x).

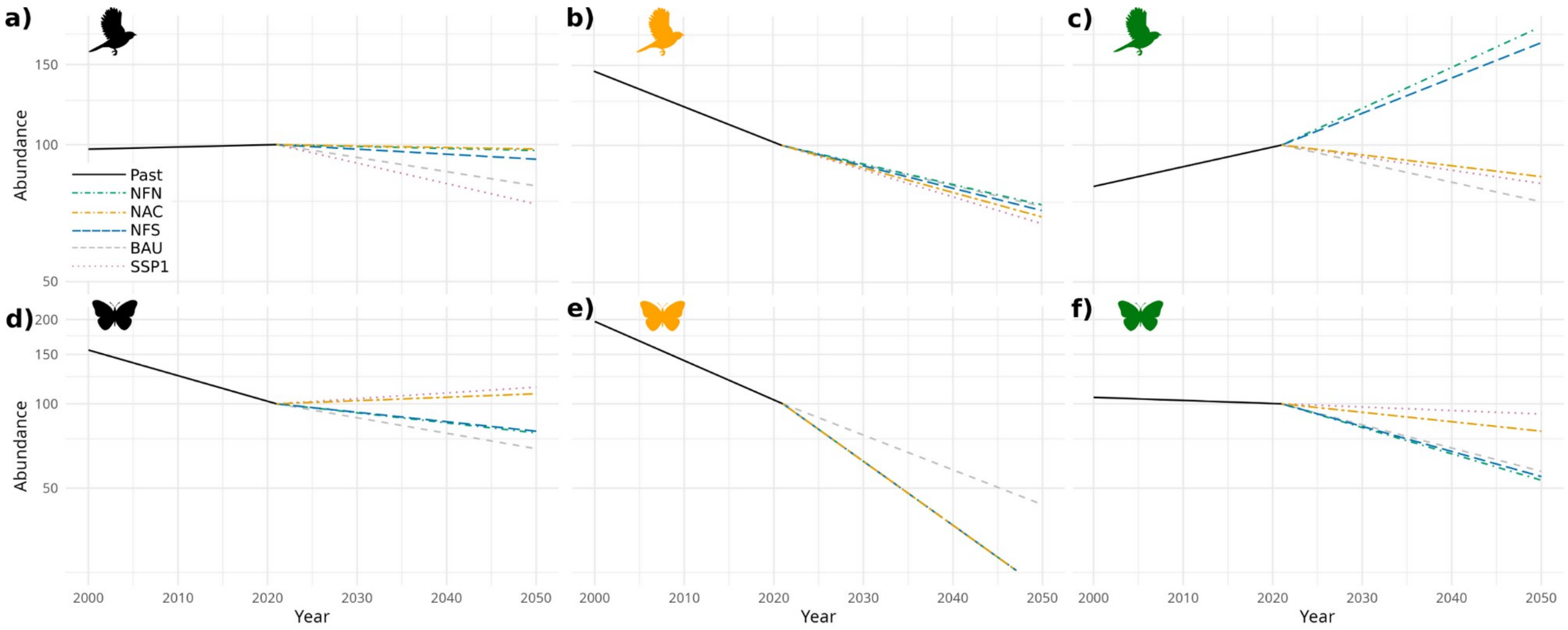

Figure S1w: Predicted abundance trends of habitat generalist species for a) all common birds (ABiI) (n = 38), b) farmland species (FaBiI) (n = 11), c) forest species (FoBiI) (n = 4), d) all common butterflies (ABuI) (n = 60), e) grassland species (GBuI) (n = 7) and f) woodland species (WBuI) (n = 14). Observed trend (2000-2021) is displayed in black, Business As Usual (BAU) in dotted grey, the “Taking the Green Road” Shared Socioeconomic Pathway (SSP1) in purple, Nature For Nature (NFN) in green, Nature For Society (NFS) in blue and Nature As Culture (NAC) in yellow (see geometric standard errors in Extended Fig. 1). We use 2021 as a pivotal year, i.e. a reference to express the relative population abundance with a value of 100 in 2021.

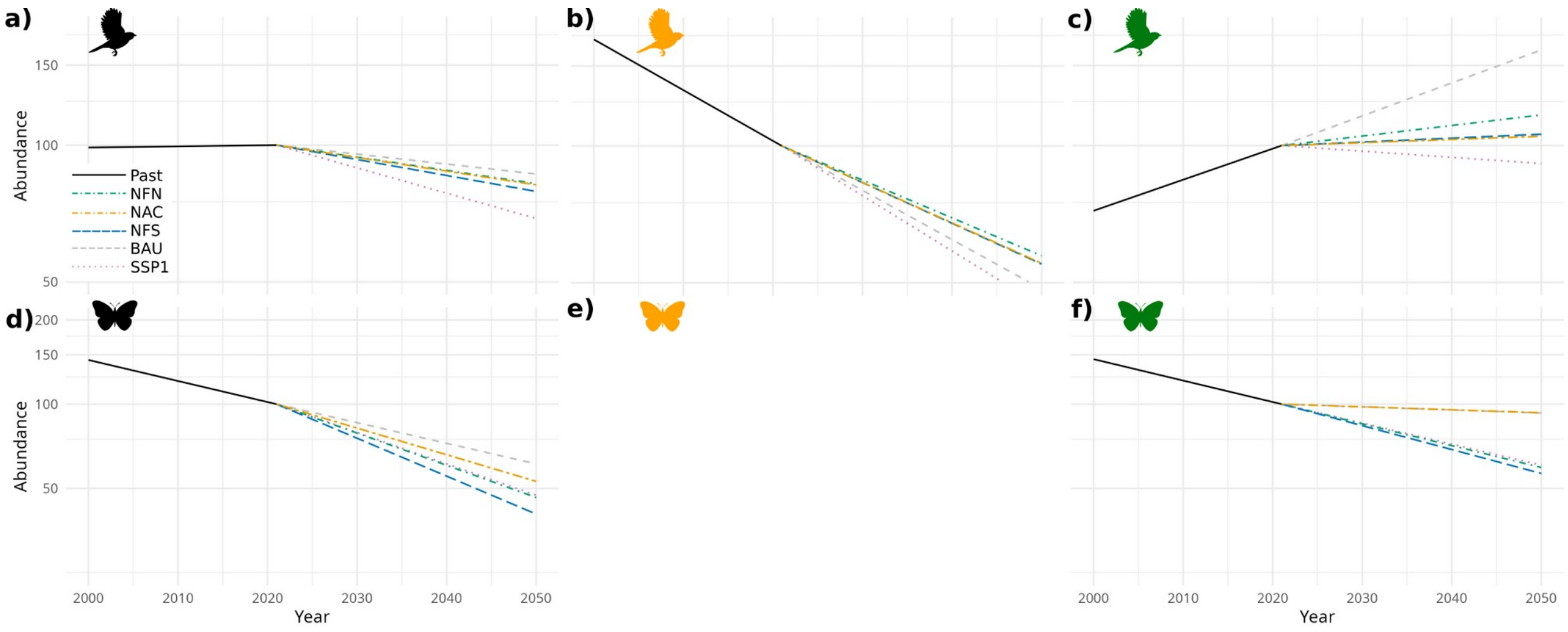

Figure S1x: Predicted abundance trends of habitat specialist species for a) all common birds (ABiI) (n = 78), b) farmland species (FaBiI) (n = 18), c) forest species (FoBiI) (n = 18), d) all common butterflies (ABuI) (n = 35), e) grassland species (GBuI) (n = 1, not represented) and f) woodland species (WBuI) (n = 8). Observed trend (2000-2021) is displayed in black, Business As Usual (BAU) in dotted grey, the “Taking the Green Road” Shared Socioeconomic Pathway (SSP1) in purple, Nature For Nature (NFN) in green, Nature For Society (NFS) in blue and Nature As Culture (NAC) in yellow (see geometric standard errors in Extended Fig. 1). We use 2021 as a pivotal year, i.e. a reference to express the relative population abundance with a value of 100 in 2021.

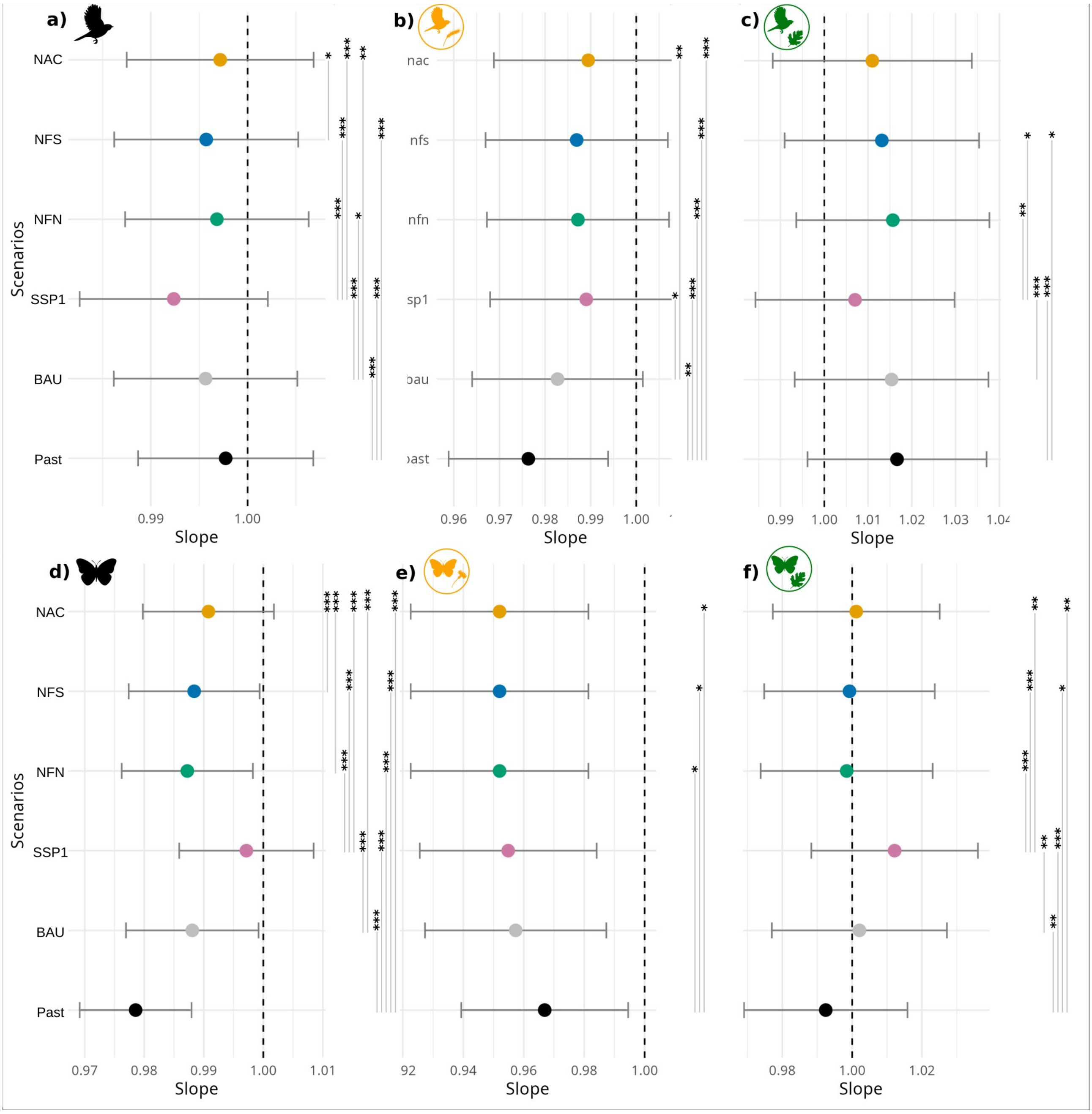

Figure S1y: Geometric mean of species temporal trend calculated with significant parameters only between 2000 and 2021 (past) and in the different scenarios (BAU, SSP1, NFN, NFS, NAC), for a) All Birds index, b) Farmland Bird index, c) Forest Bird index, d) All Butterfly index, e) Grassland Butterfly index and d) Woodland Butterfly index. 95% confidence intervals from geometric standard error and the significant differences between scenarios are displayed (* p-value < 0.05, ** p-value < 0.01, *** p-value < 0.001).

**Supplementary material 2 Species lists**

See csv files SM2_bird_table.csv and SM2_butterfly_table.csv

It should be noted that four butterfly species were not included in the species list, even though they are part of the EU butterfly indicators, as they were observed in fewer than 1% of the transects: Dusky large blue (*Phengaris nausithous*, Grassland Butterfly Indicator) recorded in 0.9% of the transects, Scarce heath (*Coenonympha hero*, Woodland Butterfly Indicator) recorded in 0.7% of the transects, Scarce fritillary (*Euphydryas maturna*, Woodland Butterfly Indicator) recorded in 0.3% of the transects and Rock grayling (*Hipparchia hermione*, Woodland Butterfly Indicator) recorded in 0.9% of the transects.

###
Explanation of columns:
###

Species: species scientific name
Name: species English name
Habitat.indicator: is the species is listed in Farmland Bird Index (FaBI), Forest Bird Index (FoBI), Grassland Butterfly Index (GBI) or Woodland Butterfly Index (WBI)
All.bird/butterfly.index: is the species is listed in BiI or BuI
Habitat.indicator.used: is the species included in the FaBI, FoBI, GBI or WBI computed in the article
trend_past: additive trend (natural logarithm of the multiplicative trend) between 2000 and 2021 at the European scale

##### Covariates effect on local abundance (beta coefficient of the GAM)

open_vs_forest: effect of open land category versus forest category on local abundance
urban_vs_forest: effect of urban category versus forest category on local abundance
spring_mean_temperature: effect of mean temperature between March and June on local abundance
spring_rainfall: effect of rainfall amount between March and June on local abundance
landscape_diversity: effect of Shannon landscape diversity (based on CORINE classes) on local abundance
primary_productivity: effect of drymatter production (i.e. local primary production) on local abundance

##### Effect of covariate trends on species trend (alpha coefficient of the GAM)

urbanisation_increase_effect_on_trend: change in yearly slope due to increase in urbanisation
treedensity_increase_effect_on_trend: change in yearly slope due to increase in tree density
non_intensively_managed_forest_effect_on_trend: change in yearly slope when amount of non intensively managed forests is high

intensively_managed_forest_effect_on_trend: change in yearly slope when amount of intensively managed forests is high
farmland_increase_effect_on_trend: change in yearly slope due to increase in farmland area
low_intensively_managed_farmland_effect_on_trend: change in yearly slope when amount of low intensively managed farmland is high
high_intensively_managed_farmland_effect_on_trend: change in yearly slope when amount of high intensively managed farmland is high
temperature_increase_effect_on_trend: change in yearly slope due to increase in spring temperature
temperature_variation_increase_effect_on_trend: change in yearly slope due to increase in variance of spring temperature
rainfall_increase_effect_on_trend: change in yearly slope due to increase in spring rainfall
landscape_diversity_increase_effect_on_trend: change in yearly slope due to increase in shannon landscape diversity
protected_area_effect_on_trend: change in yearly slope when amount of protected area is high

##### Standard deviation of the effect of covariate trends on species trend

Same column names with "_sd" suffix.

dev_exp: deviance explained by the model

**Supplementary material 3 Biodiversity data and biophysical regions**

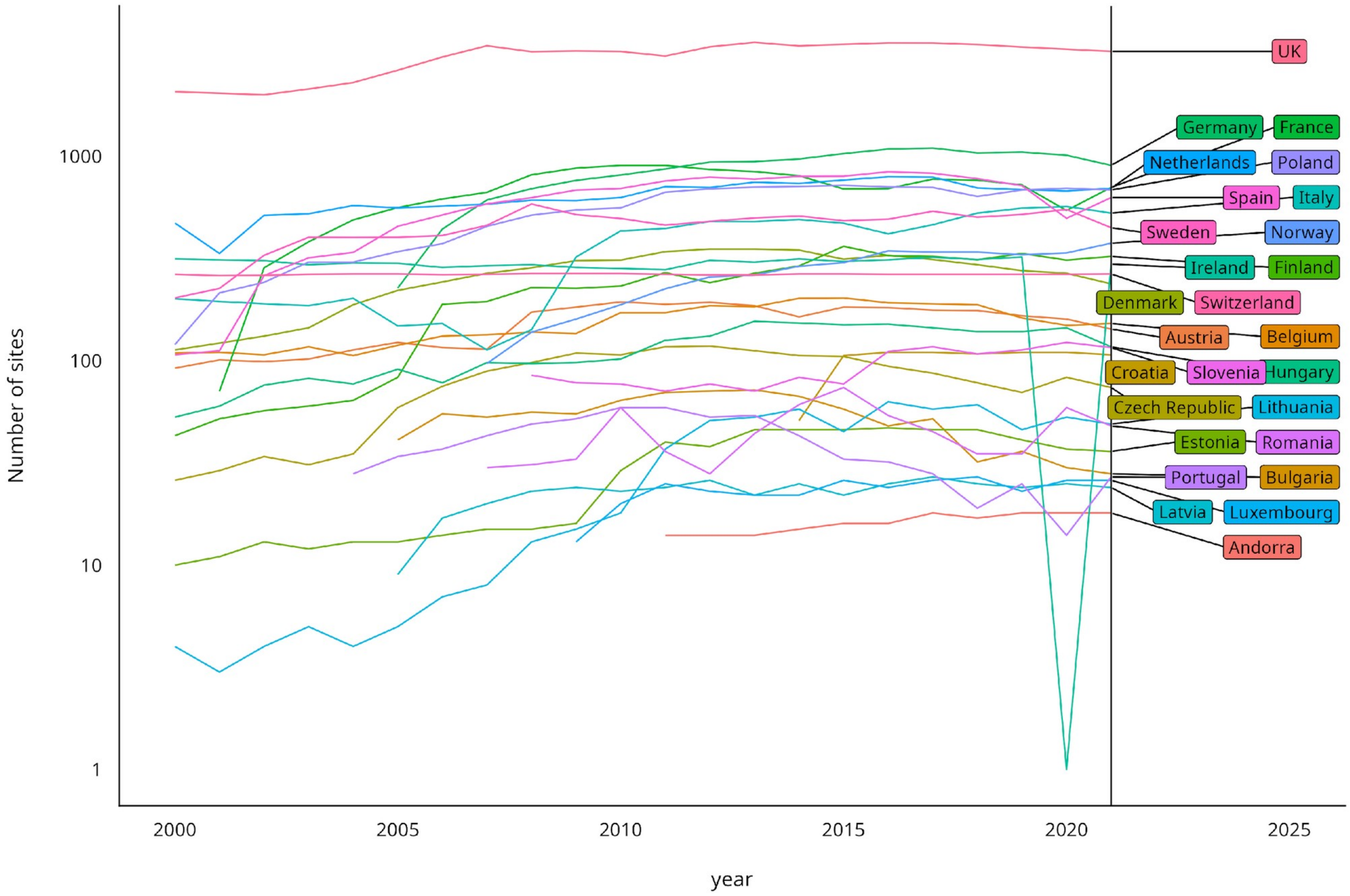

Figure S3a: Number of selected sites/transects monitored for birds by scheme between 2000 and 2021.

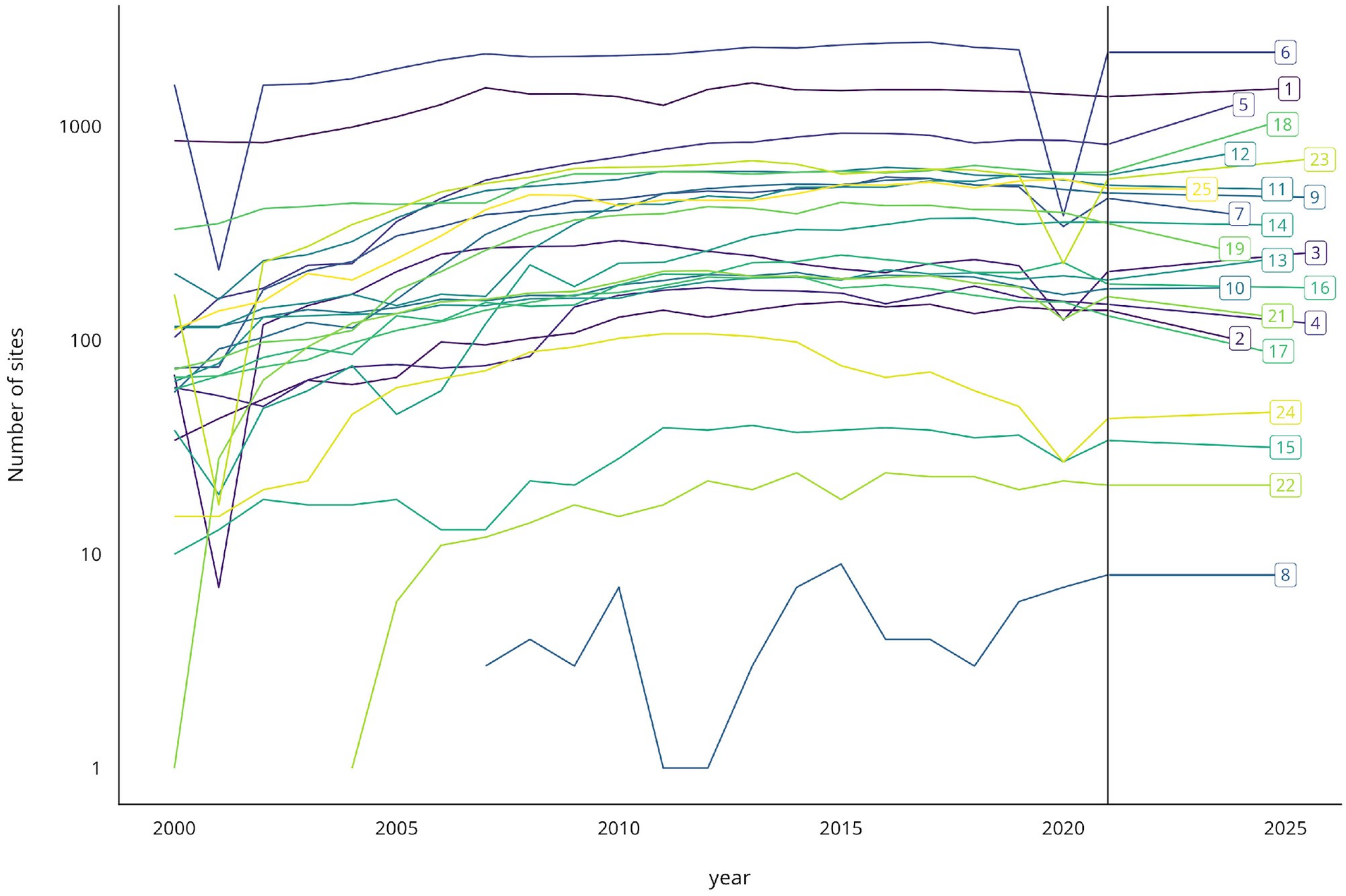

Figure S3b: Number of selected sites/transects monitored for birds by biophysical region between 2000 and 2021. See the map of biophysical regions in Fig. S1f.

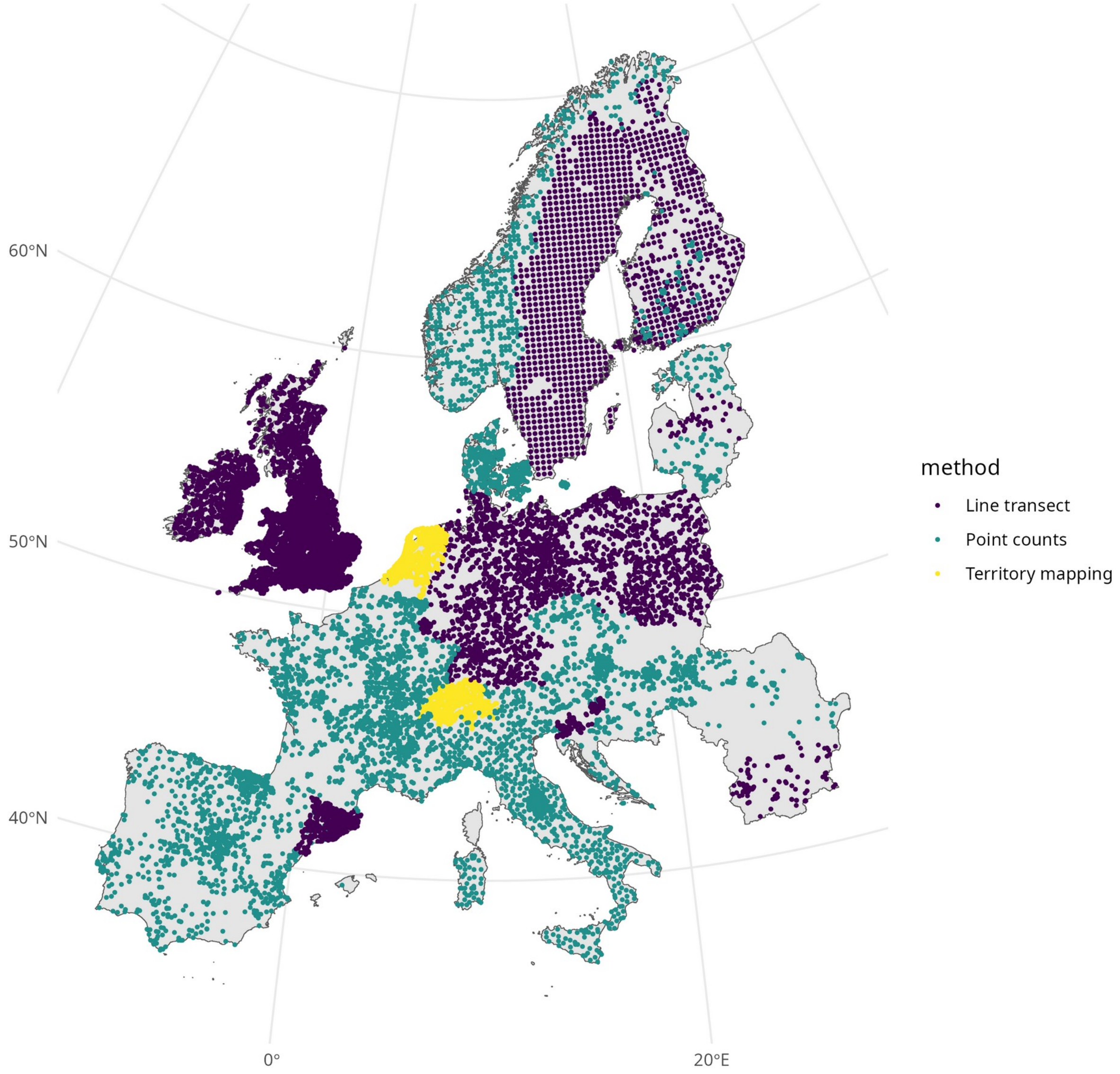

Figure S3c: Selected sites from bird monitoring schemes differentiated by scheme methods. Most countries typically use one scheme, with some exceptions: for Czechia and Latvia, we used respectively point count data and transect count data; for Finland, we included both point count data and line transect data that have a longer temporal and larger spatial coverage; for Spain we included the regional line transect scheme in Catalonia; and for Belgium we used schemes for Brussels and Wallonia.

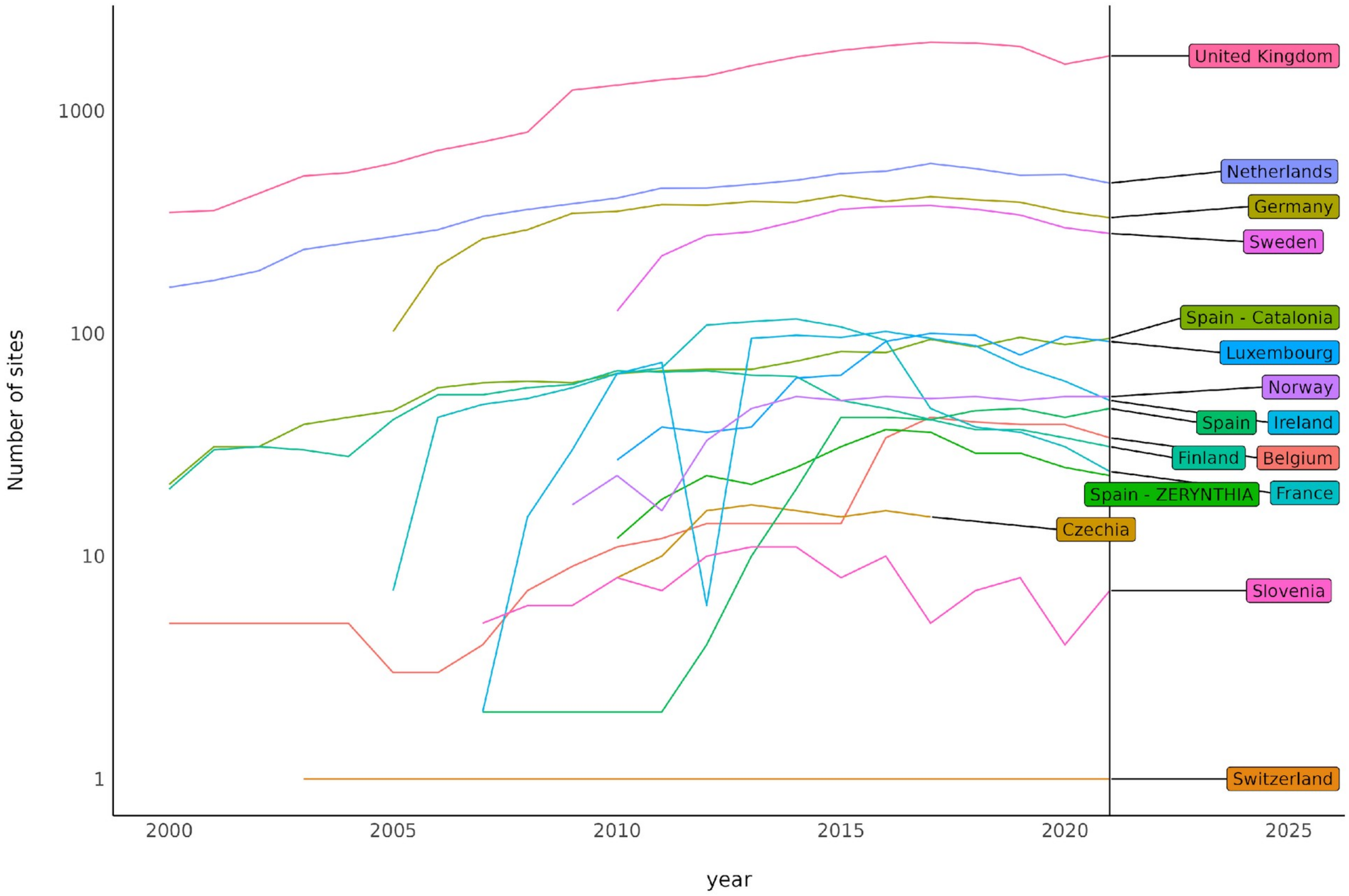

Figure S3d: Number of selected transects monitored for butterflies by scheme between 2000 and 2021.

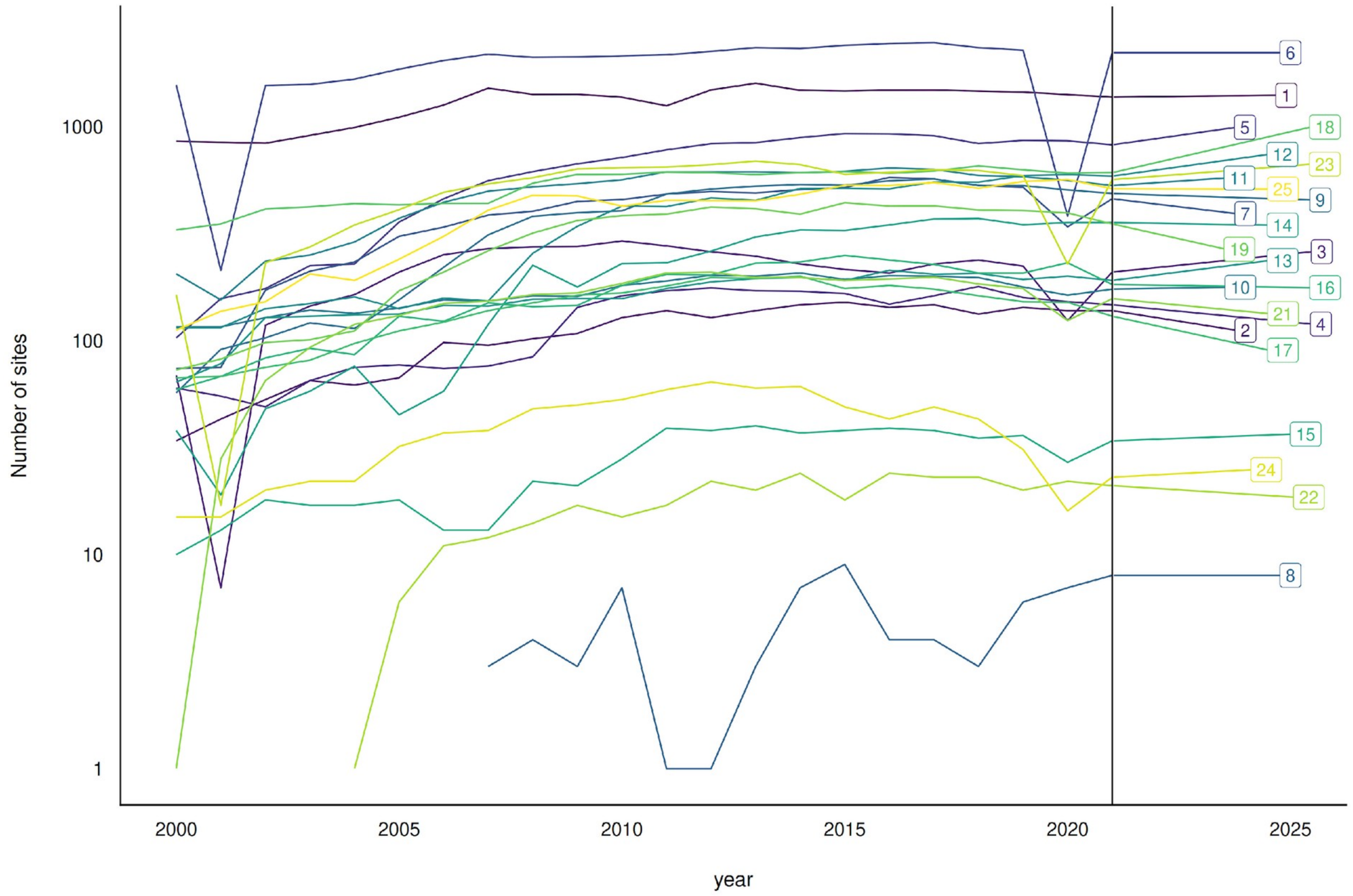

Figure S3e: Number of selected sites/transects monitored for butterflies by biophysical region between 2000 and 2021. See the map of biophysical regions in Fig. S1f.

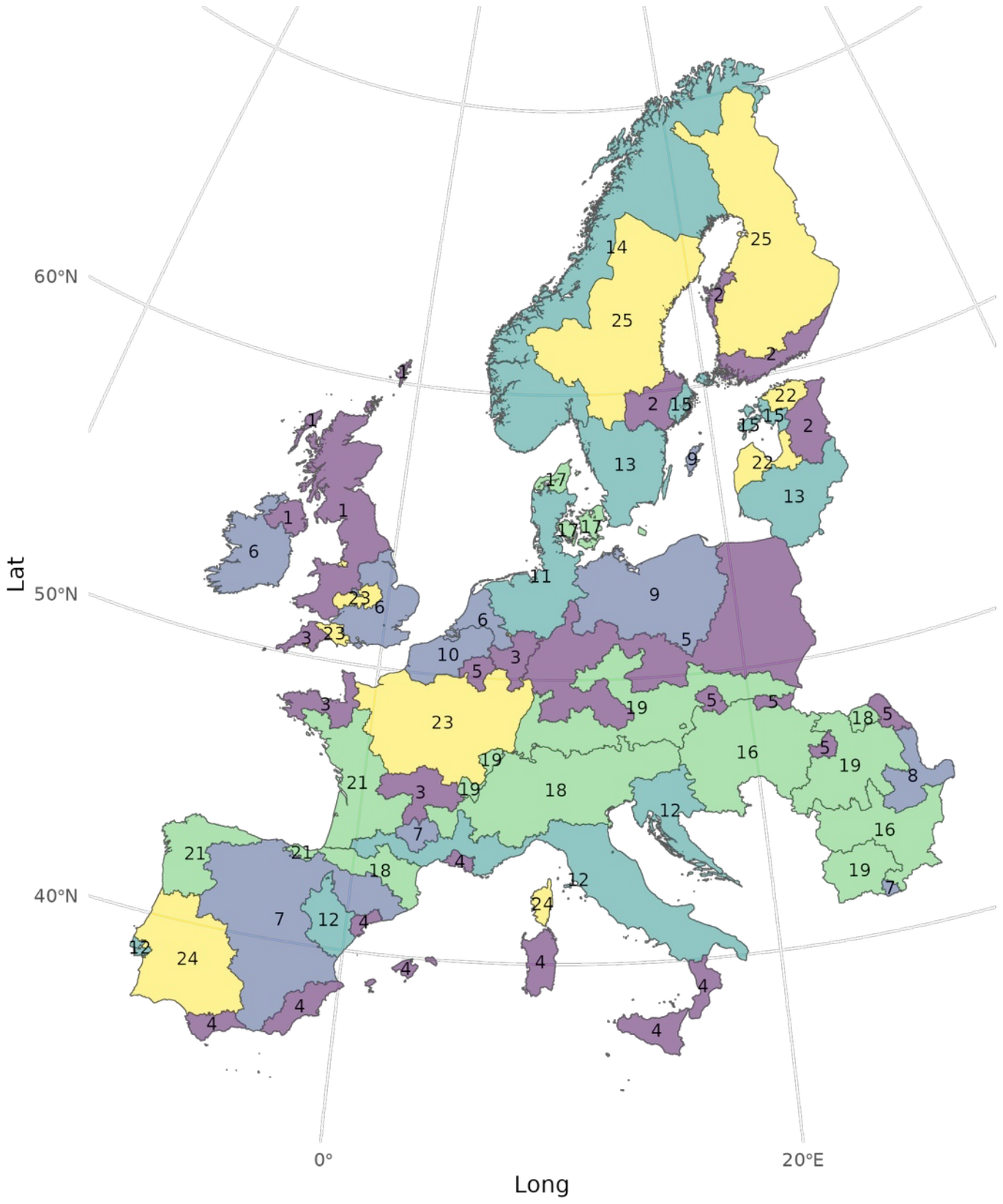

Figure S3f: Map of the 24 biophysical regions in Europe based on the primary landscape structure (climate, topography and soil) from the Spatial Regional Reference Framework[1]. From the original map, we excluded Greece and Malta, as their biophysical regions were not covered by the available biodiversity data and were not spatially continuous with other covered countries. We also included Croatia which was covered by available biodiversity monitoring data. To do so, we divided Croatian NUTS3 regions between the two adjacent biophysical regions, based on their proximity to neighbouring biophysical regions. This resulted in 24 biophysical regions covering the study region.

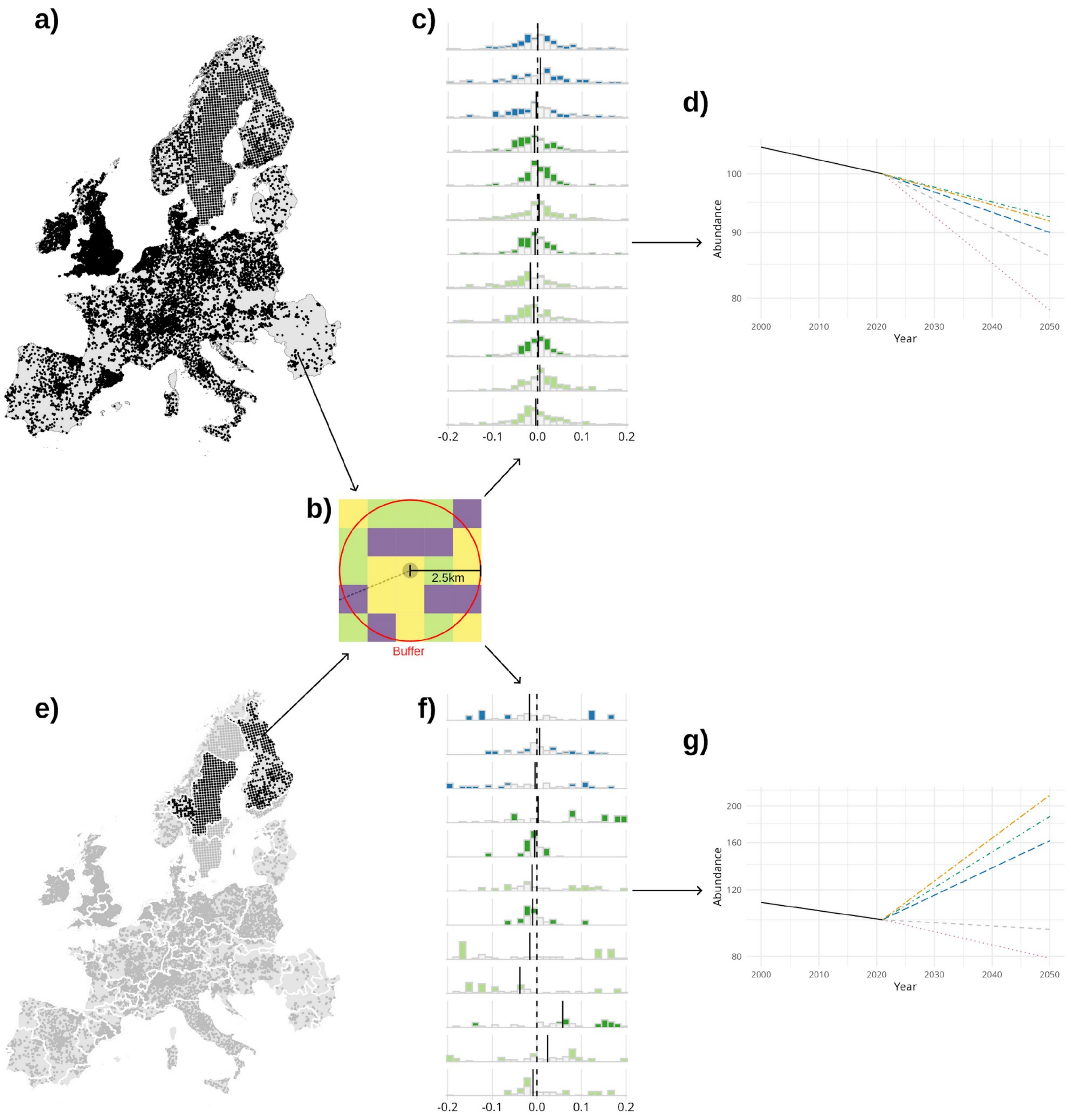

Figure S3g: Workflow across spatial levels. a) Biodiversity data (PECBMS bird data in this example) are analysed with b) environmental and driver data available at a 1 × 1 km average in a 2.5km buffer around each bird monitoring site. c) This analysis provides an estimate for the effect of each driver on each species at the European level, which is used together with driver scenarios at the European level to produce d) scenarios of bird relative abundance. e-g) A similar framework is applied for each biophysical region (region 25 in this example).

**Supplementary material 4 Model check**

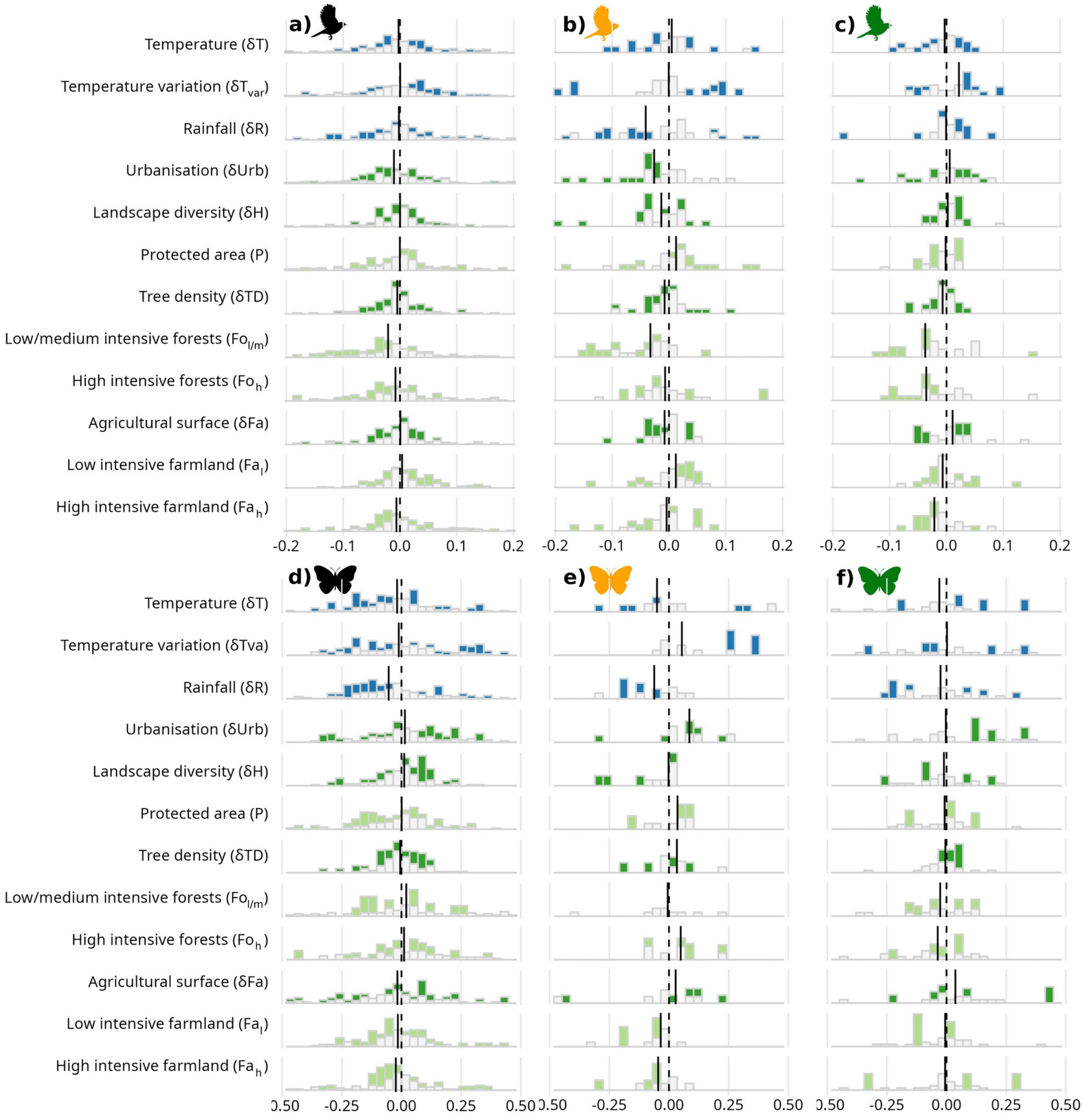

Figure S4a: Model estimate with driver estimated in a buffer of 5-km radius. Effects of climate variables (in blue), land-use variables (in dark green) and land-use intensity variables (in light green) on temporal trends at the European level for all bird species (n = 223), b) farmland bird species (n =  34), c) forest bird species (n = 29), d) all butterfly species (n = 135), e) grassland butterfly species (n =  12) and f) woodland butterfly species (n = 28).

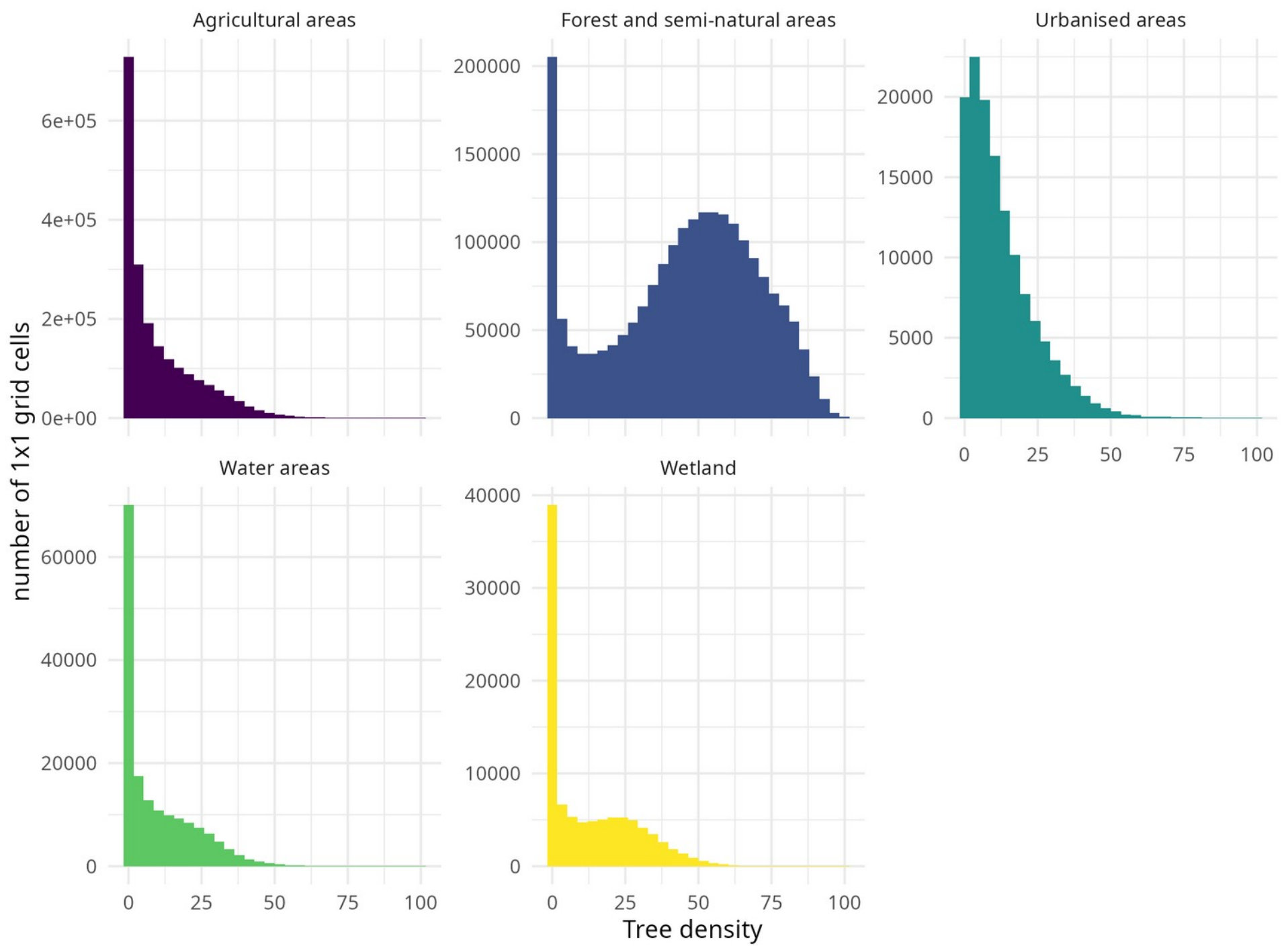

Figure S4b: Distribution of 1 × 1 km grid cells according to tree density (year 2012) from High Resolution Layer and CORINE Land-cover first order classification (year 2000).

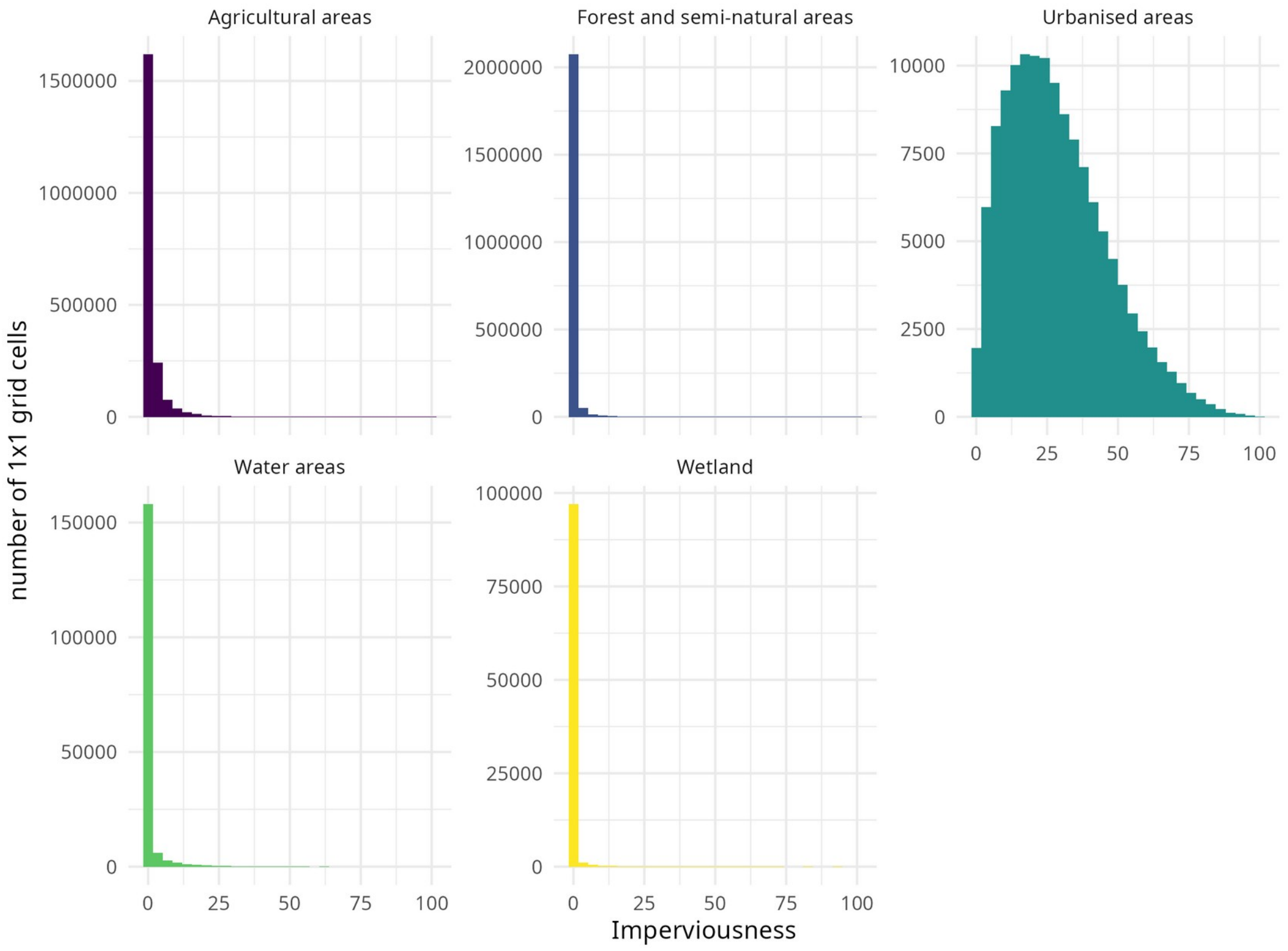

Figure S4c: Distribution of 1 × 1 km grid cells according to imperviousness (year 2006) from High Resolution Layer and CORINE Land-cover first order classification (year 2000).

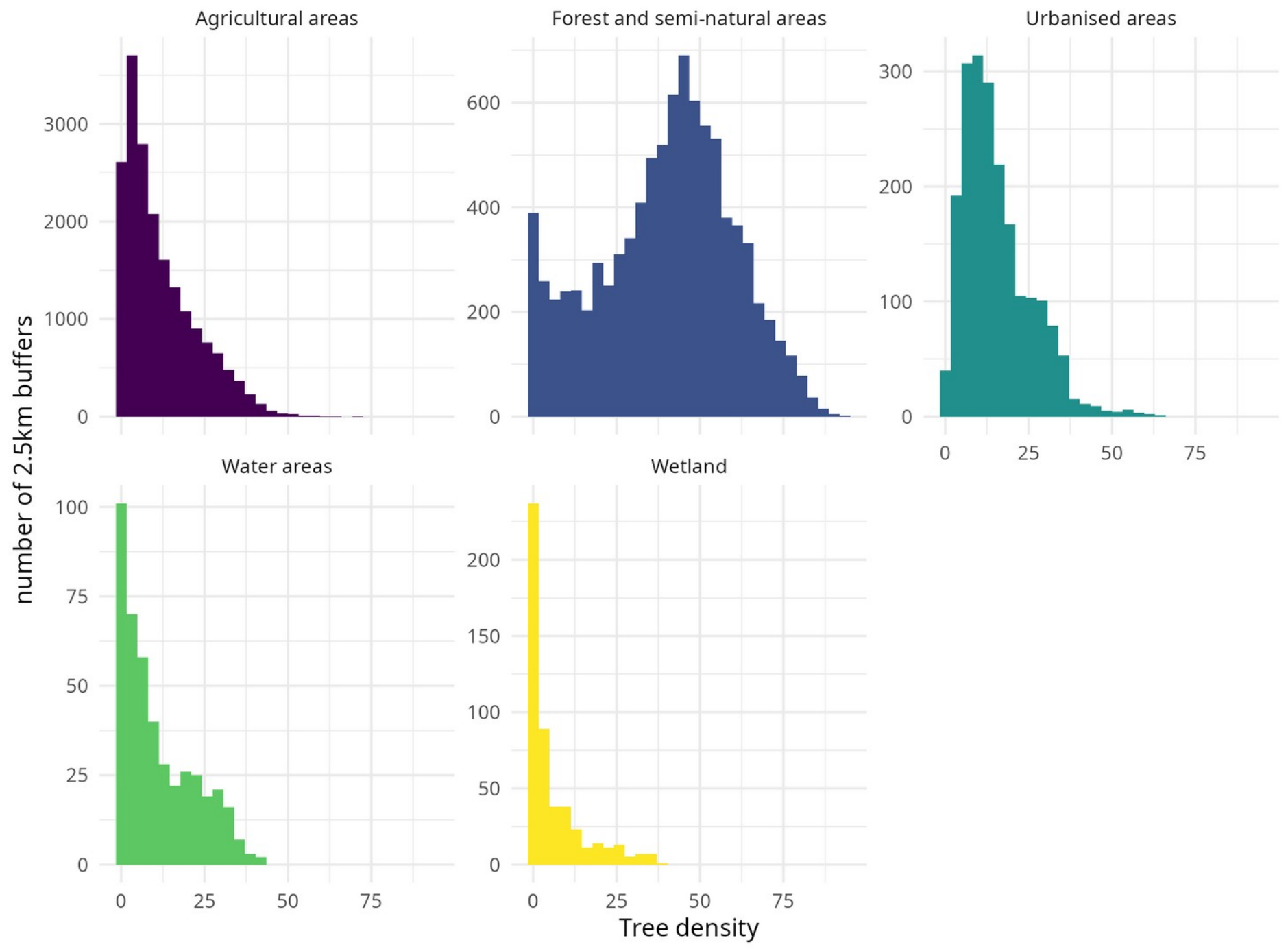

Figure S4d: Distribution of values in 2.5 km buffers around monitoring sites according to tree density (year 2012) from High Resolution Layer and CORINE Land-cover first order classification (year 2000).

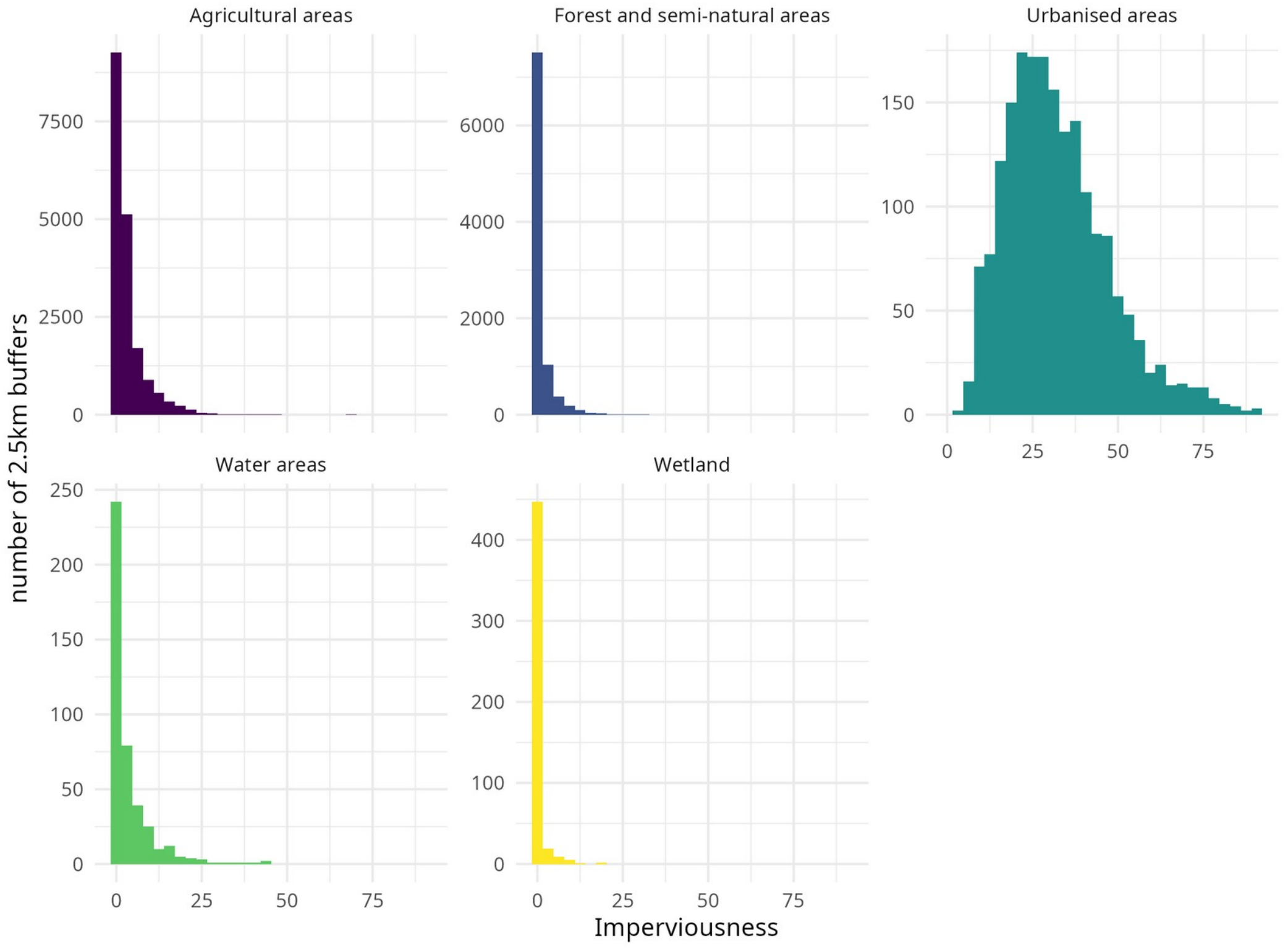

Figure S4e: Distribution of values in 2.5 km buffers around monitoring sites according to imperviousness (year 2006) from High Resolution Layer and CORINE Land-cover first order classification (year 2000).

| Type | Name | Spatial scale | Frequency | Temporal coverage | Source |
|---|---|---|---|---|---|
| Land-cover | Imperviousness density | 100 × 100 m | Every 3 years | 2006-2018 | CLMS portfolio |
| | Tree cover | 100 × 100 m | Every 3 years | 2012-2018 | CLMS portfolio |
| | CORINE acconting layer | 100 × 100 m | Every 6 years | 2000-2018 | EEA Datahub https://www.eea.europa.eu/en/datahub/datahubitem-view/a55d9224-a326-4cb1-9b9c-3a324520341a |
| Land-use intensity | European Land System map | 1 × 1 km | 1 snapshot | 2015 | https://doi.org/10.34894/XNC5KA |
| | World Database on Protected Areas | Vector | Updated yearly | | http://protectedplanet.net/ |
| Productivity | Dry matter productivity | 1 × 1 km | Every 10 days | 1999-2020 | CLMS portfolio |
| Climate | Mean temperature<br>Min temperature<br>Max temperature<br>Sum of precipitations<br>Relative humidity | 0.1° grid cells | Daily | 1950-2023 | E-Obs https://surfobs.climate.copernicus.eu/dataaccess/access_eobs.php |

Table S4: Spatial scale and temporal coverage of data used to assess environment and drivers.

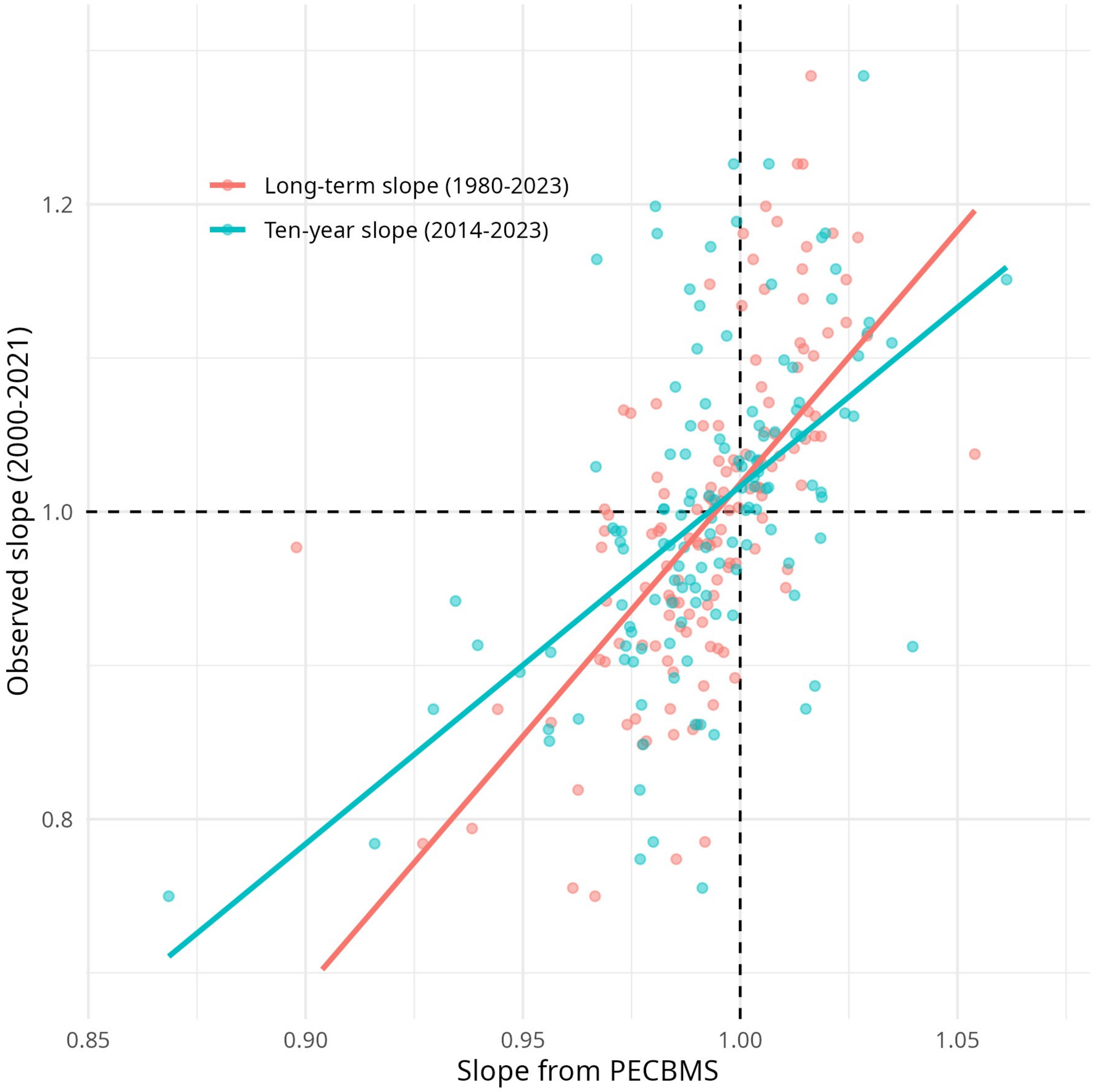

Figure S4f: Comparison between the temporal slope (i.e. multiplicative trend depicting an average percentage change per year) for the 145 bird species calculated by our model on data between 2000 and 2021 with the slope provided by the PanEuropean Common Bird Monitoring Scheme (https://pecbms.info/trends-and-indicators/species-trends/) for long term slope (1980-2023 for most species, see details on PECBMS website) and ten-year slope (2014-2023 for most species, see details on PECBMS website).

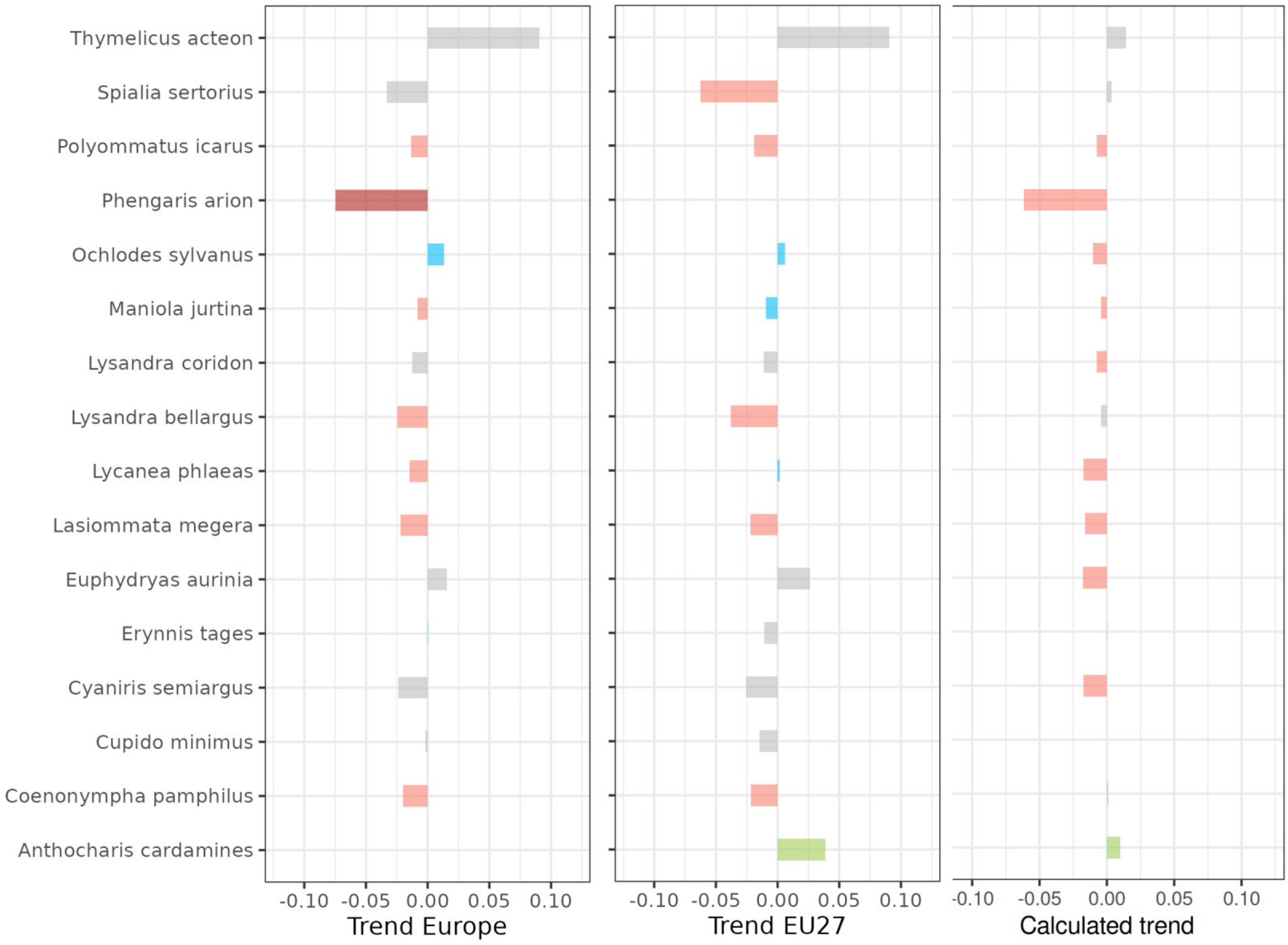

Figure S4g: Comparison between the temporal slope (i.e. multiplicative trend depicting an average percentage change per year) for 16 butterfly species calculated by our model with the slope provided by the *European Grassland Butterfly Indicator 1990-2020: Technical Report* (see this report for the detail of periods)[2]. Dark red indicate strong decline, light red indicates decline (moderate in Trend Europe and Trend EU27, significant in Calculated trend), green indicates increase (moderate in Trend Europe and Trend EU27, significant in Calculated trend), grey indicates uncertain trends (uncertain in Trend Europe and Trend EU27, non-significant in Calculated trend) and blue indicates stable trends.

See S4h_bird_model_check.pdf for model check for linearity, response distribution and spatial autocorrelation of residuals for birds at the European scale.

See S4i_butterfly_model_check.pdf for model check for linearity, response distribution and spatial autocorrelation of residuals for butterflies at the European scale.

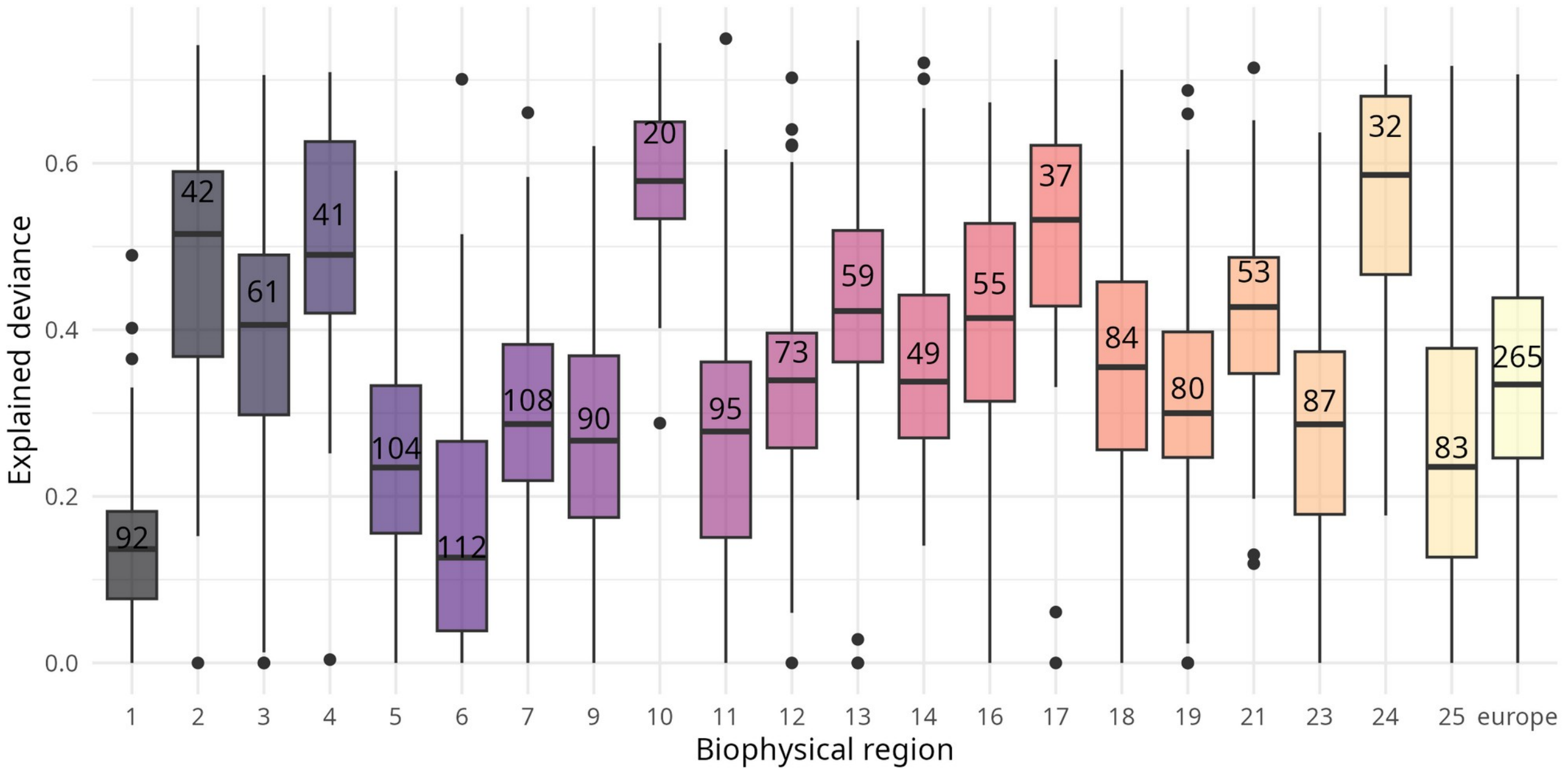

Figure S4j: Boxplot of the percentage of deviance explained for the 265 bird species per biophysical region (the number of species per biophysical region is indicated above the median) and at the European scale. Biophysical regions with not enough data to run the model are not represented.

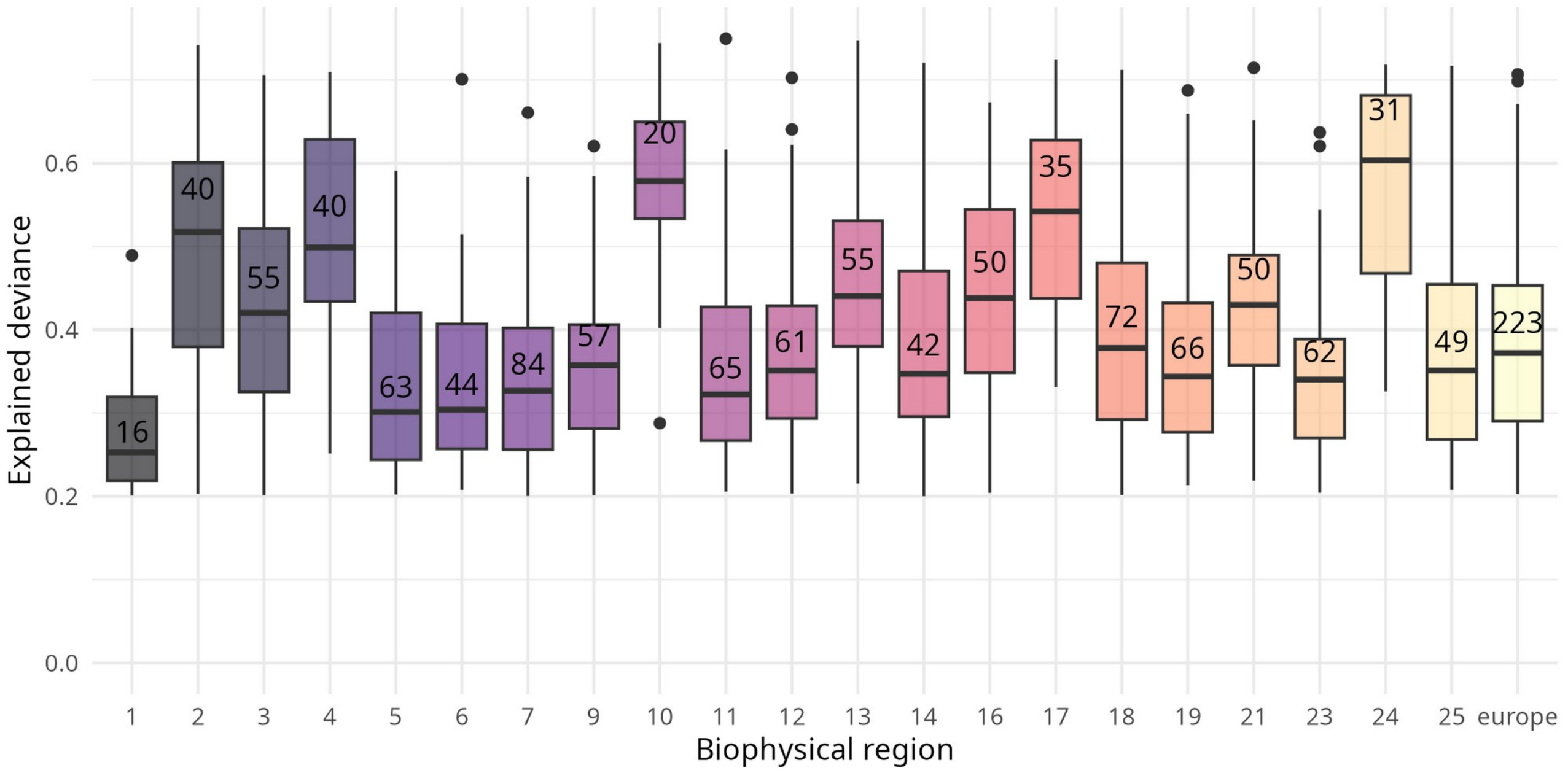

Figure S4k: Boxplot of the percentage of deviance explained after the selection step (above 20% of explained deviance) for bird species per biophysical region (the number of species is indicated above the median) and at the European scale. Biophysical regions with not enough data to run the model are not represented.

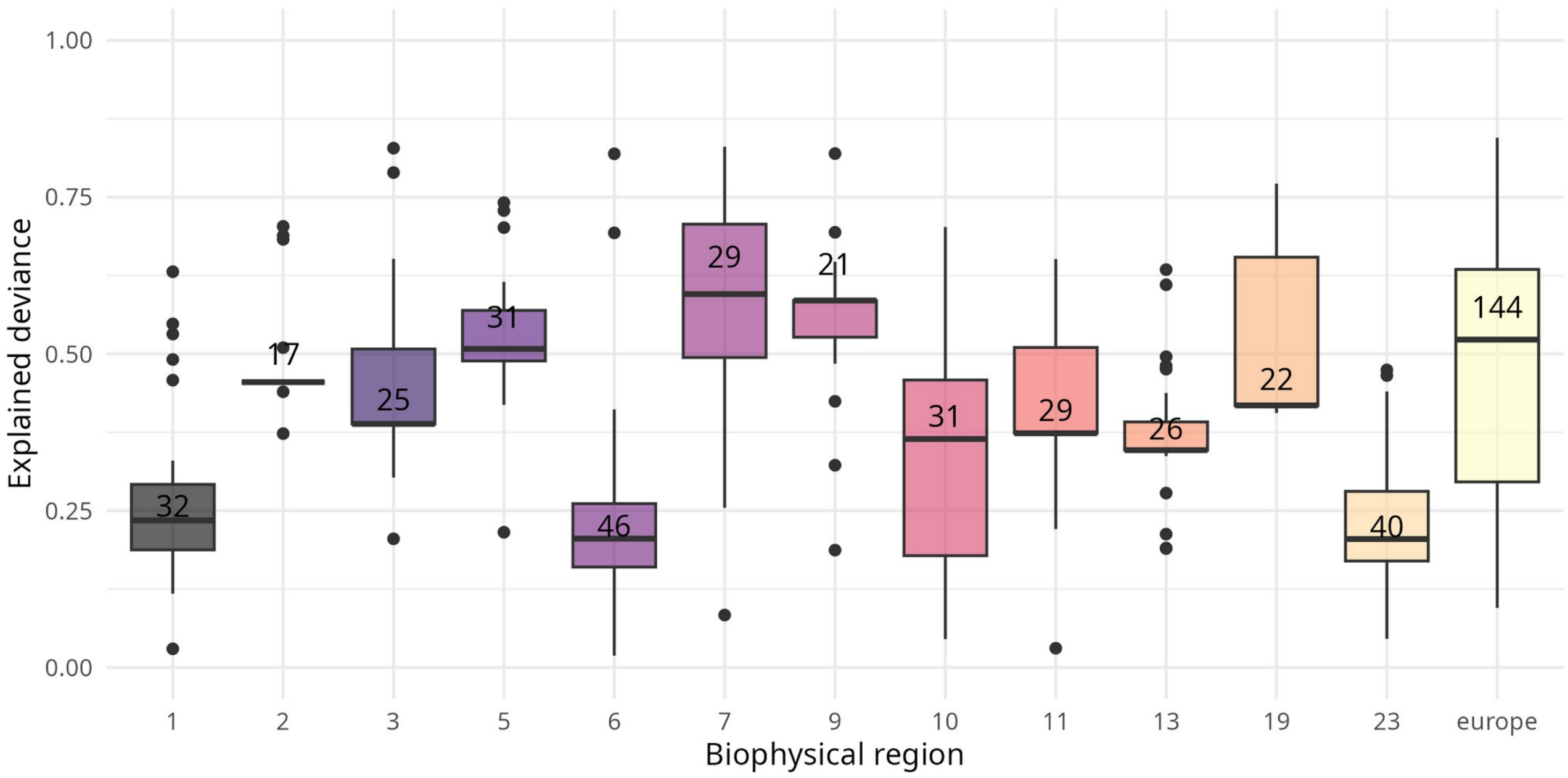

Figure S4l: Boxplot of the percentage of deviance explained for the 144 butterfly species per biophysical region (the number of species per biophysical region is indicated above the median) and at the European scale. Biophysical regions with not enough data to run the model are not represented.

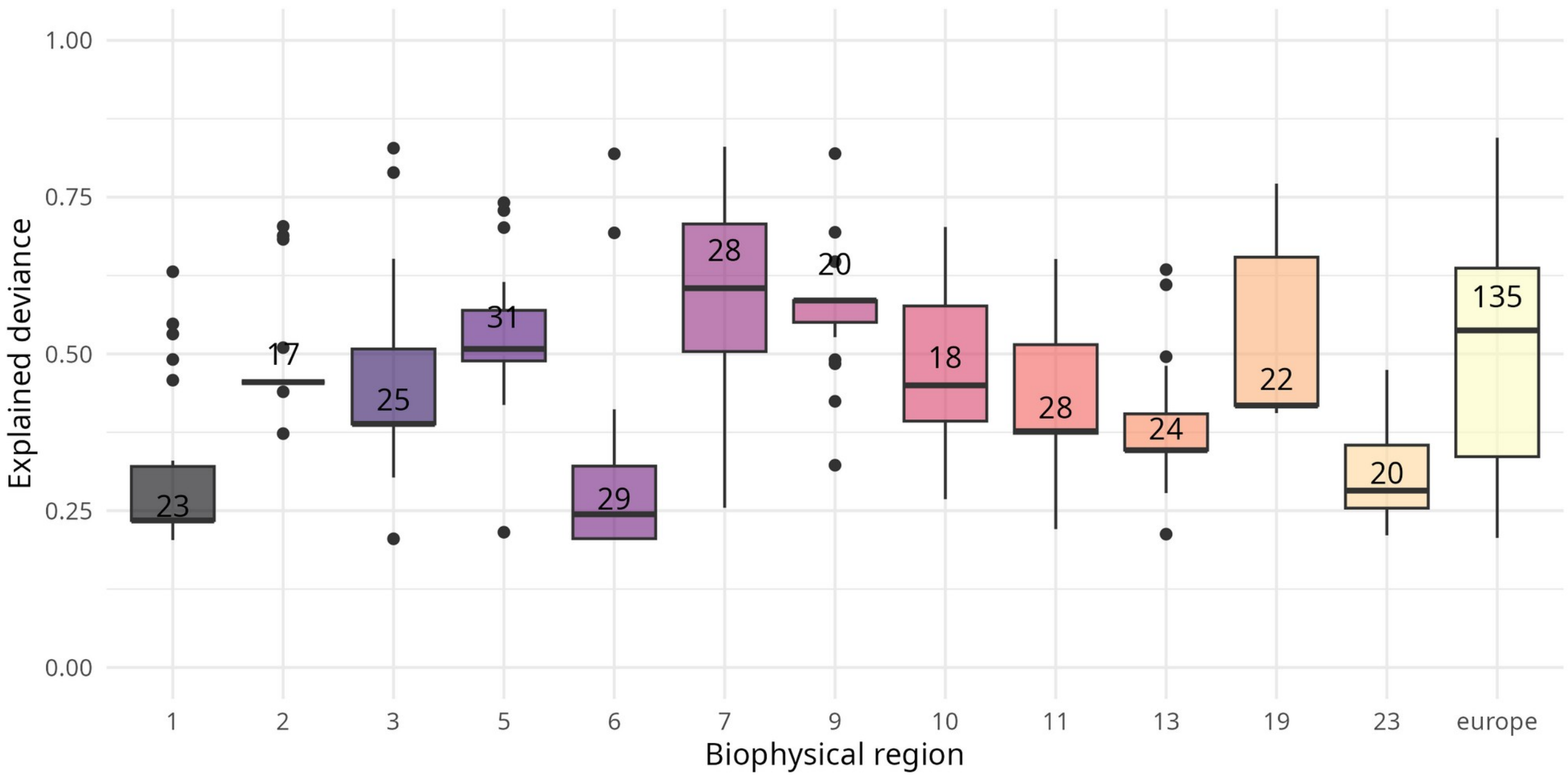

Figure S4m: Boxplot of the percentage of deviance explained after the selection step (above 15% of explained deviance) for butterfly species per biophysical region (the number of species is indicated above the median) and at the European scale. Biophysical regions with not enough data to run the model are not represented.

Supplementary references: